\documentclass[]{jfm}

\usepackage[utf8]{inputenc}			% allows direct input of ä,ö,ü,ß
\usepackage[T1]{fontenc}			% better font encoding
\usepackage[british]{babel}	        % british hyphenation, chapter names, ...
\usepackage{amsmath}				% mathematical environment
\usepackage{amssymb}				% mathematical symbols
\usepackage{mathtools}				% allow use of \mathclap macro
\usepackage{bm}						% allow writing bold mathematical symbols; see also: https://tex.stackexchange.com/questions/3238/bm-package-versus-boldsymbol
\usepackage{graphicx}				% allows including of graphics
\usepackage{comment} 				% allows comments over several lines
\usepackage{array}					% allow, amongst other things, vertical alignment in tables by m{...}
\usepackage{tabularx}				% allow for better tables
\usepackage[english=british]{csquotes} % british style of quotes: enquote, foreignquote, blockquote
\usepackage{gensymb}        % Degree Symbol
\usepackage{soul}           % strikethrough-text

\usepackage{hyperref}		% make cross references clickable
\hypersetup{hidelinks} 		% make boxes around links invisible
\hypersetup{
    colorlinks = true,
    urlcolor   = blue,
    citecolor  = black,
}

\usepackage{ragged2e}
\usepackage{newtxtext}
\usepackage{newtxmath}
\usepackage{natbib}

\usepackage{multirow}
\usepackage{booktabs}

\newcommand{\RomanNumeralCaps}[1]
\nolinenumbers

\newcommand{\rev}[1]{\textcolor{black}{#1}}

\renewcommand{\Pr}		{Pr} % overwrite command from amsmath
\newcommand{\Ra}   		{Ra}
\newcommand{\St}   		{St}
\newcommand{\Nu}   		{Nu}
\renewcommand{\Re}		{Re} % overwrite command from amsmath
\newcommand{\tauf}   	{\tau_f} % free-fall time scale
\renewcommand{\Gamma}   {\varGamma} % Gamma by default in italics
\renewcommand{\Phi}     {\varPhi}   % Phi by default in italics
\renewcommand{\Theta}   {\varTheta} % Theta by default in italics

\newcommand{\MSD}       {\mathrm{MSD}}
\newcommand{\RSS}       {\mathrm{RSS}}
\newcommand{\NuL}   	{Nu^L}
\newcommand{\NuVt}   	{\left \langle Nu \right \rangle_{V,t}}
\newcommand{\Rlambda}   {R_{\lambda}}
\newcommand{\accvar}    {\langle a^2 \rangle}
\newcommand{\epsnu}     {\langle \epsilon_{\nu} \rangle}
\newcommand{\epsT}      {\langle \epsilon_{T} \rangle}
\newcommand{\Pan}       {P_{a_n}}
\newcommand{\Bz}        {|B_z|}
\newcommand{\nsnp}      {n_{\mathrm{snp}}}
\newcommand{\rfac}      {r_{\mathrm{fac}}}
\newcommand{\JcD}        {\langle J_c \rangle_D}
\newcommand{\Jc}        {\langle | J_c | \rangle}

\usepackage[dvipsnames]{xcolor}
\usepackage{url}

\begin{document}

\title{Lagrangian single-particle, multi-particle and topological analyses in turbulent Rayleigh--Bénard convection}

\author{
Matti Ettel\aff{1},
Roshan J. Samuel\aff{1},
Michael Chertkov\aff{2}
 and
Jörg Schumacher\aff{1}
}
\affiliation{
\aff{1}Institute of Thermodynamics and Fluid Mechanics, Technische Universität Ilmenau, Postfach 100565, D-98684 Ilmenau, Germany.
\aff{2}Program in Applied Mathematics \& Department of Mathematics, University of Arizona, Tucson, AZ 87521, USA.} 

\date{\today}

\maketitle

\begin{abstract}
We present three-dimensional direct numerical simulations of turbulent Rayleigh--B\'enard convection in the Lagrangian frame \rev{for $\Pr=0.7$, spanning} $10^5\leq \Ra \leq10^{10}$ in a \rev{laterally periodic domain of} aspect ratio \rev{$\Gamma=4$, complemented by dense multi-particle cloud simulations for $10^4 \leq \Ra \leq 10^8$ at $\Gamma=16$. Using single-particle} accelerations, Lagrangian heat transfer, $Q$--$R$ invariant topology, pair dispersion, scale-dependent Lagrangian eddy viscosity, and principal component analysis of particle clouds, we \rev{establish a hierarchical, scale-bridging framework of convective transport. Controlling spectral-element signatures via cubic smoothing splines yields highly intermittent distributions and viscous Heisenberg-Yaglom scaling in the core, while a mixed viscous-thermal model reveals that thermal dissipation significantly drives vertical bulk and horizontal boundary-layer accelerations. Evaluating exact local temperature gradients along tracer trajectories reveals extreme intermittency, where particles carry convective heat fluxes up to $250$ times the global Eulerian mean. Dust-devil-like convective vortices, isolated via an ANOVA-validated filtering algorithm at $\Ra=10^{10}$, dominate the vortex-stretching quadrant of the $Q$--$R$ invariant map and drive local heat transfer up to $30$ times the global Eulerian mean. Our results do not support a Bolgiano-Obukhov (BO59) cascade in the Lagrangian frame. Complementing pair dispersion with scale-dependent Lagrangian eddy viscosity isolates initial-separation and viscous catch-up effects at the viscous-to-inertial crossover, revealing an extended K41--Richardson shoulder and bulk behaviour increasingly reminiscent of homogeneous isotropic turbulence with rising $\Ra$. Analysing particle clouds via logarithmic shape invariants, mean squared dispersion, and the logarithmic geometric mean radius extends this picture to three-dimensional Richardson-like volumetric expansion, showing that plume entrainment induces uniaxial stretching that reduces kinematic catch-up rather than sustaining a BO59 cascade.}
\end{abstract}

\clearpage

%------------------------------------------------------------------------------------------
\section{Introduction}
\label{sec:Introduction}

Thermal convection is the primary driver of turbulence in various natural systems ranging from atmospheric cloud formation \citep{Mapes1993}, through planetary atmospheres like on Jupiter \citep{Young2017}, to stellar convection zones \citep{Schumacher2020}. Rayleigh--Bénard convection (RBC) represents the canonical, simplest framework to study these buoyancy-driven flows \citep{Ahlers2009, Chilla2012, Verma2018, Alam2025}. Here, a fluid characterised by the Prandtl number $\Pr$ is confined between two horizontal plates, one uniformly heated from below and the other cooled from above. The resulting thermal instability is quantified by the Rayleigh number $\Ra$, balancing buoyant driving and viscous dissipation \citep{Rayleigh1916}. In contrast to homogeneous isotropic turbulence (HIT), RBC is driven by the temperature field, an active scalar that is intrinsically coupled to the velocity field. By virtue of gravity, the resulting flow is not only anisotropic but also inhomogeneous. As a result, coherent structures emerge from the boundary layers near the walls. Thermal plumes detach from the near-wall layers, form a self-similar network \citep{Samuel2024b, Samuel2025, Shevkar2025b} and self-organize in large-scale circulation (LSC) patterns that sweep across the bulk of the fluid \citep{Ahlers2009, Pandey2018, Stevens2018}. These intricate dynamics complicate the classical framework introduced by \citet{Kolmogorov1941}, in short K41, for shear-driven flows and suggest alternative buoyancy-related scalings.

\rev{\citet{Bolgiano1959} and \citet{Obukhov1959} (BO59) originally introduced their scaling theory for stably stratified turbulence (SST), where buoyancy acts as a sink that removes kinetic energy in the inertial range. Since then, numerous studies have proposed the existence of an analogous scaling regime in unstably stratified RBC \citep{Procaccia1989, Lvov1991, Lvov1992}. However -- contrary to SST -- buoyancy in three-dimensional RBC actively injects energy across the inertial range \citep{Kumar2014, Verma2017}, implying down-scale accumulation of this buoyant energy. As discussed by \citet{Lohse2010}, the geometric sum of this cascading energy transport converges to a scale-independent flux above the viscous bottleneck, fundamentally contradicting the scale-dependent flux proposed in the SST-derived BO59 theory. Additionally, the Bolgiano length scale $L_B$ -- marking the crossover from K41 to buoyancy-driven scaling -- is highly inhomogeneous, attaining small values only in close proximity to the walls due to intense thermal energy dissipation \citep{Calzavarini2002}. In the bulk, it shrinks globally with rising $\Ra$, actively compressing the spatial scale separation in the inertial range \citep{Kunnen2008, Lohse2010}. Consequently, the existence of global BO59-scaling in three-dimensional RBC has been scrutinised \citep{Lohse2010} and ultimately refuted via direct energy-flux spectra analyses \citep{Kumar2014, Verma2017}. Crucially, true BO59-scaling has only been observed in two-dimensional settings \citep{Celani2002, Xie2022, Samuel2024a}, where the energy cascade is inverted \citep{Kraichnan1967}. In this two-dimensional framework, kinetic energy is transported up-scale, while buoyant energy cascades down-scale, aligning with the down-scale sink of SST \citep{Kumar2014, Alam2019}.}

The Eulerian perspective of RBC has been extensively studied over several decades, establishing robust scaling laws for global heat transfer $\Nu(\Ra,\Pr)$ and momentum transport $\Re(\Ra,\Pr)$ \citep{Ahlers2009, Chilla2012, Lohse2024}. However, as turbulence is inherently a transport phenomenon, the dynamics of individual massless tracer particles (which follow the flow perfectly) must be resolved to create a complete picture of turbulent mixing, dispersion of substances and heat exchange processes. This Lagrangian perspective is well-established in HIT \citep{Yeung2002, Toschi2009, Sreenivasan2018}. \rev{Because Lagrangian particles are advected by thermal plumes in RBC where the local Bolgiano length scale is small, strong buoyant accelerations may locally exceed classical shear-driven dispersion rates \citep{Richardson1926}. However, traditional scalar pair dispersion is unable to distinguish between a three-dimensional BO59 cascade, since transient one-dimensional kinematic plume stretching and viscous initial separation (`catch-up') effects as unconditioned global domain averages mask the heavily localised impact of thermal plumes at different scales. This requires a refined and extended analysis which is a central motivation of the following work.} 

The Lagrangian formulation has been especially influential in passive-scalar turbulence, where equal-time Eulerian scalar correlation functions can be represented in terms of the stochastic dynamics of groups of Lagrangian particles. In the Kraichnan-model-type setting, this reformulation made it possible to connect non-Gaussian scalar statistics and anomalous scaling of scalar structure functions to the geometry and statistics of multi-particle configurations \citep{Chertkov1995, Gawedzki1995, Shraiman1998}. This line of work concerns passive-scalar intermittency and should be distinguished from the later tetrad phenomenology of turbulent material deformation. Motivated in part by this Lagrangian viewpoint, \citet{Chertkov1999} proposed a minimal four-particle description (`tetrad', or infinitesimal ellipsoid) of local turbulent deformation in three-dimensional flow, using tetrad shape dynamics as a compact diagnostic of stretching, rotation, and alignment with the local velocity-gradient topology. For the inhomogeneous and anisotropic case of RBC, the corresponding links between particle-pair dispersion, multi-particle cloud geometry, and coherent convective structures remain only partially explored. This will be discussed in the present work.

The statistical geometry of Lagrangian clusters was developed further in the inertial-range shape framework of \citet{Pumir2000} and in DNS of four-particle dispersion by \citet{Biferale2005b}. Related multi-particle geometry has also been used to diagnose entrainment in strongly inhomogeneous turbulent interfaces \citep{Watanabe2016}. Most recently, a complementary finite-cloud construction distinguished a particle-weighted mass/gyration ellipsoid from a minimum-area outer-envelope ellipsoid in a controlled rough-flow setting \citep{Chertkov2026}. The present RBC study uses the mass/gyration side of this finite-cloud description and extends it to dense three-dimensional clouds in a realistic buoyancy-driven flow.

\citet{Gasteuil2007} pioneered Lagrangian laboratory experiments in convection, discovering the intermittency of \rev{local Lagrangian convective heat transfer as well as its distinct positive skewness, reflecting the strong correlation between velocity and temperature in buoyancy-driven systems. Multiple studies have corroborated these results since then \citep{Schumacher2009, Liot2016a, Käufer2024}}. Soon after, \citet{Schumacher2008, Schumacher2009} \rev{conducted the first Lagrangian DNS in convection and detected acceleration PDFs with exponential tails as known from previous studies in HIT \citep{Voth2002, Mordant2004, Bec2006}. Additionally, a Richardson-type scaling of Lagrangian particle pair dispersion with $t^3$ was confirmed numerically in the horizontal direction, consistent with previous results in HIT \citep{Richardson1926, Boffetta1999, Boffetta2002}, which has since then been scrutinised by both numerical and experimental investigations \citep{Emran2010, Ni2013, Kim2018, Liot2019}. Following these seminal studies, the microstructure of the flow has been examined to verify the Heisenberg--Yaglom relationship of acceleration statistics \citep{Ni2012}; the model was probed later by introducing a mixed model with an analogous thermal Heisenberg--Yaglom relationship $\accvar = a_{0}^{(T)}(\alpha g)^{3/2} \epsT^{3/4} \kappa^{1/4}$, achieving very good agreement with experimental measurements of acceleration variance \citep{Li2021}}. Furthermore, the topological organisation of a turbulent flow including barriers of material transport were investigated by identifying Lagrangian Coherent Structures \citep{Haller2015} or Almost Invariant Sets \citep{Dellnitz2009}. In these frameworks, one can identify coherent flow regions that contribute least to the turbulent transport by spectral clustering of Lagrangian trajectories \citep{Schneide2018, Schneide2022, Vieweg2024} or the so-called inflated generator framework \citep{Badza2026}. Most recently, \citet{Weiss2024} and \citet{Shevkar2025a} have employed particle tracking algorithms (e.g., Shake-The-Box) to further investigate pair dispersion and velocity gradients in experimental settings of RBC.

The objective of the present work is therefore not only to extend Lagrangian statistics in RBC to higher Rayleigh numbers, but also to use particle trajectories, pair dispersion, velocity-gradient topology, and multi-particle cloud deformation as mechanism-resolving diagnostics of convective transport. Global Eulerian quantities such as $\Nu(\Ra,\Pr)$ and $\Re(\Ra,\Pr)$ compress the dynamics into system-level scaling laws. In contrast, particle trajectories, multi-particle cloud deformation, and local velocity-gradient topology retain information about where, when, and via which coherent structures heat and momentum are transported. This distinction is essential in convection, where boundary layers, thermal plumes, large-scale circulations, and small-scale turbulent background fluctuations coexist and interact.

While experimental Lagrangian studies have successfully probed acceleration statistics up to $\Ra \approx 10^{11}$ \citep{Ni2012}, the high velocities and required temporal resolution in this regime prohibit the simultaneous tracking of particle pairs needed to investigate relative dispersion. Meanwhile, Eulerian direct numerical simulations (DNS) in three dimensions have reached $\Ra \lesssim 10^{13}$ for aspect ratios of about 1 to probe heat transfer scalings \citep{Stevens2011, Tiwari2025}. Lagrangian particle analyses in DNS for aspect ratios equal to or larger than 1 are so far limited to Rayleigh number $\Ra \lesssim 10^9$ \rev{\citep{Schumacher2009, Emran2010, Barta2025}}. Furthermore, progressing to the range $\Ra \gtrsim 10^9$ alters the near-wall flow dynamics in a new way: recent experimental and numerical studies by \citet{Lösch2021}, \citet{Giersch2021}, and \citet{Kaestner2023} have shown the occurrence of transient swirling near-wall vortex filaments in connection with the plume detachment -- denoted as `dust devils' -- that are known from planetary atmospheres \citep{Bickel2025}. Their specific trace in the velocity gradient dynamics and impact on the heat transfer from a Lagrangian perspective still needs to be analysed, further motivating the present work.

Finally, the traditional characterisation of turbulent mixing relies primarily on the scalar metric of Lagrangian two-particle dispersion, $\langle R^2(t) \rangle$. While this quantity robustly captures separation regimes -- ballistic, Richardson-like, or diffusive -- it does not retain the geometry of a deforming material volume. After \citet{Chertkov1999} introduced the tetrad as a local shape diagnostic in HIT, \citet{Schumacher2009} brought them into RBC. However, a systematic analysis of larger particle clouds, beyond the minimal tetrad representation, remains missing. In particular, PCA-based reconstruction of the full spatio-temporal deformation of Lagrangian particle clouds has not yet been used to quantify anisotropic deformation mechanisms in turbulent convection.

The missing step is therefore not merely to extend two-particle dispersion to more particles, but to recover the geometry of a deforming material volume. This is essential in RBC, where a thermal plume may entrain a cloud coherently, stretch it into a filament, pierce it, or tear it apart. This study uses dense particle clouds and PCA-based ellipsoid diagnostics to make this material-volume deformation measurable, and to separate plume-driven buoyant ejection from subsequent shear-driven turbulent spreading.

In the present work, we address these open questions by performing three-dimensional DNS of Lagrangian turbulence in a plane-layer RBC setup that spans five orders of magnitude, \rev{$\Ra \in [10^5, 10^{10}]$} for $\Pr=0.7$ in a Cartesian domain of aspect ratio $\Gamma=L/H=4$ with periodic boundary conditions along both lateral directions; $L$ is the periodicity length scale. In a primary simulation series, we analyse the trajectories of $2^{20}$ randomly-seeded particle pairs to investigate pair dispersion, single-particle and convective heat transfer statistics along the Lagrangian particle trajectories which are advanced in time together with Eulerian fields of velocity and temperature. Crucially, this series is extended by computing particle accelerations at every integration time step for meshes of the same spectral polynomial order. This decouples the physical Lagrangian intermittency from the numerical aliasing signatures across $C^0$-continuous element boundaries, ensuring well-resolved statistics of the flow microstructure. Furthermore, local flow topologies and the occurrence of dust-devil-like convective vortices at $\Ra=10^{10}$ are studied by employing a local Lagrangian analysis of the $Q$--$R$ velocity gradient invariants\rev{, executing an ANOVA-validated trajectory filtering algorithm, and visualising a targeted dense tracer cloud seeded in the eye of a vortex.} Completing this high $\Ra$ data set, we address the geometric gap in \rev{two} simulation series in a larger domain of aspect ratio $\Gamma=16$ for \rev{$\Ra \in [10^4, 10^8]$}. Here, we track the evolution of dense particle clusters; specifically, initially localised ensembles of $1,\!000$ particles that are seeded in spheres of radius \rev{$r_0 \in \{ 1 \eta_K, 4 \eta_K \}$}. By combining PCA with a conditioned sampling approach to these multi-particle clusters, we resolve the full spatio-temporal deformation of fluid volumes, unveiling turbulent transport dynamics and resolving the interplay of shear- and buoyancy-driven dispersion laws. Finally, we emphasise that we consider massless Lagrangian tracer particles only, simply denoted as `particles' in the following.

The manuscript is organised as follows. Section \ref{sec:Theoretical_and_numerical_framework} presents the necessary theoretical and numerical methodology. Section \ref{sec:Lagrangian_single-particle_dynamics} discusses the single-particle analysis, while Section \ref{sec:Lagrangian_dual_particle_dynamics} is primarily dedicated to two-particle dynamics. Section \ref{sec:Lagrangian_multi-particle_dynamics} describes the evolution of the particle cloud geometry. We conclude with a summary and an outlook in Section \ref{sec:Discussion_and_perspective}.

%-----------------------------------------------------------------------------
\section{Theoretical and numerical framework}
\label{sec:Theoretical_and_numerical_framework}

\subsection{Governing equations}
\label{subsec:Governing_equations}

We consider an incompressible fluid confined between two horizontal plates separated by height $H$ in a Rayleigh--Bénard convection (RBC) setup. Applying the Oberbeck--Boussinesq approximation \citep{Oberbeck1879, Boussinesq1903}, we assume that only the mass density varies linearly with temperature in the buoyancy term, $\rho(T) \simeq \rho_{\mathrm{ref}}[ 1-\alpha( T-T_{\mathrm{ref}}) ]$. All other material properties are constant; $\alpha$ is the thermal expansion coefficient at constant pressure. The system is non-dimensionalised based on the free-fall velocity $U_f = \sqrt{\alpha g \Delta T H}$, free-fall time $\tauf = H / U_f$ and non-dimensional temperature difference $\Delta T = T(z=0) -T(z=1)$, leading to the non-dimensional governing equations as
\begin{align}
\label{eq:CE}
\nabla \cdot \bm{u} &= 0 \,, \\
\label{eq:NSE}
\frac{\partial \bm{u}}{\partial t} + \left( \bm{u} \cdot \nabla \right) \bm{u} &= - \nabla p + \sqrt{\frac{\Pr}{\Ra}}  \nabla^{2} \bm{u} + T \bm{e}_{z} \,, \\
\label{eq:EE}
\frac{\partial T}{\partial t} + \left( \bm{u} \cdot \nabla \right) T &= \frac{1}{\sqrt{\Ra \Pr}}  \nabla^{2} T \,.
\end{align}
The Rayleigh number $\Ra$ and Prandtl number $\Pr$ represent the control parameters of the system,
\begin{equation}
\label{eq:def_Rayleigh_number_Prandtl_number}
\Ra = \frac{\alpha g \Delta T H^{3}}{\nu \kappa} \qquad \textrm{and} \qquad \Pr = \frac{\nu}{\kappa}\,,
\end{equation}
where $\nu$ is the kinematic viscosity, $\kappa$ the thermal diffusivity, and $g$ the acceleration due to gravity. Specifying the non-dimensional temperatures at the top and bottom plates according to $T \left( z=1 \right) = 0$ and $T \left( z=0 \right) = 1$ defines thermal Dirichlet type boundary conditions. Mechanical no-slip boundary conditions according to $\bm{u} \left(z\in\{0,1\} \right) = \bm{0}$ apply at these interfaces while the domain is periodic in the lateral direction, such that any variable $\Phi$ satisfies $\Phi( \bm{x}) = \Phi(\bm{x} + n_x L \bm{e}_{x} + n_y L \bm{e}_{y})$ with $n_x, n_y \in \mathbb{Z}$ in the domain of aspect ratio $\Gamma = L / H$.

%------------------------------------------------------------------------------------------
\subsection{Numerical methods}
\label{sec:Numerical_methods}

We use the GPU-accelerated solver \textit{nekRS} to solve the governing equations \eqref{eq:CE} -- \eqref{eq:EE}, employing the spectral element method (SEM) \citep{Fischer2022}. The spatial domain is split into $N_e$ elements. 
Within each element, the velocity and temperature fields are represented by Lagrange polynomials of order $p$ based on Gauss-Lobatto-Legendre (GLL) quadrature points, allowing for exponential spatial convergence and minimising numerical dispersion, which is crucial for resolving fine-scale intermittency of high $\Ra$ turbulence. Temporal integration is performed via operator splitting, treating the non-linear convective terms explicitly using second-order extrapolation while solving the linear viscous and pressure terms implicitly via second-order backward differentiation.

In addition to solving the Eulerian fields, we track the properties of massless Lagrangian particles evolved according to the kinematic equation
\begin{equation}
    \label{eq:LPM}
    \frac{d\bm{x}_i}{dt} = \bm{u} \left( \bm{x}_i, t \right).
\end{equation}
Consistent with the accuracy of the Eulerian solver, a second-order Adams--Bashforth (AB2) scheme is employed: 
\begin{equation}
    \bm{x}_i^{n+1} = \bm{x}_i^n + \Delta t \left( \frac{3}{2} \bm{u}(\bm{x}_i^n, t^n) - \frac{1}{2} \bm{u}(\bm{x}_i^{n-1}, t^{n-1}) \right).
\end{equation}
This explicit multi-step scheme provides the computational benefit of requiring only one spectral evaluation per time step. Crucially, interpolation is performed spectrally; the velocity field $\bm{u}$ and its velocity gradient tensor $\nabla \bm{u}$ are evaluated based on the underlying spectral basis functions rather than low-order linear or cubic splines. This high-fidelity interpolation allows preserving small-scale velocity gradients required for $Q$--$R$ and acceleration analysis. 

%------------------------------------------------------------------------------------------
\subsection{Simulation series}
\label{subsec:Simulation_series}

\begin{table}
\begin{center}
\begin{tabular}{lccccccccc}
Case & $\Ra$ & $\Gamma$ & Elements ($N_e$) & Poly. ($p$) & $N_{\rm{BL}}$ & $\left(\frac{\Delta x}{\eta_K }\right)_{\rm{avg,BL}}$ & $\left( \frac{\Delta x}{\eta_K} \right)_{\rm{avg,Bulk}}$ & $\left( \frac{\Delta z}{\eta_K} \right)_{\rm{max,BL}}$ & $\left( \frac{\Delta z}{\eta_K} \right)_{\rm{max,Bulk}}$ \\

\multicolumn{10}{l}{\textit{Series A: Acceleration Statistics}} \\
A5 & $10^5$ & $4$ & $100 \times 100 \times 64$ & 5 & 67 & 0.30 & 0.18 & 0.08 & 0.19 \\
A6 & $10^6$ & $4$ & $140 \times 140 \times 90$ & 5 & 54 & 0.49 & 0.29 & 0.11 & 0.23 \\
A7 & $10^7$ & $4$ & $180 \times 180 \times 115$ & 5 & 41 & 0.85 & 0.52 & 0.18 & 0.39 \\
A8 & $10^8$ & $4$ & $210 \times 210 \times 135$ & 5 & 27 & 1.67 & 1.01 & 0.35 & 0.70 \\
A9 & $10^9$ & $4$ & $270 \times 270 \times 174$ & 5 & 18 & 3.05 & 1.85 & 0.63 & 1.15 \\
A10 & $10^{10}$ & $4$ & $560 \times 560 \times 280$ & 5 & 17 & 3.40 & 2.01 & 0.79 & 1.64 \\

\multicolumn{10}{l}{\textit{Series P: Pair Statistics}} \\
P5 & $10^5$ & $4$ & $100 \times 100 \times 64$ & 5 & 67 & 0.30 & 0.18 & 0.08 & 0.19 \\
P6 & $10^6$ & $4$ & $100 \times 100 \times 64$ & 7 & 58 & 0.49 & 0.30 & 0.11 & 0.32 \\
P7 & $10^7$ & $4$ & $100 \times 100 \times 64$ & 9 & 42 & 0.86 & 0.51 & 0.19 & 0.54 \\
P8 & $10^8$ & $4$ & $150 \times 150 \times 96$ & 7 & 24 & 1.69 & 1.00 & 0.40 & 0.90 \\
P9 & $10^9$ & $4$ & $150 \times 150 \times 96$ & 9 & 23 & 3.06 & 1.83 & 0.50 & 1.33 \\
P10 & $10^{10}$ & $4$ & $400 \times 400 \times 200$ & 7 & 17 & 3.48 & 2.07 & 0.80 & 1.66 \\

\multicolumn{10}{l}{\textit{Series \rev{C$_k$ with $k=1,4$}: Cloud Geometry}} \\
C$_k$4 & $10^4$ & $16$ & $128 \times 128 \times 8$  & 5 & 13 & 0.39 & 0.24 & 0.39 & 0.72 \\
C$_k$5 & $10^5$ & $16$ & $128 \times 128 \times 8$  & 5 & 8  & 0.93 & 0.57 & 0.64 & 1.63 \\
C$_k$6 & $10^6$ & $16$ & $128 \times 128 \times 8$  & 9 & 8  & 1.18 & 0.71 & 0.73 & 2.04 \\
C$_k$7 & $10^7$ & $16$ & $256 \times 256 \times 32$ & 7 & 17 & 1.69 & 1.02 & 0.46 & 1.69 \\
C$_k$8 & $10^8$ & $16$ & $384 \times 384 \times 64$ & 5 & 27 & 3.62 & 2.16 & 0.31 & 2.04 \\
\end{tabular}
    \caption{\justifying{Simulation parameters for series A, P\rev{, C$_1$ and C$_4$. The latter two series contain 5 runs each (C$_k$4 to C$_k$8), differing by their initial particle cloud radii of $r_0=1\eta_K$ and $r_0=4\eta_K$, respectively.} $N_e$ denotes the number of spectral elements of the used mesh, $p$ is the polynomial order. $N_{\rm{BL}}$ grid points lie within the thermal boundary layer height $\delta_T$. The horizontal resolution $(\Delta x / \eta_K)_{\rm{avg}}$ is computed using the average horizontal grid spacing and the strictest (minimum) local Kolmogorov scale $\eta_K \left(z\right)$ found within the respective region (boundary layer or bulk). Conversely, the vertical resolution quality $(\Delta z / \eta_K)_{\rm{max}}$ is determined using the absolute maximum local grid spacing to guarantee strict sub-Kolmogorov resolution in the region where intense vertical gradients and plume detachment occur.
    }}
\label{tab:simulation_parameters}
\end{center}
\end{table}

We perform three distinct series of DNS with Lagrangian particle tracking to investigate specific aspects of turbulent transport in RBC. These are denoted as series P, A\rev{, C$_1$ and C$_4$ in the following which stand for pair, acceleration, and particle cloud analysis (the latter two with distinct initial seeding radii)}, respectively.

{\em Series P for particle pair statistics.} We initialize $2^{20}$ pairs of Lagrangian particles, separated by an initial distance $R_0=\eta_K$, at random positions in a statistically stationary flow field of aspect ratio $\Gamma=4$ for $\Ra \in [10^5, 10^{10}]$. In addition to pair dispersion, the large ensemble size also permits converged statistics for single-particle dispersion and heat transfer intermittency as well as topological analysis.

{\em Series A  for particle accelerations.} Using the same statistically stationary flow fields from series P, we seed $2^{21}$ particles at random positions. We explain the evaluation of the acceleration components in detail in Appendix \ref{sec:Appendix_Acceleration_filtering_via_cubic_smoothing_splines}. 

{\em Series C for cloud geometry.} We consider a wider domain of $\Gamma=16$ for $\Ra \in [10^4, 10^8]$ to investigate how convective turbulence disperses dense clusters of particles. We seed \rev{$256$} independent clusters of $1,\!000$ particles each in three different horizontal planes to capture distinct physics\rev{, where the two series C$_1$ and C$_4$ solely differ in the initial seeding radius of particle clouds ($r_0 \in \{1\eta_K, 4\eta_K\}$)}: the \rev{thermal boundary layer ($z_0 = \delta_T$), the plume root zone ($z_0 = 2\delta_T$), and the core ($z_0 = 0.5$). The spatial extent of individual clouds, particularly at $r_0=4\eta_K$, alongside the requirement of a statistically converged ensemble size ($N_c=256$ clouds per layer), necessitates this wider domain.}

Resolving fine-scale structures is crucial for DNS, particularly for Lagrangian tracking, where fluid properties are evaluated at positions between grid points. Small-scale intermittency is intrinsically linked to local dissipation rates
\begin{equation}
    \label{eq:epsilon}
    \epsilon_{\nu} ({\bm x},t) = \frac{\nu}{2} \left( \nabla \bm{u} + \nabla \bm{u}^T \right)^2 \quad \mathrm{and} \quad \epsilon_T \left( \bm{x}, t \right) = \kappa \left( \nabla T \right)^2\,.
\end{equation}
To obtain a reliable estimate of the effective resolution, the Kolmogorov and Batchelor scales must be defined based on the local flow topology \citep{Scheel2013}. They are given by
\begin{equation}
    \label{eq:Kolmogorov_Batchelor_local}
    \left \langle \eta_K \right \rangle_{A,t} \left(z \right) = \frac{\nu^{3/4}}{\left \langle\epsilon_{\nu}\right \rangle_{A,t}^{1/4}} \quad \mathrm{and} \quad \left \langle \eta_B \right \rangle_{A,t} \left(z \right) = \frac{\left \langle \eta_K \right \rangle_{A,t}}{\sqrt{\Pr}}\,.
\end{equation}
For $\Pr<1$, the Kolmogorov scale $\eta_K$ represents the smallest active length scale and governs the resolution requirement. Table \ref{tab:simulation_parameters} summarises the grid parameters, whereas table \ref{tab:Series_scales} reports fundamental scales.

Because of the uniform horizontal distribution of grid elements, we report the average horizontal resolution $(\Delta x / \eta_K)_{\rm{avg}}$. In the vertical direction, we evaluate the absolute maximum local grid spacing to capture the intense near-wall dynamics. These resolution metrics are evaluated independently within the boundary layers and the bulk to separate the dynamics in the two domains.

%------------------------------------------------------------------------------------------
\section{Lagrangian single-particle dynamics}
\label{sec:Lagrangian_single-particle_dynamics}
%------------------------------------------------------------------------------------------

\subsection{Acceleration statistics}
\label{subsec:Acceleration_statistics}

\begin{table}
\begin{center}
\begin{tabular}{lcccccc}
Case & $\Ra$ & $\left \langle Nu \right\rangle_{V,t}$ & $\delta_T$ & $\eta_K$ & $t_{\rm run}$ & $n_{\rm snp}$ \\
\hline
\\[-8pt]
\multicolumn{7}{l}{\textit{Series A: Acceleration Statistics}} \\
A5 & $10^{5}$ & 4.24 & 0.118 & 0.0351 & 215.3 & 57,444 \\
A6 & $10^{6}$ & 8.12 & 0.062 & 0.0162 & 44.4 & 21,022 \\
A7 & $10^{7}$ & 15.48 & 0.032 & 0.0076 & 18.7 & 16,927 \\
A8 & $10^{8}$ & 30.37 & 0.016 & 0.0036 & 25.8 & 41,922 \\
A9 & $10^{9}$ & 59.65 & 0.008 & 0.0017 & 13.3 & 30,139 \\
A10 & $10^{10}$ & 126.53 & 0.004 & 0.0008 & 1.7 & 9,154 \\
\\[-5pt]
\multicolumn{7}{l}{\textit{Series P: Pair Statistics \rev{($R_0=1\eta_K$}})} \\
P5 & $10^{5}$ & 4.25 & 0.118 & 0.0350 & 851.8 & 8,730 \\
P6 & $10^{6}$ & 8.16 & 0.061 & 0.0162 & 571.9 & 11,760 \\
P7 & $10^{7}$ & 15.63 & 0.032 & 0.0076 & 302.4 & 6,710 \\
P8 & $10^{8}$ & 30.29 & 0.017 & 0.0036 & 232.0 & 8,947 \\
P9 & $10^{9}$ & 60.27 & 0.008 & 0.0017 & 234.4 & 6,816 \\
P10 & $10^{10}$ & 122.26 & 0.004 & 0.0008 & 127.2 & 13,026 \\
\\[-5pt]
\multicolumn{7}{l}{\rev{\textit{Series C$_1$: Cloud Geometry ($r_0=1\eta_K$})}} \\
\rev{C$_1$4} & \rev{$10^{4}$} & \rev{2.29} & \rev{0.218} & \rev{0.0785} & \rev{850.0} & \rev{1,435} \\
\rev{C$_1$5} & \rev{$10^{5}$} & \rev{4.28} & \rev{0.117} & \rev{0.0350} & \rev{450.0} & \rev{1,317} \\
\rev{C$_1$6} & \rev{$10^{6}$} & \rev{8.13} & \rev{0.062} & \rev{0.0162} & \rev{300.0} & \rev{1,493} \\
\rev{C$_1$7} & \rev{$10^{7}$} & \rev{15.52} & \rev{0.032} & \rev{0.0076} & \rev{200.0} & \rev{1,489} \\
\rev{C$_1$8} & \rev{$10^{8}$} & \rev{30.35} & \rev{0.016} & \rev{0.0036} & \rev{150.0} & \rev{1,500} \\
\\[-5pt]
\multicolumn{7}{l}{\rev{\textit{Series C$_4$: Cloud Geometry ($r_0=4\eta_K$})}} \\
\rev{C$_4$4} & \rev{$10^{4}$} & \rev{2.27} & \rev{0.220} & \rev{0.0788} & \rev{850.0} & \rev{1,395} \\
\rev{C$_4$5} & \rev{$10^{5}$} & \rev{4.28} & \rev{0.117} & \rev{0.0349} & \rev{450.0} & \rev{1,327} \\
\rev{C$_4$6} & \rev{$10^{6}$} & \rev{8.13} & \rev{0.061} & \rev{0.0162} & \rev{300.0} & \rev{1,693} \\
\rev{C$_4$7} & \rev{$10^{7}$} & \rev{15.54} & \rev{0.032} & \rev{0.0076} & \rev{200.0} & \rev{1,488} \\
\rev{C$_4$8} & \rev{$10^{8}$} & \rev{30.35} & \rev{0.016} & \rev{0.0036} & \rev{150.0} & \rev{1,508} \\
\end{tabular}
\caption{\justifying{Fundamental scales for series A, P\rev{, C$_1$ and C$_4$}. The Eulerian Nusselt number $\left \langle Nu \right\rangle_{V,t}$, the thermal boundary layer thickness $\delta_T$ \rev{and} the Kolmogorov length scale $\eta_K$ \rev{are reported}. The total time in free-fall times over which Lagrangian particles were tracked $t_{\rm run}$ and the number of evaluated snapshots $n_{\rm snp}$ serve as metrics for statistical convergence. \rev{$L=2^{20}$ particle pairs were seeded with an initial separation of $R_0=1\eta_K$ in series P. Initial radii of $N_c=3 \times 256$ clouds of $N_p=1,\!000$ particles are $r_0=1\eta_K$ and $r_0=4\eta_K$ in series C$_1$ and C$_4$, respectively.}}}
\label{tab:Series_scales}
\end{center}
\end{table}

%------------------------------------------------------------------------------------------
\begin{figure}
\centering
\includegraphics[width = \textwidth]{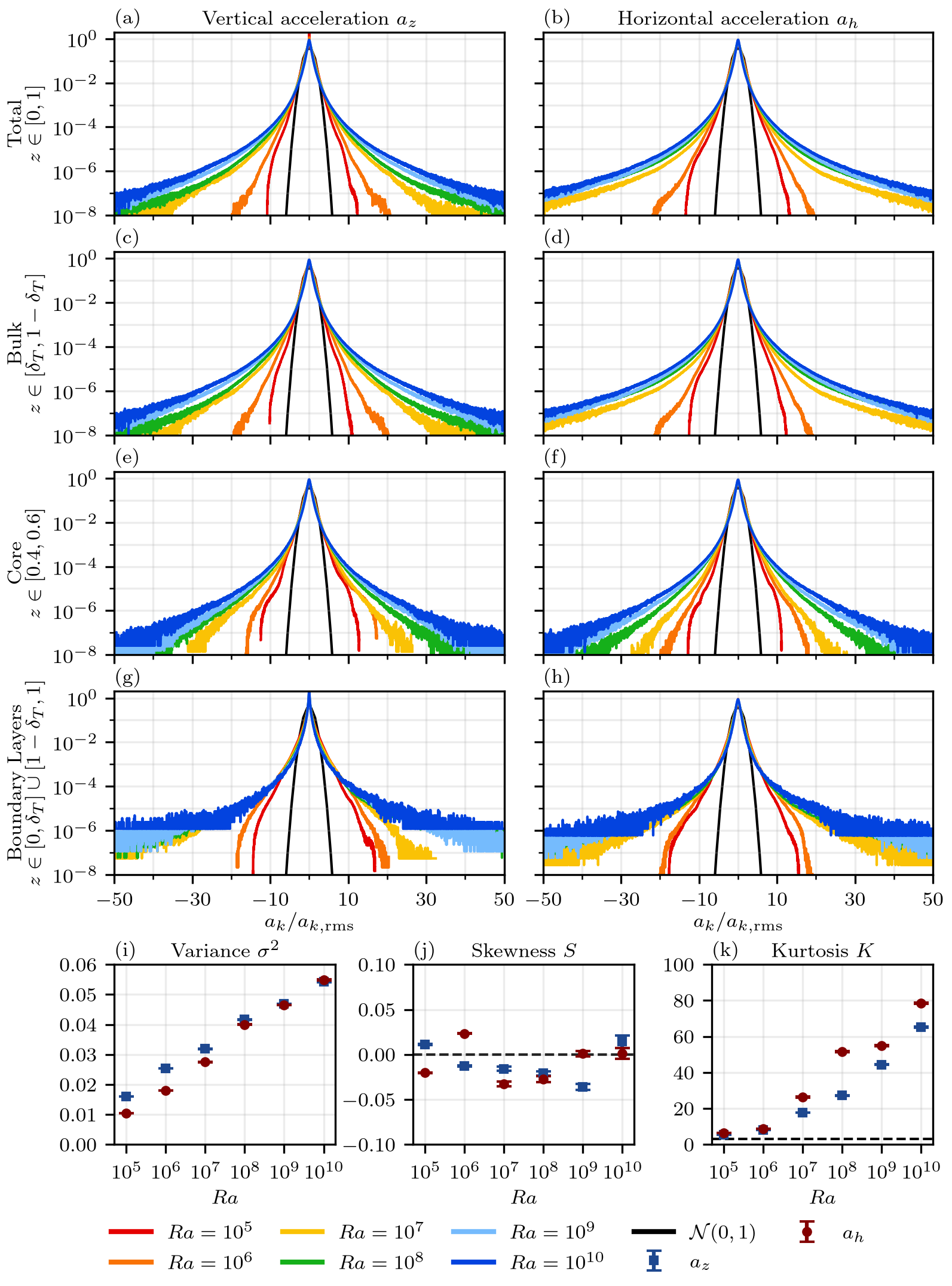}
\caption{\justifying{
Lagrangian particle acceleration statistics. Probability density functions (PDFs) of the vertical ($a_z$) and horizontal ($a_h$) accelerations for the $2^{21}$ Lagrangian particles, normalised by their respective root-mean-square values $a_{k,\mathrm{rms}}$. Statistics are shown for the total domain (a,b), the bulk region $z \in [\delta_T, 1-\delta_T]$ (c,d)\rev{, the core $z \in [0.4, 0.6]$ (e, f)}, and the boundary layers $z \in [0, \delta_T] \cup [1-\delta_T, 1]$ \rev{(g, h)}. \rev{Second- to fourth-order moments of vertical and horizontal PDFs for the total domain are shown in panels (i--k). The Gaussian reference values for skewness ($S=0$) and kurtosis ($K=3$) are indicated by dashed lines in panels (j) and (k).}}}
\label{fig:accPDFs}
\end{figure}
%------------------------------------------------------------------------------------------

%------------------------------------------------------------------------------------------
\begin{table}
\color{black}
\begin{center}
\begin{tabular}{l r@{\;$\pm$\;}l r@{\;$\pm$\;}l c r@{\;$\pm$\;}l r@{\;$\pm$\;}l r@{\;$\pm$\;}l r@{\;$\pm$\;}l c}
Case & \multicolumn{2}{c}{$\sigma^2_z \times 10^2$} & \multicolumn{2}{c}{$\sigma^2_h \times 10^2$} & $\sigma^2_z / \sigma^2_h$ & \multicolumn{2}{c}{$S_z \times 10^2$} & \multicolumn{2}{c}{$S_h \times 10^2$} & \multicolumn{2}{c}{$K_z$} & \multicolumn{2}{c}{$K_h$} & $K_z / K_h$  \\
\hline
\\[-8pt]
A5  & $1.59$ & $0.00$ & $1.04$ & $0.00$ & $1.54$ & $1.12$ & $0.04$ & $-2.01$ & $0.03$ & $5.6$ & $0.0$ & $6.4$ & $0.0$ & $0.88$ \\
A6  & $2.54$ & $0.00$ & $1.81$ & $0.00$ & $1.41$ & $-1.26$ & $0.08$ & $2.34$ & $0.06$ & $8.1$ & $0.0$ & $8.7$ & $0.0$ & $0.94$ \\
A7  & $3.19$ & $0.00$ & $2.75$ & $0.00$ & $1.16$ & $-1.59$ & $0.24$ & $-3.27$ & $0.28$ & $17.7$ & $0.1$ & $26.4$ & $0.2$ & $0.67$ \\
A8  & $4.16$ & $0.00$ & $4.00$ & $0.00$ & $1.04$ & $-2.10$ & $0.26$ & $-2.75$ & $0.33$ & $27.3$ & $0.1$ & $51.5$ & $0.3$ & $0.53$ \\
A9  & $4.68$ & $0.00$ & $4.65$ & $0.01$ & $1.01$ & $-3.59$ & $0.32$ & $0.11$ & $0.30$ & $44.4$ & $0.2$ & $55.0$ & $0.3$ & $0.81$ \\
A10 & $5.43$ & $0.01$ & $5.49$ & $0.02$ & $0.99$ & $1.43$ & $0.70$ & $0.12$ & $0.61$ & $65.2$ & $0.3$ & $78.5$ & $0.3$ & $0.83$ \\
\end{tabular}
    \caption{\justifying{\rev{Higher-order moments of acceleration PDFs. Variance $\sigma^2$, skewness $S$ and kurtosis $K$ in vertical ($z$) and horizontal ($h$) directions are computed for $a_k/a_{k,\mathrm{rms}} \in [-50,50]$ and reported alongside their batched standard errors. Skewness values are multiplied by $10^2$. The variance ratio $\sigma_z^2 / \sigma_h^2$ converges to unity as $\Ra$ is increased whereas the kurtosis ratio $K_z/K_h<1$ indicates increased intermittency in horizontal rather than vertical direction.}}}
\label{tab:moments}
\end{center}
\end{table}
%------------------------------------------------------------------------------------------

The data from series A were obtained at every integration time step and subsequently filtered using calibrated cubic smoothing splines to provide physically well-resolved acceleration statistics, as detailed in Appendix \ref{sec:Appendix_Acceleration_filtering_via_cubic_smoothing_splines}. We compute the probability density functions (PDFs) of the vertical ($a_z$) and horizontal ($a_h$) acceleration components. The latter is averaged over the two homogeneous lateral directions $x$ and $y$ to improve statistical convergence. Both accelerations are normalised by their respective root-mean-square (rms) value $a_{k,\mathrm{rms}}$. Figure \ref{fig:accPDFs} illustrates these normalised acceleration PDFs for the total domain, (figure \ref{fig:accPDFs}a,b), the bulk $z \in [\delta_T, 1-\delta_T]$ (figure \ref{fig:accPDFs}c,d)\rev{, the core $z \in [0.4, 0.6]$ \ref{fig:accPDFs}e,f)} and the thermal boundary layers $z \in [0, \delta_T] \cup [1-\delta_T, 1]$ (figure \ref{fig:accPDFs}\rev{g,h)}. We also plot the \rev{second-,} third- and fourth-order central moments \rev{variance $\sigma^2$,} skewness $S=\langle a^3\rangle/\langle a^2\rangle^{3/2}$ and kurtosis $K=\langle a^4\rangle/\langle a^2\rangle^2$ \rev{for the total domain in figure \ref{fig:accPDFs}(i--k). Figure \ref{fig:accPDFs}(i) reveals that the ratio of vertical to horizontal acceleration variances, $\sigma^2_z/\sigma^2_h$, approaches unity with increasing $Ra$, indicating a progressive resemblance to homogeneous isotropic turbulence (HIT).}

% Acceleration PDFs
All normalised acceleration PDFs exhibit a distinct, symmetrical shape and become increasingly fat-tailed with rising $\Ra$, significantly surpassing the Gaussian baseline $\mathcal{N}(0,1)$ beyond \rev{$50\,a_{\mathrm{rms}}$}. This is quantified by a uniform skewness $S\approx 0$ in figure \ref{fig:accPDFs}(j) and monotonically increasing kurtosis $K$ in figure \ref{fig:accPDFs}(k), indicating an elevated probability of extreme particle acceleration events, especially as the flow transitions beyond $\Ra \geq 10^7$. 

% Kurtosis
\rev{Calculating the kurtosis of Lagrangian acceleration statistics is notoriously difficult because their wide, stretched exponential tails require an enormous data set to ensure convergence \citep{Yakhot2005, Emran2010}. We thus limit the calculation of higher-order moments to the presented region $a_k / a_{k,\mathrm{rms}} \in [-50, 50]$ as discussed in detail in Appendix \ref{sec:Appendix_Higher-order_moments_of_Lagrangian_acceleration}. The traditional definition of the standard error $SE (m_n) = \sigma_{m_n} / \sqrt{N}$ requires the 8th-order central moment for kurtosis. Furthermore, accelerations along Lagrangian trajectories are not independent events but correlated in time. The standard errors for variance, skewness and kurtosis are thus obtained via a batched approach: here, the data set of $2^{21}$ particle trajectories is split into $M=4\times32\times44=5,\!632$ batches of $N_p\approx372$ trajectories each. The $n$-th order moments $m_n \in \{\sigma^2,S, K\}$ are computed batch-wise to obtain the batched standard error}
\begin{equation}
    \rev{SE (m_n) = \sqrt{\frac{1}{M(M-1)} \sum_{i=1}^M (m_{n,i} -\langle m_n\rangle)^2}\,.}
\end{equation}
\rev{Evidently, the horizontal kurtosis for $\Ra=10^8$ exceeds the vertical trend far beyond its statistical error bars (figure \ref{fig:accPDFs}k). Having performed rigorous comparative simulations and due to our numerical resolution (see Section \ref{subsec:Simulation_series}), we can rule out numerical artefacts. Instead, the horizontal PDFs suggest that this disparity originates from thermal plumes rather than turbulent background noise, given a near collapse of the distributions at  $\Ra=10^9$ in the bulk, see figure \ref{fig:accPDFs}(d), and the thermal boundary layers in figure \ref{fig:accPDFs}(h), but not in the core of the domain as shown in figure \ref{fig:accPDFs}(f). This distinct result at $\Ra \approx 10^8$ corroborates earlier studies by \citet{Schumacher2009}, who found $K_h=63.9 \pm 16$ and $K_z=30.3 \pm 11$ under similar boundary conditions when sampling up to $\pm 60 \, a_{\mathrm{rms}}$. Our complete values of higher-order moments are reported in table \ref{tab:moments}.}

\rev{Highlighting the high sensitivity of kurtosis, \citet{Emran2010} reported a kurtosis value of $K_z = 89 \pm 18$ for $\Ra=10^7, \Pr=0.7$ in a closed cylindrical cell of $\Gamma=1$ for $4\times10^9$ events, significantly exceeding the value of $K_z = 17.7 \pm 0.2$ (or $K_z = 18.6 \pm 0.3$ when considering the global domain) we found at the same $\Ra$ for approximately $6 \times 10^{10}$ events. This discrepancy might be linked to the inclusion of statistically undersampled extreme events, again highlighting the difficulty of achieving convergence for these higher-order moments. Furthermore, the volume in \citet{Emran2010} was a closed cylinder and not a plane layer with periodic boundaries. Finally, \citet{Bec2006} computed higher-order moments for varying $\St$ in HIT obtained via DNS. Their reported values at the lowest investigated $\St=0.16$ of $K ( \Rlambda\in \{ 65, 105, 185\}) = \{19.1, 32.9, 48.1 \} $ align very well with our $K_z (\Ra \in \{ 10^7, 10^8, 10^9 \}) = \{17.7, 27.3, 44.4 \}$, roughly matching their respective $\Rlambda$ as well as the experimental results by \citet{Voth2002}. However, their values for ideal Lagrangian tracers at $\St=0$, obtained via artificial trajectories, are significantly larger with $K ( \Rlambda \in \{ 65, 105, 185\}) = \{52.8, 133, 288 \} $.}

% PDF shape
This characteristic shape of Lagrangian acceleration PDFs with stretched exponential tails \rev{matches numerous experimental and numerical studies in both HIT and RBC \citep{Voth2002, Mordant2004, Bec2006, Schumacher2008, Emran2010, Ni2012, Liot2016b, Kim2018, Li2021}. In an RBC setup, researchers have found contradicting results regarding the tail behaviour of vertical and lateral acceleration PDFs. While \citet{Schumacher2008} first discovered wider tails in lateral direction, presumably caused by vertical vortices induced by thermal plumes, other studies found either no significant difference \citep{Ni2012, Liot2016b, Li2021} or wider vertical PDFs \citep{Emran2010, Kim2018}. The root cause of this discrepancy seems to lie in the employed mechanical boundary conditions. The setup of \citet{Schumacher2008} featured a box of $\Gamma=2$ with free-slip boundary conditions at the top and bottom plate and periodic boundary conditions in the lateral direction. Evidently, this allows lateral accelerations to manifest significantly more freely than in a more constrained domain, as is the case in many experimental setups like those of \citet{Ni2012, Liot2016b, Kim2018, Li2021}, which are no-slip conditions in closed convection cells.} 

\rev{The aspect ratio also plays a role: while the vertical PDFs exhibited wider tails at $\Gamma=1$, the lateral and vertical PDFs converged once the domain was increased to $\Gamma=8$ in the study of \citet{Emran2010}. In the present work, we are able to recover the findings of \citet{Schumacher2008} in our periodic domain of $\Gamma=4$ -- allowing unrestricted manifestation of lateral accelerations in the bulk region, which can be quantified by the kurtosis ratio $K_z/K_h < 1$ in all cases (see table \ref{tab:moments}). These differences are smaller in the core of the domain for $z \in [0.4, 0.6]$ as seen in figure \ref{fig:accPDFs}(e,f), revealing nearly isotropic behaviour in this well-mixed, plume-sparse region.} Furthermore, while the total domain and the bulk exhibit small differences, the wider tails within the thermal boundary layers reveal that the Lagrangian intermittency is especially strong within this near-wall region. Progressively thinner thermal boundary layers lead to fewer traceable events as $\Ra$ is increased. Finally, the variance ratio $\sigma_z^2 / \sigma_h^2$ monotonically approaches unity with rising $\Ra$, indicating progressively isotropic behaviour at large $\Ra$, resembling the behaviour of HIT, see again figure \ref{fig:accPDFs}(i) and table \ref{tab:moments}.

\subsection{\rev{Acceleration variance and Heisenberg--Yaglom relationship}}
\label{subsec:Acceleration_variance_and_Heisenberg--Yaglom_relationship}

%------------------------------------------------------------------------------------------
\begin{figure}
\centering
\includegraphics[width = 0.95\textwidth]{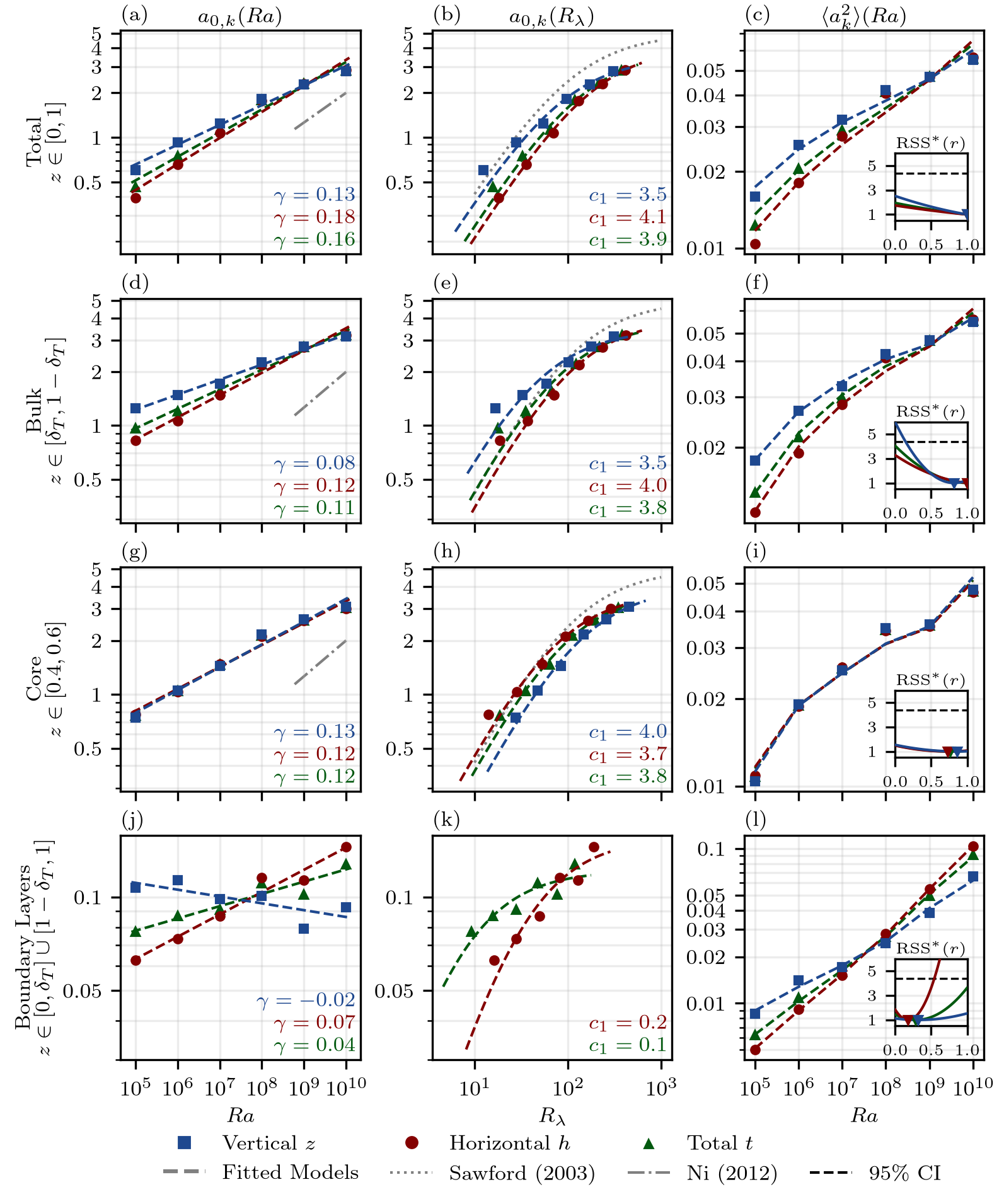}
\caption{\justifying{
\rev{Conditioned Heisenberg constants $a_{0,k}$ as a function of $\Ra$ (column 1) and $\Rlambda$ (column 2) alongside acceleration variance $\langle a_k^2 \rangle$ as a function of $\Ra$ (column 3). Index $k$ stands for vertical ($z$), combined horizontal ($h$), and all components ($t$). Power-law fits $a_{0} = C \Ra^{\gamma}$ (dashed) are compared against $a_0=0.02 \Ra^{0.20}$ (dash-dotted grey) obtained by \citet{Ni2012} (a,d,g,j). The vertical and horizontal components of $a_0$ collapse in the core (g), suggesting local isotropy. The empirical fit for HIT by \citet{Sawford2003} $a_0 = 5(1+110\Rlambda^{-1})^{-1}$ (dotted grey) accurately captures $a_0$ in the total domain (b), the bulk (e) and the core (h), but breaks down in the boundary layers (k), suggesting mechanisms beyond small-scale viscous dissipation. Acceleration variance is fitted using the mixed scaling model of equation \eqref{eq:mixed_hy_model_fit}. Insets show the normalised residual sum of squares $\RSS^*=\RSS/\RSS_{min}$ alongside the $95\%$ confidence threshold (dashed black). The minima reveal that while viscous dissipation dominates in the core, thermal plumes in the bulk (vertical, f) and intense temperature gradients (horizontal, l) exhibit significant thermal contributions beyond $95\%$ confidence.}
}}
\label{fig:a0}
\end{figure}
%------------------------------------------------------------------------------------------

% Heisenberg--Yaglom
\rev{The Heisenberg--Yaglom relationship, rooted in Kolmogorov's K41 theory for HIT \citep{Kolmogorov1941}, dictates that accelerations are driven solely by small-scale fluctuations \citep{Heisenberg1948, Yaglom1949}}
\begin{equation}
    \label{eq:Heisenberg_Yaglom}
    \rev{\langle a_k^2 \rangle = a_0 \langle \epsilon_{\nu}\rangle^{3/2} \nu^{-1/2}\,,}
\end{equation}
\rev{where $a_0$ was originally postulated to be a universal constant. However, numerous experimental and numerical investigations have since then revealed various scalings with the Taylor microscale Reynolds number $\Rlambda$, ranging from $a_0 \sim \Rlambda^{1/2}$ at low $\Rlambda \lesssim 100$ \citep{Yeung1989} to $a_0 \sim \Rlambda^{0.25}$ at high $\Rlambda \lesssim 1300$ \citep{Buaria2022} or plateauing to a constant $a_0$ as $\Rlambda \gtrsim 470$ \citep{LaPorta2001, Voth2002}. \citet{Borgas1993} studied a multifractal model, suggesting $a_0\sim \Rlambda^{0.135}$, while \citet{Hill2002} proposed $a_0 \sim \Rlambda^{0.25}$. Combining experimental and numerical data, \citet{Sawford2003} introduced the empirical fit $a_0 = c_1(1+c_2\Rlambda^{-1})^{-1}$. Additionally, investigations of \citet{Vedula1999} and \citet{LaPorta2001} discovered anisotropic behaviour of $a_0$. In RBC, \citet{Ni2012} have investigated the Heisenberg--Yaglom relationship in the centre region of a cylindrical cell ($\Gamma=1$) and found the fit $a_0 = 0.02\Ra^{0.20}$.}

% ---- R_lambda ----
\rev{We compute the Taylor microscale Reynolds number $\Rlambda = u_{\mathrm{rms}} \lambda / \nu$ to contrast the scaling $a_0(\Rlambda)$ to the extensive literature in HIT. Instead of relying on the traditional HIT-derived estimate $\Rlambda = u_{\mathrm{rms}}^2 \sqrt{15 \Ra / (\Nu-1)}$ -- which does not capture spatial heterogeneity of buoyancy driven flows -- we use the Lagrangian velocity gradients to compute a volume-time averaged effective Taylor microscale}
\begin{equation}
    \label{eq:Taylor_microscale_effective}
    \rev{\left. \lambda_{k}\right|_\mathcal{R} = \sqrt{ \frac{\left.u_{\mathrm{rms}}^2 \right|_\mathcal{R}}{\left \langle \left( \left. \frac{\partial u_k}{\partial x_k} \right|_\mathcal{R} \right)^2\right \rangle_L}}}
\end{equation}
\rev{conditioned to the respective regions $\mathcal{R} \in \{ \mathrm{Total,Bulk,Core, BL}\}$ and acceleration components $k\in \{ z, h, t\}$. We note that the Taylor microscale is strictly defined under the assumption of statistical homogeneity, i.e., $\langle u_k (\partial u_k /\partial x_k) \rangle = 0$. In RBC, this assumption is violated in the wall-normal direction near the plates. Thus, we intentionally omit the definition of $\left. R_{\lambda,z} \right|_{\mathrm{BL}}$ within the highly anisotropic thermal boundary layers.}

% ---- a0 results ----
\rev{Figure \ref{fig:a0} illustrates the Heisenberg constant $a_0$ conditioned to the aforementioned regions $\mathcal{R}$ and directions $k$. We apply the predictive fits $a_0(\Ra) = C \Ra^{\gamma}$ and $a_0(\Rlambda) = c_1(1+c_2\Rlambda^{-1})^{-1}$ following \citet{Ni2012} and \citet{Sawford2003}.}

\rev{Consistent with \citet{Ni2012}, $a_{0,k}$ increases monotonically with $\Ra$ in the total (or full) domain, the bulk and the core region as seen in figure \ref{fig:a0}(a,d,g). The vertical and horizontal components $a_{0,z}$ and $a_{0,h}$ collapse in the core for all $\Ra$, indicating isotropic behaviour in the centre of the domain, see figure \ref{fig:a0}(g). Due to intense kinetic energy dissipation $\epsilon_\nu$ within the boundary layers, the Heisenberg constant is about an order of magnitude smaller than in the bulk. Furthermore, no scaling of $a_{0,z}(\Ra)$ is observed in the vicinity of the walls in BL (figure \ref{fig:a0}j), underscoring the breakdown of K41-assumptions within this region.}

\rev{Within the nearly homogeneous core (figure \ref{fig:a0}g), we extract a scaling of $a_{0,t}(\Ra)=(0.19\pm0.03)\Ra^{0.124\pm0.008}$ in comparison to $a_{0}(\Ra)=0.02\Ra^{0.20}$ by \citet{Ni2012}, who, however, considered higher Rayleigh numbers ($6\times 10^8 \leq \Ra \leq 10^{11}$) as well as different Prandtl numbers $\Pr \in \{ 4.4, 5.5, 6.1\}$. Within our investigated range of $10 \lesssim \Rlambda \lesssim 415$, the fit by \citet{Sawford2003} accurately captures the scaling of the Heisenberg constant taken with respect to all components, especially within the core of the domain, see figure \ref{fig:a0}(h). This provides further quantitative proof of the HIT-like behaviour of RBC in this well-mixed region. Conversely, this empirical HIT fit is inadequate within the thermal boundary layers. Furthermore, a clear divergence between the vertical and horizontal Heisenberg constants emerges in the bulk for low $\Ra \lesssim 10^7$, indicating acceleration-driving mechanisms beyond small-scale viscous dissipation.}

% ---- Mixed model -----
\rev{To address this breakdown of purely viscous scaling of acceleration variance in buoyancy-driven flows, \citet{Li2021} introduced a Heisenberg--Yaglom relationship based on the thermal dissipation rate, 
\begin{equation}
\accvar = a^{(T)}_{0} (\alpha g)^{3/2} \epsT^{3/4} \kappa^{1/4}\,.
\end{equation}
Combining both relationships in equal parts, $\accvar \sim \epsT^{3/8} \epsnu^{3/4}$ approximated their experimentally observed scaling of $\accvar \sim \Ra^{1.54 \pm 0.08}$ better than either mechanism alone, suggesting that both thermal and viscous dissipation contribute in plume-abundant regions. Since \citet{Li2021} relied on experimental measurements, $\epsnu$ had to be estimated using HIT-derived second-order Eulerian structure functions. In the present numerical framework, we possess the exact area-averaged fields of both viscous and thermal dissipation, $\langle \epsilon_{\nu,T} \rangle_{A}(z,t)$, allowing us to decouple the relative contributions to the acceleration variance scaling $\accvar \sim \Ra^{\gamma}$. We connect both mechanisms via a mixing ratio $r\in [0,1]$, yielding the non-dimensional model}
\begin{equation}
    \label{eq:mixed_hy_model}
    \rev{\accvar = \left( a_{0} \epsnu^{3/2} (\Ra\Pr^{-1})^{1/4} \right)^r \left( a^{(T)}_{0} \epsT^{3/4} (\Ra\Pr)^{-1/8}\right)^{1-r}\,.}
\end{equation}
\rev{Given a constant $\Pr=0.7$ and absorbing $\Ra$-dependent scaling of both Heisenberg constants into a single exponent $\gamma$, the model reduces to}
\begin{equation}
    \label{eq:mixed_hy_model_fit}
    \rev{\accvar = C \Ra^{\gamma} \epsnu^{\frac{3}{2}r} \epsT^{\frac{3}{4}(1-r)}\,.}
\end{equation}
\rev{We fit $C$ and $\gamma$ directly to our acceleration variance data using a profile likelihood grid search over $r$. The results are illustrated in figure \ref{fig:a0}(c,f,i,l), where the insets represent the normalised residual sum of squares $\RSS^*=\RSS/\RSS_{\mathrm{min}}$ across $r$ and the dashed-black line the $95\%$ confidence threshold given an F-distribution.}

\rev{Within the total domain, the bulk and the core, thermal dissipation only exhibits a statistically significant contribution to the vertical direction in the bulk with an optimal mixing ratio of $r=0.82$. This is a direct consequence of thermal plumes and confirms the findings of \citet{Li2021} in this plume-abundant region. While the remaining fits in the total domain and the bulk find their optima strictly at $r=1$, the flat $\RSS^*$ does not provide sufficient evidence for the presence of thermal contribution in the core.}

\rev{In the thermal boundary layers, the most pronounced divergence from purely viscous dissipation occurs -- revealing a dominating contribution of thermal dissipation to horizontal acceleration variance with $r=0.18$. With $95\%$ confidence, the viscous contribution is bounded by $r\in [0,0.54]$. This is directly driven by intense temperature gradients ($\epsT \sim |\nabla T|^2$) close to the isothermal walls, demonstrating why the purely viscous HIT model fails in this region \citep{He2007, Emran2008}. To conclude this section, our data supports a mixed model of viscous and thermal dissipation driving acceleration events in an RBC setup in plume-abundant regions and within the thermal boundary layers.}

\subsection{Local Lagrangian heat transfer}
\label{subsec:Local_Lagrangian_heat_transfer}

%------------------------------------------------------------------------------------------
\begin{figure}
\centering
\includegraphics[width = \textwidth]{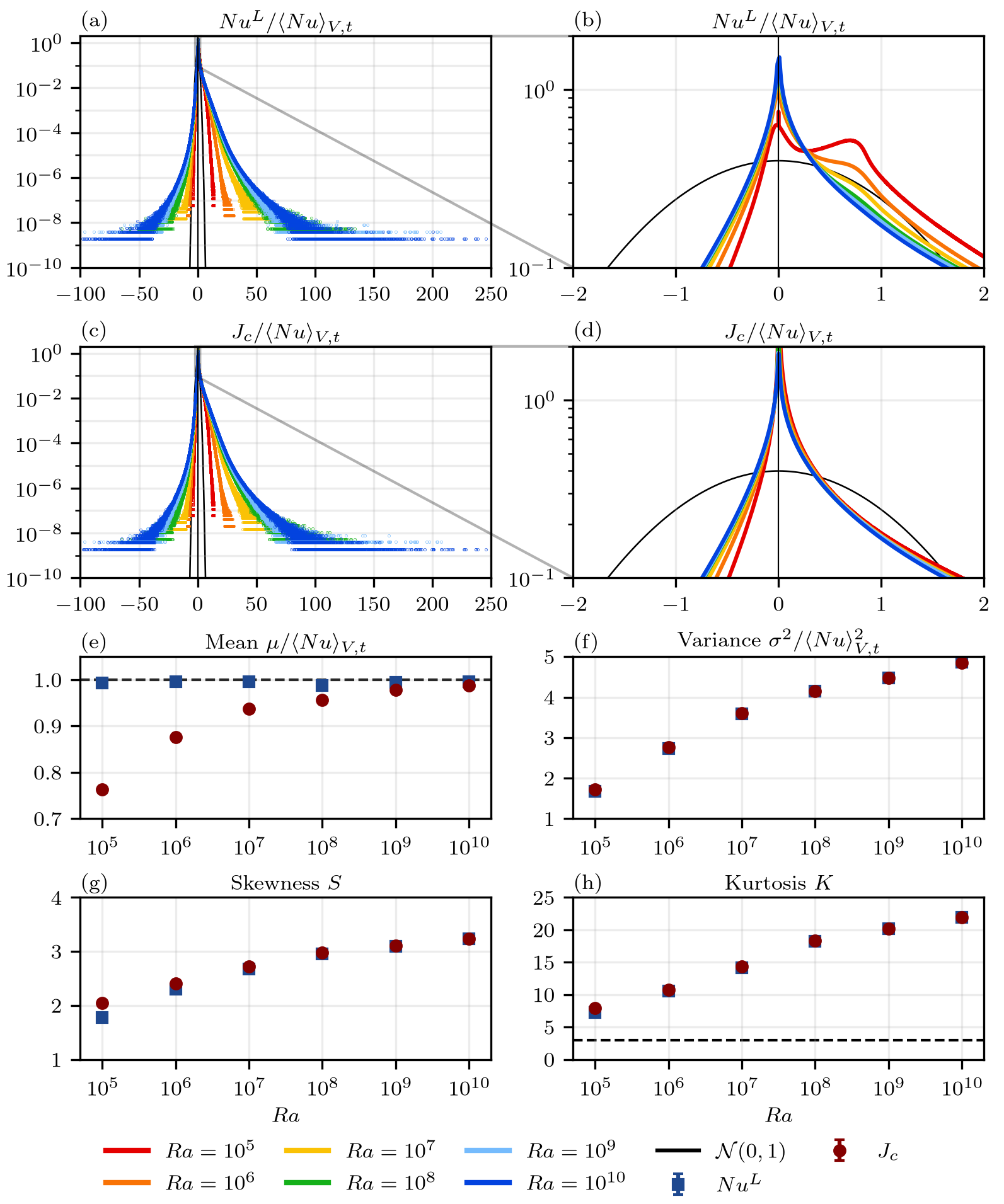}
\caption{\justifying{
Lagrangian \rev{heat transfer} statistics. Probability density functions (PDFs) of the Lagrangian Nusselt number $\NuL$ (a) and \rev{the purely convective heat transfer} $J_c$ (c) are shown, alongside zoom-ins of the PDFs around zero (b,d). The PDFs exhibit an asymmetric shape, underlining the dominance of gradient-aligned convective heat transfer. The structural difference between $\NuL$ and $J_c$ is only apparent in the zoom-ins, where the diffusive transfer term $- \partial T^L / \partial z$ \rev{blunts} the sharp peak at $\NuL=0$. Higher values of $\Ra$ lead to significantly wider tails, underscoring the increasing prevalence of rare but very intense heat flux fluctuations\rev{, as quantified by the normalised variance $\sigma^2 / \NuVt^2$ in panel (f) and kurtosis $K$ in panel (h). The normalised mean $\mu / \NuVt$ of $\NuL$ equals unity, confirming the conservation of macroscopic heat flux, while the normalised mean of $J_c$ monotonically approaches $1$, highlighting the predominant impact of convection at high $\Ra$. The positive, monotonically rising skewness in panel (g) quantifies the strong correlation between vertical velocity and temperature deviations (i.e., hot fluid rising, cold fluid falling) in buoyancy-driven systems.}
}}
\label{fig:NuL}
\end{figure}
%------------------------------------------------------------------------------------------

For the remainder of this section, we consider the extended trajectories of series P. We define the Lagrangian Nusselt number $\NuL$ in analogy to the traditional definition of the Eulerian Nusselt number by comparing the total (Lagrangian) heat flux of the system to its baseline conduction:
\begin{equation}
    \label{eq:NuL_dimensional}
    \NuL_{dim} (z,t) := \frac{\bm{J}^L \cdot \bm{e}_z}{\left \langle \bm{J} \cdot \bm{e}_z \right \rangle_A} = \frac{u_z^L \rev{\Theta^L} - \kappa \frac{\partial T^L}{\partial z}}{\kappa \frac{\Delta T}{H}}\,,
\end{equation}
where the superscript $L$ denotes quantities evaluated from a moving particle perspective. Note that the denominator is defined by the pure conduction profile without fluid motion in the Eulerian point-of-view. Using free-fall time units (see Section \ref{subsec:Governing_equations}), we obtain the non-dimensional Lagrangian Nusselt number and its purely convective component, the \rev{non-dimensional convective heat transfer}:
\begin{equation}
    \label{eq:NuL}
    \NuL (z,t) := \sqrt{\Ra \Pr}\, u_z^L \rev{\Theta^L} - \frac{\partial T^L}{\partial z} \qquad \textrm{and} \qquad \rev{J_c} (z,t) := \sqrt{\Ra \Pr}\, u_z^L \rev{\Theta^L}\,,
\end{equation}
where we omit the tildes for better readability. Crucially, the vertical particle velocity $u_z^L$, the particle \rev{temperature deviation $\Theta^L=T^L - \langle T \rangle_{A}(z^L)$}, and the vertical temperature gradient $- \partial T^L / \partial z$ are evaluated precisely at the continuously updated positions of the $2 \times 2^{20}$ individual Lagrangian particles, rather than at fixed Eulerian grid points. \rev{In contrast to prior Lagrangian studies, we compute the exact vertical temperature gradient along all Lagrangian trajectories instead of estimating $- \partial T^L / \partial z =1$ \citep{Gasteuil2007, Schumacher2009, Emran2010, Liot2016a}.} Computing a PDF of these \rev{metrics} thus allows for excellent statistical convergence. We normalise the PDFs of both quantities by the global Eulerian Nusselt number $\NuVt$ (see table \ref{tab:Series_scales}) to directly relate the instantaneous heat transfer achieved by a single particle to the global mean heat transfer of the system.

Figure \ref{fig:NuL}(a,c) illustrates the PDFs of the individual particles across thousands of snapshots for $\Ra \in [10^5, 10^{10}]$, contrasting them with the standard Gaussian distribution $\mathcal{N}(0,1)$ (black line). \rev{The observed stretched-exponential shape aligns well with prior experimental and numerical studies in convection, while our sample size and the resulting resolution of the far-tails are unprecedented \citep{Shang2003, Gasteuil2007, Schumacher2009, Liot2016a, Käufer2024}}. As $\Ra$ is increased, the distributions exhibit significantly wider tails, spanning as far as 250 times the global mean for the highest studied Rayleigh number. This indicates that the heat transfer of a single fluid parcel caught in a violent thermal plume can exceed the \rev{system's global mean 250-fold.} However, these extreme fluctuations are rare, with the vast majority of particles contributing very little to the overall global heat transfer at any given instant. This explains the sharp peaks of the PDFs around zero. Furthermore, a distinct asymmetry of the PDFs can be observed, quantified by the monotonically rising skewness of the distributions in figure \ref{fig:NuL}(g). All curves are heavily skewed into the positive domain, highlighting the strong correlation between velocity and temperature in the buoyancy-driven system (i.e., hot fluid rising, cold fluid sinking)\rev{, corroborating prior studies \citep{Shang2003, Gasteuil2007, Schumacher2009, Liot2016a, Käufer2024}}. Conversely, negative values correspond to counter-gradient heat transfer (i.e., cold fluid caught in an updraft). Typically, the majority of heat transfer occurs in the gradient-aligned direction, although counter-gradient mixing can be substantial, especially at higher $\Ra$.

Interestingly, the PDFs of \rev{Lagrangian Nusselt number $\NuL$ and the convective heat transfer $J_c$} are visually almost indistinguishable -- highlighting that the vast majority of heat transfer occurs via convection, whereas local microscopic diffusion $- \partial T^L / \partial z$ is statistically negligible over the domain volume. Figure \ref{fig:NuL}(b,d) presents zoom-ins of the probability peaks\rev{, where the impact of the diffusive term becomes apparent. While the purely convective heat transfer (see figure \ref{fig:NuL}d) exhibits a sharp peak precisely at $J_c=0$, this peak for the total $\NuL$ is blunted by diffusion, as shown in figure \ref{fig:NuL}(b). Especially at low $\Ra \lesssim 10^7$, this positively skewed second peak emerges, indicating that large-scale circulation rolls contribute significantly to the baseline heat transport in the system.} Since a large fraction of the background fluid in the bulk of the domain fluctuates about $\Theta^L \approx 0$, a sharp peak at $J_c=0$ occurs. Including the term $- \partial T^L / \partial z$ in $\NuL$ washes this central peak out via microscopic diffusion.

Figure \ref{fig:NuL}(e--h) presents \rev{the normalised mean $\mu / \NuVt$, the normalised variance $\sigma^2 / \NuVt^2$, and} higher-order moments, skewness $S$ and kurtosis $K$. The curves of $\NuL$ and $J_c$ converge across all metrics for $\Ra \gtrsim 10^8$, further underscoring the dominance of the convective heat transport \rev{in the so-called hard convective turbulent regime}. \rev{The normalised mean of $\NuL$ exactly equals unity across all $\Ra$, mathematically confirming the global conservation of macroscopic heat transfer. The normalised mean of $J_c$ similarly approaches this value at high $\Ra$, emphasising that convection predominantly drives global heat transfer as the thermal boundary layer shrinks. The normalised variance $\sigma^2 / \NuVt^2$ and kurtosis $K$ rise monotonically with $\Ra$, confirming the increase in intermittency that was previously observed for Lagrangian accelerations (see Section \ref{subsec:Acceleration_statistics}). While the existence of these intense fluctuations has been established by \citet{Shang2003, Gasteuil2007, Schumacher2009} and \citet{Emran2010}, a coherent analysis of higher-order moments of Lagrangian heat transfer has been unprecedented thus far.}

%------------------------------------------------------------------------------------------
\subsection{Velocity gradient invariant analysis and dust-devil-like convective vortices}
\label{subsec:Velocity_gradient_invariant_analysis_and_dust-devil-like_vortices}

%------------------------------------------------------------------------------------------
\begin{figure}
\centering
\includegraphics[width = \textwidth]{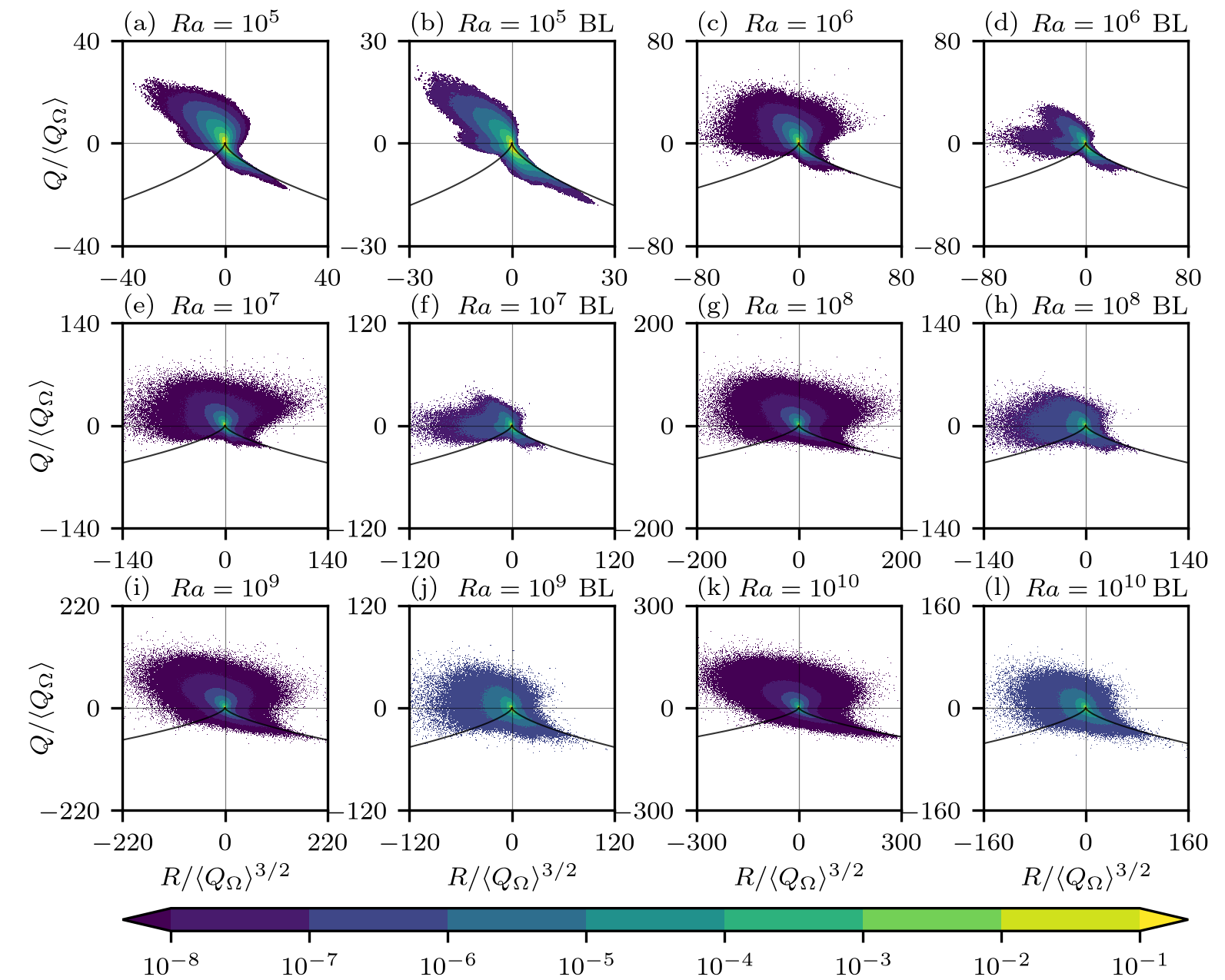}
\caption{\justifying{
Parameter plane spanned by the invariants $Q$ and $R$ of the velocity gradient tensor along Lagrangian particle trajectories. The normalised joint PDFs of the invariants of the velocity gradient tensor across $\Ra \in [10^5, 10^{10}]$ are shown for the total domain ($z \in [0, 1]$) and the thermal boundary layers (BL, $z \in [0, \delta_T] \cup [1-\delta_T, 1]$). Joint total PDFs (a,c,e,g,i,k) exhibit the classical teardrop shape and align with the Vieillefosse tail. Joint BL-PDFs (b,d,f,h,j,l) are skewed towards the upper left quadrant ($Q > 0, R < 0$), indicating an increase in vortex stretching events as particles enter the near-wall region.
}}
\label{fig:QR}
\end{figure}
%------------------------------------------------------------------------------------------

Although RBC is an idealised configuration, it contains several ingredients common to more complex convective and active-scalar flows: localised injection at boundaries, coherent plume emission, anisotropic transport, and the competition between organised structures and a turbulent background. For this reason, RBC provides a controlled setting in which Lagrangian methods developed in passive-scalar and homogeneous isotropic turbulence can be stress-tested in a more realistic, inhomogeneous, boundary-influenced environment. The subsequent velocity gradient analysis in the plane spanned by the two invariants $Q$ and $R$ bridges between the Lagrangian trajectory-based dynamics and local flow topology, the latter being characterised by the invariants of the velocity gradient tensor $A_{ij} = \partial u_i / \partial x_j$. A key advantage of the present spectral element framework is that $A_{ij}$ is computed spectrally and interpolated directly to particle positions. The tensor $A_{ij}$ can be decomposed into symmetric (rate-of-strain tensor $S_{ij}$) and antisymmetric (rotation-rate tensor $\Omega_{ij}$) parts, satisfying the Cayley--Hamilton equation
\begin{equation}
\label{eq:Cayley_Hamilton}
\bm{A}^3 - P \bm{A}^2 + Q \bm{A} -R \bm{I} = 0\,,
\end{equation}
where, for an incompressible flow, the invariants are \citep{Chong1990}
\begin{equation}
\label{eq:QR_invariants}
P = A_{ii} = 0\,, \hspace{0.3cm} Q = -\frac{1}{2}A_{ij}A_{ji} = \frac{1}{2} \left(-\Omega_{ij}\Omega_{ji} - S_{ij}S_{ji} \right), \hspace{0.3cm} R = -\frac{1}{3}A_{ij}A_{jk}A_{ki}\,.
\end{equation}
The discriminant $D = 4Q^3 + 27R^2 = 0$ separates the flow into regions of real and complex eigenvalues, highlighting distinct local flow topologies. The second invariant $Q$ quantifies the balance of enstrophy and strain, where regions of $Q>0$ are dominated by rotation whereas $Q<0$ indicates strain. $R$ separates the regions into dominance of stretching ($R<0$) or compression ($R>0$). This local topology is closely related to the tetrad-based material-deformation viewpoint of \citet{Chertkov1999}, but is directly evaluated along Lagrangian particle trajectories in this work. This Vieillefosse tail \citep{Vieillefosse1982}, $R = - \left( 4/27 Q \right)^{3/2}$, in particular highlights the prevalence of two topologies: unstable node/saddle/saddle ($Q<0, R>0$) -- associated with viscous dissipation -- and stable focus/stretching ($Q>0, R<0$), where vortex stretching dominates. The latter is of particular interest when characterising intense dust-devil-like convective vortices in the boundary layers, whose dynamics are distinctly characterised by this flow topology.

% Lagrangian $Q$--$R$ diagram
Figure \ref{fig:QR} shows Lagrangian joint PDFs of the invariants of the velocity gradient tensor ($Q$--$R$ diagrams) for $\Ra \in [10^5, 10^{10}]$ across the total domain ($z \in [0, 1]$) and conditioned to the thermal boundary layers ($z \in [0, \delta_T] \cup [1-\delta_T, 1]$). Following \citet{Ooi1999}, all diagrams were normalised by the mean second invariant of the antisymmetric rotation rate tensor $\langle Q_{\Omega}\rangle = \frac{1}{2} \langle \Omega_{ij} \Omega_{ji} \rangle$. This renders $Q / \langle Q_{\Omega}\rangle$ and $R / \langle Q_{\Omega}\rangle^{3/2}$ dimensionless, preserving the geometrical topology of the Vieillefosse tail across all Rayleigh numbers to allow direct comparison. The $Q$--$R$ diagrams of the total domain exhibit the classical teardrop shape typically obtained in HIT and Eulerian RBC, aligning with the Vieillefosse tail \citep{Chertkov1999, Dabbagh2016, Vieillefosse1982}. The $Q$--$R$ space explored by Lagrangian particles expands with growing $\Ra$, underlining the increased prevalence of rare, intermittent events. When considering the $Q$--$R$ diagrams conditioned to the thermal boundary layers, the joint PDFs become skewed towards values of $R < 0$, particularly into the upper left quadrant ($Q > 0, R < 0$), indicating an increase in vortex stretching events as particles enter the near-wall region. This distinct topological signature suggests the existence of intense convective vortices -- such as dust-devil-like ones -- in proximity to the wall.

% Dust devils

%------------------------------------------------------------------------------------------
\begin{figure}
\centering
\includegraphics[width = \textwidth]{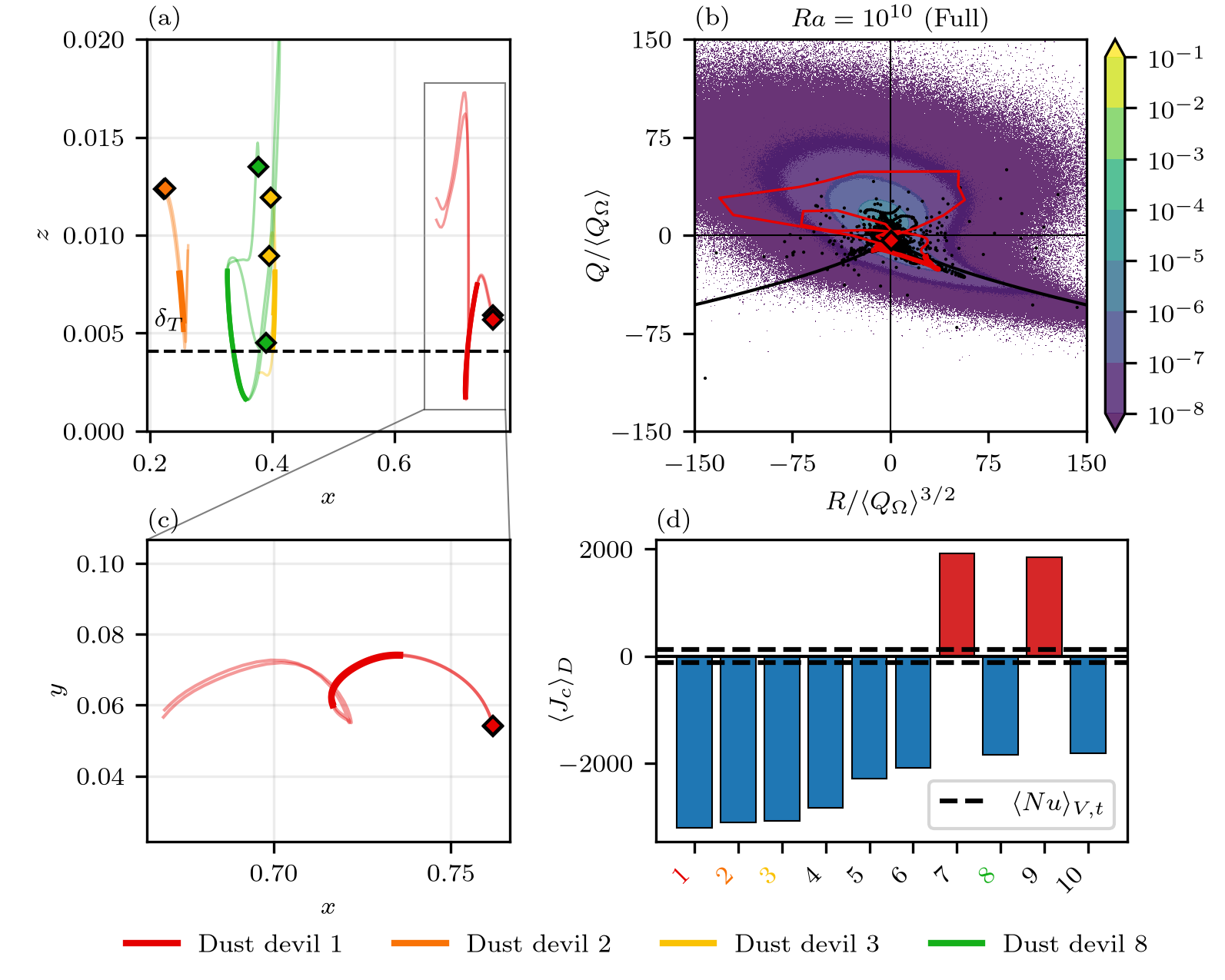}
\caption{\justifying{
Analysis of dust-devil-like convective vortices \rev{at $\Ra=10^{10}$}. (a) Projected trajectories ($x-z$ plane) of four selected individual events \rev{(dust devils \rev{1, 2, 3 and 8}) relative to the thermal boundary layer thickness $\delta_T$ (black dashed line)}. Thick line segments correspond to the active entrainment phase of Lagrangian particles within these convective vortices, while thin lines are trajectory extensions of 50 snapshots ($0.5 \tau_f$) in both directions, highlighting approach and ejection phases. \rev{Diamonds mark the trajectory origins.} (b) Dust devils in the $Q$--$R$ plane. Black dots show the instantaneous topology of all \rev{$57$} events. The continuous red trajectory exemplifies the life cycle of a single particle from event \rev{1} in the $Q$--$R$ plane, underscoring the topological footprint dust-devil-like convective vortices leave via a rapid, vigorous excursion into the upper-left quadrant ($Q>0, R<0$, vortex stretching). (c) Top-down view ($x$--$y$ plane) of dust devil \rev{1} of panel (a), highlighting the characteristic helical swirling motion. (d) Dust-devil-like convective vortices lead to significant enhancements of local convective heat flux \rev{$\JcD$}, exceeding the global Eulerian mean $\NuVt$ (black, dashed) \rev{by nearly $30$ times}.
}}
\label{fig:Dust_devils}
\end{figure}
%------------------------------------------------------------------------------------------

We treat dust-devil-like convective vortices (hereafter referred to simply as `dust devils') as a representative example of a broader diagnostic strategy: identify intense Lagrangian events, condition them on their spatial and temporal coherence, and then examine their trajectory-level topology in the $Q$--$R$ plane. This avoids relying solely on Eulerian snapshots of vortical structures and highlights whether a moving fluid parcel experiences a reproducible dynamical pathway during entrainment, swirling, and ejection. To extract Lagrangian particles that meet the signature of dust devils, we apply a conditional filtering algorithm to the data of our highest Rayleigh number, $\Ra=10^{10}$. 

A particle is considered trapped within a dust devil if it simultaneously meets the following criteria:
\begin{enumerate}
    \item \textit{Locality:} $z \leq 2 \delta_T$. The particle must reside close to the lower thermal boundary layer, where dust devils originate.
    \item \textit{Alignment:} $\rev{\Bz =} |\bm{B} \cdot \bm{e}_z| \geq \rev{0.93}$. The unit binormal vector $\bm{B}= (\bm{u} \times \bm{a}) / |\bm{u} \times \bm{a}|$ of the particle must strongly align with the vertical axis, eliminating intense horizontal stretching events.
    \item \textit{Intensity:} The centripetal particle acceleration $a_n = | \bm{u} \times \bm{a} | / |\bm{u}|$ must be in the 99th percentile of $a_n$ \rev{($\Pan = 99.0 \%$)} of all values recorded within $z \in [0, 2\delta_T]$, isolating only the most extreme rotational events.
    \item \textit{Temporality:} Persistence of the criteria (i)--(iii) for at least \rev{$\nsnp \geq 20$} consecutive snapshots (\rev{$0.20\tau_f$}) to exclude short-term turbulent fluctuations.
\end{enumerate}
Subsequently, particles that overlap both temporally and spatially are clustered into dust devils. They must coexist within a horizontal radial distance of $\sqrt{(x_1-x_2)^2+(y_1-y_2)^2} \leq 4 \delta_T$ \rev{($\rfac=4$)} for at least one snapshot. Over the entire runtime of case P10 (see table \ref{tab:simulation_parameters}), \rev{$N_D = 57$} individual events were identified. \rev{A sensitivity analysis regarding the choice of the aforementioned extraction criteria is given in Appendix \ref{subsec:Appendix_Sensitivity_of_dust-devil-like_convective_vortices}.}

Figure \ref{fig:Dust_devils} illustrates the topological and heat transfer-related impact of dust devils on the flow at $\Ra=10^{10}$. Figure \ref{fig:Dust_devils}(a) presents the projected trajectories of four selected dust devils in the $x-z$ plane. Thick line segments highlight the time frame during which the particles satisfy the extraction criteria -- the active entrainment phase -- while the thin lines extend the Lagrangian trajectories over 50 snapshots ($0.5\tau_f$) in both directions, revealing the approach and ejection phases of the dust devils. Diamonds mark the Lagrangian origins of the tracks. Figure \ref{fig:Dust_devils}(c) provides a top-down ($x$--$y$) projection of a representative dust devil (dust devil \rev{1}), explicitly confirming the helical swirling motion that is characteristic of such a convective vortex. These extended Lagrangian trajectories illustrate the full life cycle of a fluid parcel that is entrained into a dust devil: following rapid entrainment into the convective vortex, the parcel experiences intense helical swirling -- causing extreme centripetal acceleration -- before being violently ejected into the bulk of the fluid. This is highlighted by the long, straight exit trajectories.

As shown in figure \ref{fig:Dust_devils}(d), this entrainment results in intense convective heat transfer, \rev{$\JcD$}, exceeding the global Eulerian Nusselt number $\NuVt$ by \rev{nearly $30$ times} for the most intense vortices. Notably, these localised heat fluxes can be both strongly positive and negative: the strong rotation acts as a powerful vacuum for fluid parcels, capable of counteracting thermal buoyancy before forcefully releasing the fluid upward into the bulk.

Figure \ref{fig:Dust_devils}(b) illustrates the topology in the $Q$--$R$ plane. Black dots represent all instantaneous states of particles within the \rev{57} extracted dust devils during the active entrainment phase. The continuous red trajectory exemplifies the evolution of \rev{dust devil 1}. Initially located near the origin (background shear), the particle undergoes a rapid, deep excursion into the upper-left quadrant ($Q>0, R<0$, vortex stretching). Unequivocally, the entrainment in dust devils leaves a distinct topological footprint in the $Q$--$R$ plane and explains the notable shift towards vortex stretching observed in the joint PDFs conditioned to the thermal boundary layers in figure \ref{fig:QR}. Ultimately, we have confirmed the existence of intense dust-devil-like convective vortices in the high-Rayleigh-number regime at $\Ra=10^{10}$ that originate close to the lower thermal boundary layer. They leave a characteristic topological footprint in the upper-left quadrant of the $Q$--$R$ plane and contribute significantly to the convective heat flux of the system.

%------------------------------------------------------------------------------------------
\begin{figure}
\centering
\includegraphics[width = \textwidth]{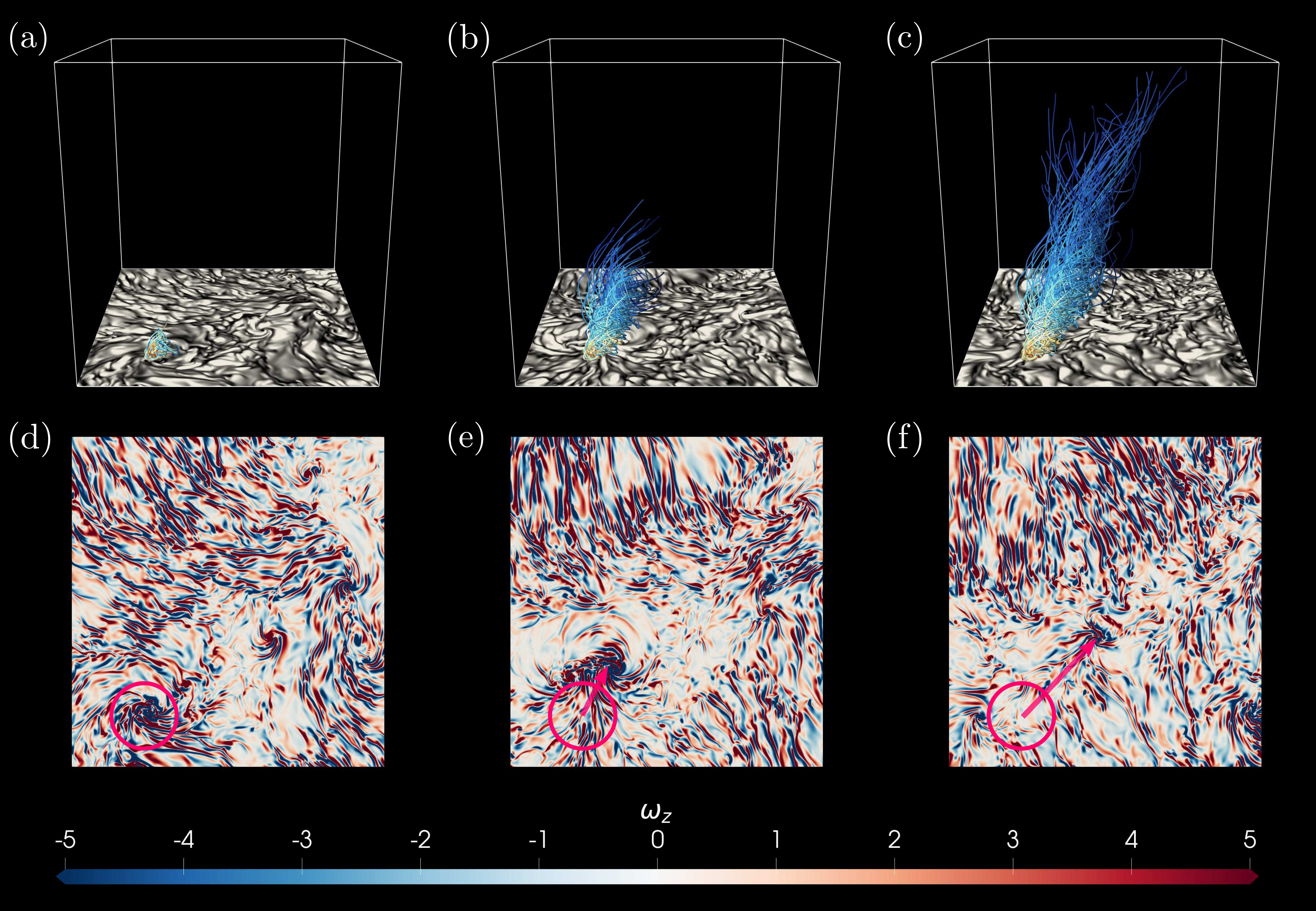}
\caption{\justifying{\rev{
Visualization of a cloud of tracers dispersed by a dust devil from the bottom wall at $\Ra=10^{10}$. The three columns from left to right span $4$ free-fall time units of evolution (at $t=2251.3$, $2253$ and $2255$, respectively) during which the dust-devil moves diagonally across the wall, ejecting the tracers into the bulk as it progresses. A spherical cloud of radius $0.02$, containing $10,\!000$ particles, is seeded in the eye of the dust devil (left column) at a height of $z=6\delta_T$. The initial position of the dust-devil was identified from a preliminary run as $(x, y) = (0.265, -0.395)$, which was chosen as the particle seeding location. The box chosen for visualization is fixed in space and spans $[-0.4, 0.4] \times [-0.6, 0.2] \times [0.0, 0.8]$ in $x$, $y$ and $z$, respectively. The tracers are coloured by temperature in (a), (b) and (c), where the temperature contours in grey at the bottom are sampled at $z=0.001$. The bottom row (d--f) shows contours of the vertical vorticity component $\omega_z$ at $z=0.001$, highlighting the initial footprint region of the dust devil via a pink circle, while the pink arrows in (e) and (f) point to subsequent positions of the maximum swirling region as the dust devil progresses following tracer cloud detachment.
}}}
\label{fig:Tornado_cloud}
\end{figure}
%------------------------------------------------------------------------------------------

\rev{The better visualise the immediate impact of dust-devil-like convective vortices, figure \ref{fig:Tornado_cloud} further demonstrates how a cloud of tracers seeded at the centre of a dust devil is dispersed into the bulk. Here, we perform a short simulation at $\Ra = 10^{10}$ spanning $4$ free-fall time units. Before this, a preliminary run at the same parameters was conducted without tracers to identify a strongly localised circulation region, indicative of a dust devil event, at the bottom plate. This region was found to be centred at $(x, y) = (0.265, -0.395)$ at $t = 2251$ and subsequently chosen to be the instantaneous eye of the dust devil to seed the particle cloud in. The spherical cloud of radius $0.02$ units, comprised of $10,\!000$ particles, is released at a height of $z=6\delta_T$. The visualization of the dispersion and corresponding movement of the dust devil on the plate is depicted in figure \ref{fig:Tornado_cloud} over a fixed window of $[-0.4, 0.4] \times [-0.6, 0.2] \times [0.0, 0.8]$ (within the full domain of $[-2, 2] \times [-2, 2] \times [0, 1]$), in $x$, $y$ and $z$, respectively. Figure \ref{fig:Tornado_cloud}(a--c) shows the particle tracks of $100$ randomly chosen tracers, coloured by temperature, at $t=2251.3$, $2253$ and $2255$, respectively. The contour slices at the bottom represent the temperature fields at $z=0.001$, indicating the plumes swept along by the near-wall flow. Correspondingly, figures \ref{fig:Tornado_cloud}(d--f) illustrate the $z$-component of vorticity, $\omega_z$, again at $z=0.001$, showing the high-circulation footprint of the dust devil as it moves across the plate. The initial position of the dust devil is marked by a fixed pink circle in all three frames, whereas the subsequent positions are marked by the pink arrows leading away from the centre of the circle. Crucially, the tracers disperse in tandem with the translating dust devil, imparting a pronounced tilt to the column as it stretches vertically and rises into the bulk.}

This result illustrates the value of Lagrangian topology for convection: the same coherent structure is seen simultaneously as a trajectory-level acceleration burst, a localised heat-flux event, and a distinctive excursion in the $Q$--$R$ plane. The convergence of these three diagnostics makes the structure physically interpretable rather than merely visually identifiable. \rev{The evolution of the tracer cloud in figure \ref{fig:Tornado_cloud} highlights that coherent near-wall ejections do not merely advect individual tracers, but rather subject finite material fluid volumes to intense, anisotropic stretching before dispersing them into the bulk. Quantifying how these localised ejections close to the thermal boundary layers and turbulent eddies in the bulk of the domain affect particle separation and three-dimensional volumetric deformation across scales necessitates an expansion from single-particle trajectories to two-particle pair dispersion (Section \ref{sec:Lagrangian_dual_particle_dynamics}) and the statistical geometry of multi-particle clouds (Section \ref{sec:Lagrangian_multi-particle_dynamics}).}

\section{Lagrangian dual-particle dynamics}
\label{sec:Lagrangian_dual_particle_dynamics}

\subsection{\rev{Single-particle and pair dispersion}}
\label{subsec:Single-particle_and_pair_dispersion}

%------------------------------------------------------------------------------------------
\begin{figure}
\centering
\includegraphics[width = \textwidth]{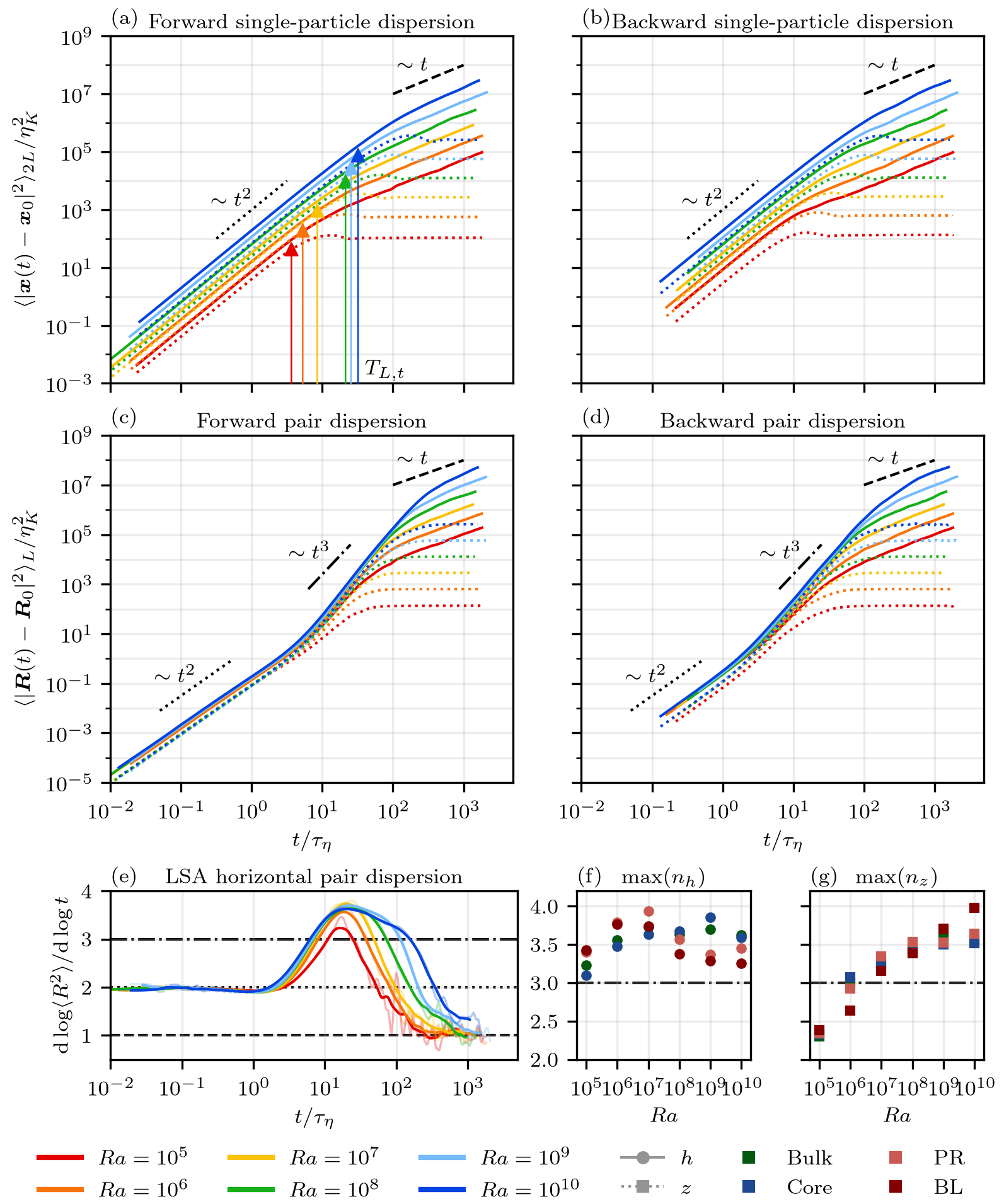}
\caption{\justifying{
\rev{Single-particle and pair dispersion. Single-particle dispersion $\langle |\bm{x}(t) - \bm{x}_0|^2 \rangle_{2L}$ (a,b) and relative particle pair dispersion $\langle | \bm{R}(t) - \bm{R}_0|^2\rangle_L$ (c,d) are shown forward (a,c) and backward (b,d) in time for horizontal (solid lines) and vertical (dotted lines) components, using $L=2^{20}$ particle pairs seeded randomly throughout the laterally periodic domain ($\Gamma=4$) with an initial separation of $R_0=\eta_K$. All metrics are normalised by the Kolmogorov scales $\eta_K$ and $\tau_{\eta}$. The total Lagrangian integral time $T_{L,t}$ is indicated by coloured arrows in panel (a). While single particles exclusively exhibit two scaling regimes (ballistic $\sim t^2$ and diffusive $\sim t$), relative pair dispersion reveals the intermediate inertial subrange and faster separation in the backward direction. A local slope analysis of horizontal forward pair dispersion (e) (thin lines represent raw data, thick lines a Savitzky--Golay filter) reveals that Richardson scaling ($\sim t^3$) is transiently exceeded for an initial separation of $R_0=\eta_K$. The peak scaling exponents $\max(n)$ within $t^n$ in the horizontal (f) and vertical (g) directions are shown for conditioned  particle seeding in the bulk ($z \in [\delta_T, 1-\delta_T]$, dark green), the core ($z \in [0.4, 0.6]$, blue), the plume root zone ($z \in [\delta_T, 2\delta_T] \cup [1-2\delta_T, 1-\delta_T]$, PR, salmon) and the thermal boundary layers ($z \in [0, \delta_T] \cup [1-\delta_T, 1]$, BL, oxblood).}
}}
\label{fig:Dispersion}
\end{figure}
%------------------------------------------------------------------------------------------

% --- 1. Introduction --- 
%------------------------------------------------------------------------------------------
Having investigated small-scale intermittency for individual particles, we proceed to probe distinct dispersion regimes in turbulence by examining \rev{absolute single-particle dispersion $\langle | \bm{x}(t) - \bm{x}_0 |^2 \rangle_{2L}$ and} relative particle pair dispersion $\langle | \bm{R}(t) - \bm{R}_0|^2 \rangle_L$. \rev{Here,} $\bm{R}(t) = \bm{x}_2(t) - \bm{x}_1(t)$ \rev{represents} the separation vector between two particles of a given pair, $\left \langle \cdot \right \rangle_L$ denotes the ensemble average over all $L=2^{20}$ particle pairs \rev{and the subscript $0$ denotes the initial value at $t=0$}. 

\rev{In homogeneous isotropic turbulence, \citet{Richardson1926} empirically discovered that the mean squared separation between two particles, $\langle R^2(t) \rangle$ with $R=|\bm{R}|=|\bm{x}_2-\bm{x}_1|$, does not grow diffusively ($\sim t$), but rather follows a $t^3$-scaling until their trajectories become uncorrelated. The dimensional framework of \citet{Kolmogorov1941} (K41) reveals that this inertial-range dispersion is solely governed by the turbulent kinetic energy dissipation rate $\epsilon_{\nu}$ and scales according to}
\begin{equation}
    \label{eq:Richardson_scaling_t}
    \rev{\langle R^2(t)\rangle \sim \langle\epsilon_{\nu}\rangle t^3\,.}
\end{equation}
\rev{This classical scaling is directly derived from the velocity structure function $\delta u(R) \sim \langle \epsilon_{\nu} \rangle^{1/3} R^{1/3}$. While accurately describing shear-driven mixing, it does not account for the active role of buoyancy. Consequently, \citet{Bolgiano1959} and \citet{Obukhov1959} (BO59) independently extended the model to account for the effects of thermal dissipation $\epsilon_T$ and buoyancy via the coupling constant $\beta = \alpha g$ in stably stratified turbulence (SST). In a buoyancy-dominated subrange, the velocity difference $\delta u(R) \sim \beta^{2/5} \epsilon_T^{1/5} R^{3/5}$ presumably scales more steeply, implying the Bolgiano--Obukhov scaling}
\begin{equation}
    \label{eq:Bolgiano_Obukhov_scaling_t}
    \rev{\langle R^2(t)\rangle \sim \beta^2 \epsilon_T t^5\,.}
\end{equation}
\rev{The Bolgiano length}
\begin{equation}
    \label{eq:Bolgiano_length}
    \rev{L_B(\bm{x}) = \frac{\epsilon_{\nu}^{5/4} (\bm{x})}{\epsilon_T^{3/4}(\bm{x})\beta^{3/2}}} \qquad \rev{\mathrm{or}} \qquad \rev{L_B=\frac{\left(\Nu -1 \right)^{5/4}}{\Nu^{3/4} \Ra^{1/4} \Pr^{1/4}}}
\end{equation}
\rev{nominally marks the transition point between the inertial and buoyancy-driven regimes. While at smaller separations, $10\eta_K < R < L_B$, shear and inertial forces dictate the energy cascade and thus suggest a $t^3$-scaling, buoyant forces associated with thermal plumes presumably dominate the separation dynamics for scales $L_B < R < L$ and lead to $t^5$-scaling \citep{Lohse2010, Effinger1987}.}

\rev{While the existence of a continuous, global BO59 cascade has been refuted in the Eulerian framework of three-dimensional RBC (see Section \ref{sec:Introduction}), the temporal and spatial scaling predictions in equations \eqref{eq:Bolgiano_Obukhov_scaling_t} and \eqref{eq:Bolgiano_length} remain important mathematical benchmarks in the Lagrangian frame. Because particles are actively advected by thermal plumes where $L_B$ is locally small, Lagrangian dispersion may exhibit transient, super-diffusive scaling reminiscent of BO59 behaviour.}

\rev{In this section, we examine this phenomenon and isolate it from finite initial separation effects and scale-separation constraints. To this end, we investigate $\langle | \bm{R}(t) - \bm{R}_0|^2 \rangle_L$ rather than dispersion from a point source ($R_0=0$) as originally postulated by \citet{Richardson1926}, since flow inhomogeneity implies $\left \langle | \bm{R}(t) - \bm{R}_0|^2 \right \rangle_L \neq \langle R^2 \rangle_L - R_0^2$ for $R_0\neq0$ \citep{Ouellette2006, Salazar2009}. Furthermore, we decompose dispersion into horizontal ($h$) and vertical ($z$) components to probe the anisotropic behaviour of convection flows \citep{Schumacher2009, Emran2010, Ni2013}. Finally, to fully disentangle true turbulent dispersion from one-dimensional plume entrainment, we complement temporal pair dispersion with scale-conditioned Lagrangian eddy viscosity in Section \ref{sec:Lagrangian_eddy_viscosity} and extend our framework to multi-particle dispersion and a topological analysis of dense particle clouds in Section \ref{sec:Lagrangian_multi-particle_dynamics}.}

% --- 2. Single-particle dispersion --- 
%------------------------------------------------------------------------------------------
\begin{table}
\color{black}
\begin{center}
\begin{tabular}{lccccccccccc}
Case & $\Ra$ & $T_{L,h}$ & $T_{L,z}$ & $T_{L,t}$ & $\tau_{\eta}$ & $B$ & $L_B$ & $r_0$ & $T_{L,t} / \tau_{\eta}$ & $B/r_0$ & $L_B/\eta_K$  \\
\hline
\\[-8pt]
\multicolumn{12}{l}{\textit{Series P: Pair Statistics \rev{($R_0=1\eta_K$}})} \\
P5 & $10^{5}$ & 2.01 & 1.30 & 1.69 & 0.464 & 0.765 & 0.091 & 0.035 & 3.6 & 21.8 & 2.6 \\
P6 & $10^{6}$ & 2.08 & 1.22 & 1.65 & 0.313 & 0.877 & 0.084 & 0.016 & 5.3 & 54.2 & 5.2 \\
P7 & $10^{7}$ & 2.48 & 1.29 & 1.86 & 0.219 & 0.936 & 0.071 & 0.008 & 8.5 & 123.0 & 9.3 \\
P8 & $10^{8}$ & 3.73 & 2.16 & 3.29 & 0.155 & 0.967 & 0.058 & 0.004 & 21.3 & 268.9 & 16.0 \\
P9 & $10^{9}$ & 3.32 & 1.71 & 2.77 & 0.109 & 0.983 & 0.047 & 0.002 & 25.5 & 580.0 & 27.6 \\
P10 & $10^{10}$ & 2.84 & 1.75 & 2.44 & 0.076 & 0.992 & 0.038 & 0.001 & 32.1 & 1244.0 & 47.5 \\
\\[-5pt]
\multicolumn{12}{l}{\rev{\textit{Series C$_1$: Cloud Geometry ($r_0=1\eta_K$})}} \\
C$_1$4 & $10^{4}$ & 5.50 & 3.73 & 4.64 & 0.737 & 0.563 & 0.081 & 0.079 & 6.3 & 7.2 & 1.0 \\
C$_1$5 & $10^{5}$ & 5.01 & 3.05 & 4.07 & 0.462 & 0.766 & 0.091 & 0.035 & 8.8 & 21.9 & 2.6 \\
C$_1$6 & $10^{6}$ & 5.58 & 3.40 & 4.97 & 0.313 & 0.877 & 0.084 & 0.016 & 15.9 & 54.2 & 5.2 \\
C$_1$7 & $10^{7}$ & 5.14 & 3.35 & 4.65 & 0.220 & 0.936 & 0.070 & 0.008 & 21.2 & 122.7 & 9.2 \\
C$_1$8 & $10^{8}$ & 5.00 & 3.45 & 4.73 & 0.154 & 0.967 & 0.058 & 0.004 & 30.7 & 269.0 & 16.1 \\
\\[-5pt]
\multicolumn{12}{l}{\rev{\textit{Series C$_4$: Cloud Geometry ($r_0=4\eta_K$})}} \\
C$_4$4 & $10^{4}$ & 7.21 & 5.68 & 6.90 & 0.742 & 0.560 & 0.080 & 0.315 & 9.3 & 1.8 & 1.0 \\
C$_4$5 & $10^{5}$ & 4.88 & 3.04 & 4.01 & 0.462 & 0.767 & 0.091 & 0.140 & 8.7 & 5.5 & 2.6 \\
C$_4$6 & $10^{6}$ & 4.60 & 2.62 & 4.02 & 0.313 & 0.877 & 0.084 & 0.065 & 12.8 & 13.5 & 5.2 \\
C$_4$7 & $10^{7}$ & 5.05 & 3.31 & 4.57 & 0.219 & 0.936 & 0.071 & 0.030 & 20.8 & 30.7 & 9.3 \\
C$_4$8 & $10^{8}$ & 4.93 & 3.39 & 4.65 & 0.154 & 0.967 & 0.058 & 0.014 & 30.1 & 67.3 & 16.1 \\
\end{tabular}
\caption{\justifying{\rev{Temporal and spatial scale separation. The combined horizontal ($h$), vertical ($z$), and all ($t$) velocity components are used to compute the Lagrangian integral time scale as the first zero-crossing of the Lagrangian velocity autocorrelation function. These times are reported alongside the Kolmogorov time scale $\tau_{\eta}$. The macroscopic bulk depth $B=1-2\delta_T$, the global Bolgiano length $L_B$, and the initial pair separation ($R_0$, P) or cloud seeding radius ($r_0$, C) are shown. The temporal scale separation $T_{L,t} / \tau_{\eta}$ bounds the temporal window for the inertial subrange, $B/r_0$ quantifies the spatial runway between the initial seeding scale and the macroscopic domain bounds, and $L_B/\eta_K$ marks the nominal crossover scale to buoyancy-driven dispersion.}}}
\label{tab:Scale_separation}
\end{center}
\end{table}
%------------------------------------------------------------------------------------------

\rev{Figure \ref{fig:Dispersion}(a,b) illustrates single-particle dispersion forward and backward in time across $\Ra \in [10^5, 10^{10}]$ in horizontal (solid lines) and vertical (dotted lines) components.} Following the theory of \citet{Taylor1921}, single-particle dispersion exhibits two distinct regimes. At early times ($t \ll T_L$), particle velocities remain highly auto-correlated with their initial velocity $\bm{v}_0$ and disperse in a kinematic, ballistic manner ($\sim t^2$). At late times ($t \gg T_L$), once particles fully lose memory of this initial velocity, their trajectories consolidate into a macroscopic random walk, resulting in the diffusive regime ($\sim t$).

% --- 3. T_L ---
%------------------------------------------------------------------------------------------
\rev{Here, the Lagrangian integral time scale}
\begin{equation}
    \label{eq:T_L}
    \rev{T_L = \int_0^{\infty} C(\tau) \,d\tau \qquad \mathrm{with} \qquad  C(\tau) = \frac{\left \langle \bm{v}(t+\tau)\bm{v}(t)  \right \rangle_L}{\left \langle \bm{v}(t)\bm{v}(t) \right \rangle_L }}
\end{equation}
\rev{represents the time span over which Lagrangian particles achieve decorrelation from their initial velocity. We report the values for the horizontal, vertical and total directions in table \ref{tab:Scale_separation}, computed via the first zero-crossing of the Lagrangian velocity autocorrelation function $C(\tau)$ to account for periodic domain confinement and large-scale recirculation. Notably, particles remain correlated significantly longer in the horizontal than in the vertical direction, which is a direct consequence of the laterally periodic boundary conditions and vertical wall confinement. A corresponding increase of $T_{L,h}$ with aspect ratio $\Gamma$ was similarly observed by \citet{Emran2010} and is corroborated by Series C$_1$ and C$_4$ employing a larger aspect ratio of $\Gamma=16$ (see table \ref{tab:Scale_separation}). Coloured arrows in figure \ref{fig:Dispersion}(a) indicate the normalised total Lagrangian integral time scale $T_{L,t}/\tau_{\eta}$ and thus the onset of the transition toward the diffusive regime, where theoretical asymptotes intersect at $t = 2T_{L,t}$. Because walls restrict the domain in the vertical direction whereas periodic boundary conditions are applied laterally, total dispersion at late times is predominately driven by its horizontal component \citep{Schumacher2009}.}

%------------------------------------------------------------------------------------------
\begin{figure}
\centering
\includegraphics[width = \textwidth]{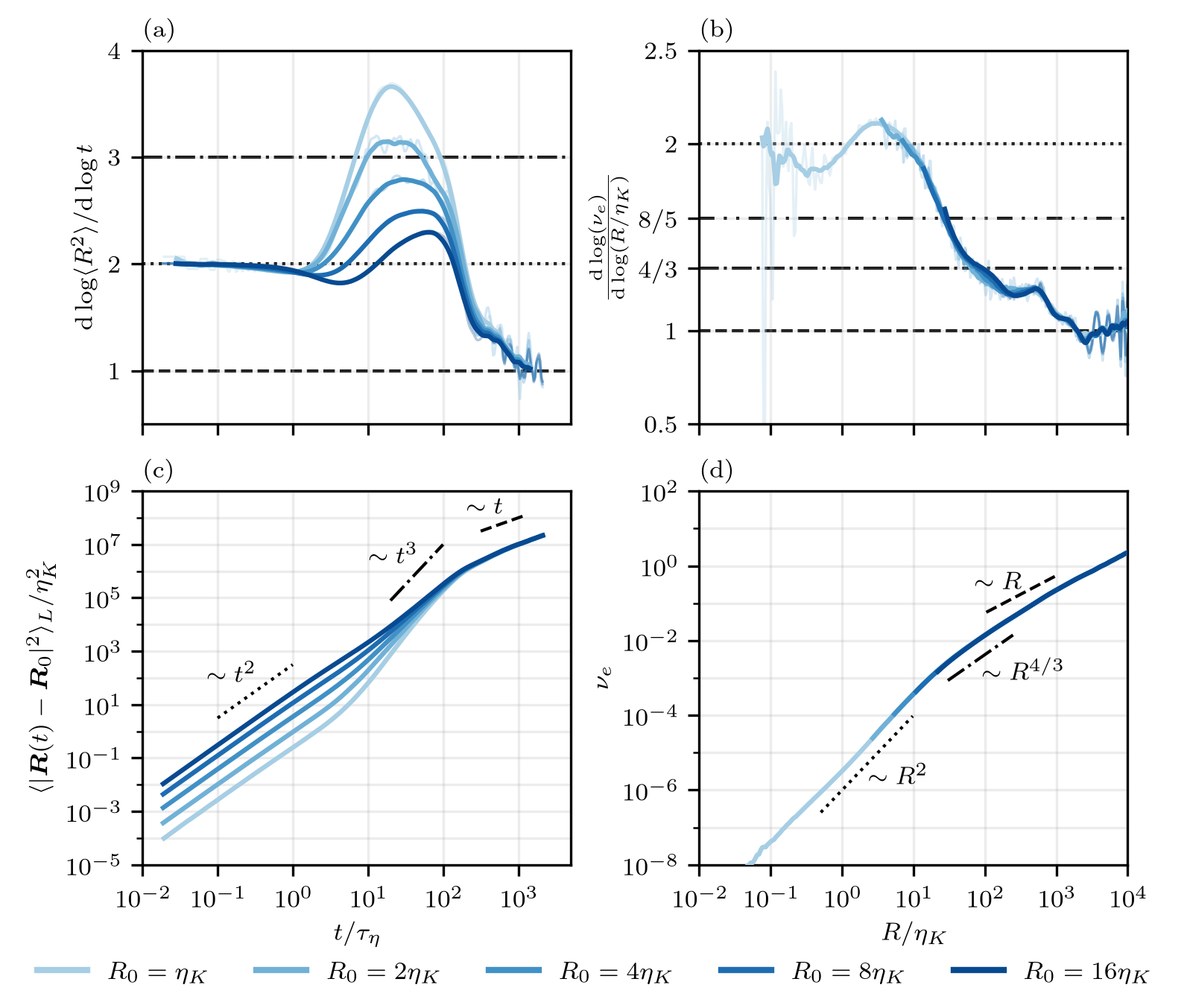}
\caption{\justifying{
\rev{Sensitivity of temporal forward pair dispersion (a,c) and scale-dependent Lagrangian eddy viscosity (b,d) to different initial separations $R_0 \in \{1\eta_K, 2\eta_K, 4\eta_K, 8\eta_K, 16\eta_K\}$ at $\Ra=10^9$. Thin lines represent raw data, thick lines a Savitzky--Golay filter. (a,b) The presence of super-diffusive Richardson scaling of pair dispersion ($\sim t^3$) is systematically influenced by the initial separation of particle pairs. Sub-Kolmogorov or near-viscous separations ($R_0 \lesssim 2 \eta_K$) induce kinematic `catch-up' effects and a transient overshoot above $n=3$ \citep{Boffetta2002, Sawford2008, Bitane2012, Liot2019}, whereas large initial separations bypass the inertial subrange entirely. (c,d) Conditioning directly to spatial separation via scale-dependent Lagrangian eddy viscosity removes temporal seeding bias once particles align with the local strain field ($R \gtrsim 1.3 R_0$), causing a collapse of all curves across different $R_0$. The pre-alignment regimes ($R < 1.3 R_0$) are omitted in $\nu_e(R)$ to exclude inward-turnaround and initial strain-alignment transients.}
}}
\label{fig:Initial_separations}
\end{figure}
%------------------------------------------------------------------------------------------

Crucially, single-particle dispersion does not exhibit a super-diffusive inertial subrange ($n>2$ within $\sim t^n$) \rev{as proposed by Richardson's law for pair dispersion}, because individual Lagrangian particle velocities are strictly bounded by the underlying Eulerian velocity field. Furthermore, the single-particle dispersion curves in the forward and backward directions do not differ, as absolute dispersion lacks the two-point spatial conditioning required to reveal the temporal asymmetry. This fundamental limitation of single-particle dispersion underlines why relative dispersion is a vital tool to investigate the intricate dynamics of convective Lagrangian turbulence.

% --- 4. Pair Dispersion ---
Figure \ref{fig:Dispersion}(c) illustrates forward pair dispersion across $\Ra \in [10^5, 10^{10}]$. Initially seeded at \rev{$R_0=\eta_K$}, the velocities of the paired particles remain highly correlated. Separation at early times is governed by the local initial velocity difference $\delta \bm{u}^L_0 = \bm{u}^L_2(0) - \bm{u}^L_1(0)$, leading to a kinematic ballistic regime \rev{with} $t^2$-scaling for $t < \tau_{\eta}$ (see \citealt{Bourgoin2015}). As time exceeds the Kolmogorov time scale $\tau_{\eta}$, particles separate beyond the dissipative scales $\eta_K$ and are subjected to turbulent eddies as they enter the inertial subrange. In homogeneous isotropic turbulence (HIT), shear forces dominate the energy cascade, resulting in Richardson scaling $\sim t^3$ \eqref{eq:Richardson_scaling_t}. Finally, at times sufficiently longer than the Lagrangian integral time scale ($t \gg T_L$), the trajectories of particle pairs become fully uncorrelated, and transition into a random walk characterised by the diffusive regime ($\sim t$).

% --- 5. Backward dispersion ---
%------------------------------------------------------------------------------------------
Figure \ref{fig:Dispersion}(d) complements the forward pair dispersion in figure \ref{fig:Dispersion}(c) with the backward-in-time perspective. While forward dispersion tracks the future separation of particles initially separated by \rev{$R(0)=\eta_K$}, backward dispersion identifies particle pairs separated by \rev{$R(t_{\rm end})=\eta_K$} at a late time $t=t_{\rm end}$ and tracks their trajectories backward to their spatial origins. Comparing the two panels reveals a distinct temporal asymmetry: particle pairs separate significantly faster backward than forward in time.

This is a direct consequence of the irreversible nature of the turbulent energy cascade \citep{Xu2014}: in flows evolving forward in time, regions of strain frequently lead to the compression of fluid elements, driving particles into close proximity; a feature that we quantified via a $Q$--$R$-analysis in Section \ref{subsec:Velocity_gradient_invariant_analysis_and_dust-devil-like_vortices}. Conditioning pairs to be close together at $t=t_{\rm end}$ introduces an inherent bias towards the trajectories of particles that have just undergone intense compression. Following their trajectories backward in time essentially means unravelling these vigorous events, leading to statistically enhanced separation.

The statistical ensemble curves of forward and backward pair dispersion would look identical if turbulence were fully reversible. However, the dominance of vortex stretching over compression (see Section \ref{subsec:Velocity_gradient_invariant_analysis_and_dust-devil-like_vortices}) dictates that converging trajectories are statistically favoured forward in time, leading to a relative acceleration of dispersion in the backward direction \citep{Jucha2014}.

% --- 6. LSA ---
%------------------------------------------------------------------------------------------
\rev{To investigate scaling behaviour in the inertial range, we conduct a local slope analysis for horizontal forward pair dispersion by calculating its logarithmic derivative}
\begin{equation}
    \label{eq:LSA_PD}
    \frac{\mathrm{d} \log \left( \langle | \bm{R}(t) - \bm{R}_0 |^2\rangle_L / \eta_K^2 \right)}{\mathrm{d} \log \left( t / \tau_{\eta} \right)}\,.
\end{equation}
\rev{The result is illustrated in figure \ref{fig:Dispersion}(e), with the raw derivative shown in thin lines and a Savitzky--Golay filter applied in thick lines to clarify the underlying trend. Unlike the predictions of classical cascade theories (both Richardson and Bolgiano--Obukhov), we do not observe a pure, continuous scaling plateau across an extended period of time. Rather, the scaling exponents exhibit a pronounced overshoot above the Richardson limit ($\sim t^3$), peaking near \rev{$t \approx 20 \tau_{\eta}$} before falling monotonically toward the final diffusive limit ($\sim t$) at large times $t \gg 100 \tau_{\eta}$. With increasing $\Ra$, the peak scaling exponent rises and the duration of the overshoot exceeding $n=3$ increases. This broadening is directly governed by the broadening of temporal and spatial scale separation reported in table \ref{tab:Scale_separation}: from $\Ra=10^5$ to $\Ra=10^{10}$, the temporal window $T_{L,t}/\tau_{\eta}$ expands by nearly an order of magnitude (from $3.6$ to $32.1$), while the spatial runway $B/R_0$ rises from $21.8$ to $1244.0$. Here, $B=1-2\delta_T$ represents the macroscopic bulk depth. At low $\Ra$, the macroscopic decorrelation time $T_{L,t}$ is reached almost immediately after crossing $\tau_{\eta}$, rendering the emergence of an inertial subrange impossible.}

\rev{While it is tempting to attribute this super-diffusive overshoot ($n > 3$) at high $\Ra$ to a mixed Richardson and Bolgiano--Obukhov cascade, it is well established that the observability of inertial-range scaling in the time domain depends sensitively on the initial particle separation $R_0$. Sub-Kolmogorov and near-dissipative initial separations ($R_0 \lesssim 2 \eta_K$) place particles within the locally smooth, viscous Batchelor regime, systematically delaying separation and inducing a kinematic `catch-up' effect as particle pairs traverse the viscous bottleneck into the inertial subrange \citep{Falkovich1994, Boffetta2002, Sawford2008, Bitane2012, Liot2019}. To disentangle this dependence, we investigate total forward pair dispersion alongside its local slopes across different $R_0 \in \{1\eta_K, 2\eta_K, 4\eta_K, 8\eta_K, 16\eta_K\}$ at $\Ra=10^9$ in figure \ref{fig:Initial_separations}(a,c). Crucially, the transient overshoot above $n = 3$ vanishes once the initial separation is moved beyond the immediate dissipative range ($R_0 \gtrsim 4 \eta_K$). Consequently, if $R_0$ is too large ($R_0 = 16\eta_K$), the remaining spatial runway $B/R_0$ is too small to allow the exertion of an inertial subrange altogether. Thus, the  global overshoot ($n > 3$) at $R_0 = \eta_K$ represents a viscous catch-up rather than a manifestation of BO59 scaling. This result is corroborated for multi-particle clouds in Section \ref{subsec:Spatio-temporal_evolution_of_cloud_ensembles}.}

\rev{Regardless, regional conditioning of particle pairs seeded at $R_0 = \eta_K$ reveals distinct structural anisotropy. Figure \ref{fig:Dispersion} provides the peak scaling exponents $\max(n)$ exhibited in the horizontal, see figure \ref{fig:Dispersion}(f), and vertical, see figure \ref{fig:Dispersion}(g), directions for particle pairs seeded within the bulk ($z \in [\delta_T, 1-\delta_T]$, dark green), the core ($z \in [0.4, 0.6]$, blue), the plume root zone ($z \in [\delta_T, 2\delta_T] \cup [1-2\delta_T, 1-\delta_T]$, PR, salmon) and the thermal boundary layers ($z \in [0, \delta_T] \cup [1-\delta_T, 1]$, BL, oxblood) across $\Ra \in [10^5, 10^{10}]$. Here, $\delta_T=1/(2\Nu)$ is the mean thermal boundary layer thickness. Notably, although the absolute values of these peak scaling exponents remain artificially enhanced by the catch-up effect, a directional comparison reveals a distinct physical divergence: while the horizontal exponents $\max(n_h)$ exhibit no systematic trend and the vertical exponents $\max(n_z)$ in the core, the bulk and the plume root region asymptotically plateau near $\max(n_z) \approx 3.5$, particle pairs originating within the thermal boundary layers exhibit a monotonic increase in $\max(n_z)$ with rising $\Ra$. This is a direct consequence of a progressive thinning of the thermal boundary layers and plume detachment: particles seeded inside the boundary layers remain trapped in close proximity to the walls before being ejected via detaching thermal plumes, amplifying the kinematic catch-up effect. Crucially, because this steepening is strictly confined to the vertical direction of the pairs seeded within the thermal boundary layers and is absent in the bulk, it reflects localised, one-dimensional plume ejection rather than a three-dimensional BO59 cascade. Because temporal pair dispersion samples different spatial scales at any given time $t$ and is fundamentally unable to disentangle the impact of local structures from three-dimensional volumetric dispersion, we next perform scale-conditioned statistics in Section \ref{sec:Lagrangian_eddy_viscosity} and introduce a conditioned sampling approach applied to dense clouds of particles in Section \ref{subsec:Spatio-temporal_evolution_of_cloud_ensembles}.}

%------------------------------------------------------------------------------------------
\subsection{Lagrangian eddy viscosity}
\label{sec:Lagrangian_eddy_viscosity}

%------------------------------------------------------------------------------------------
\begin{figure}
\centering
\includegraphics[width =\textwidth]{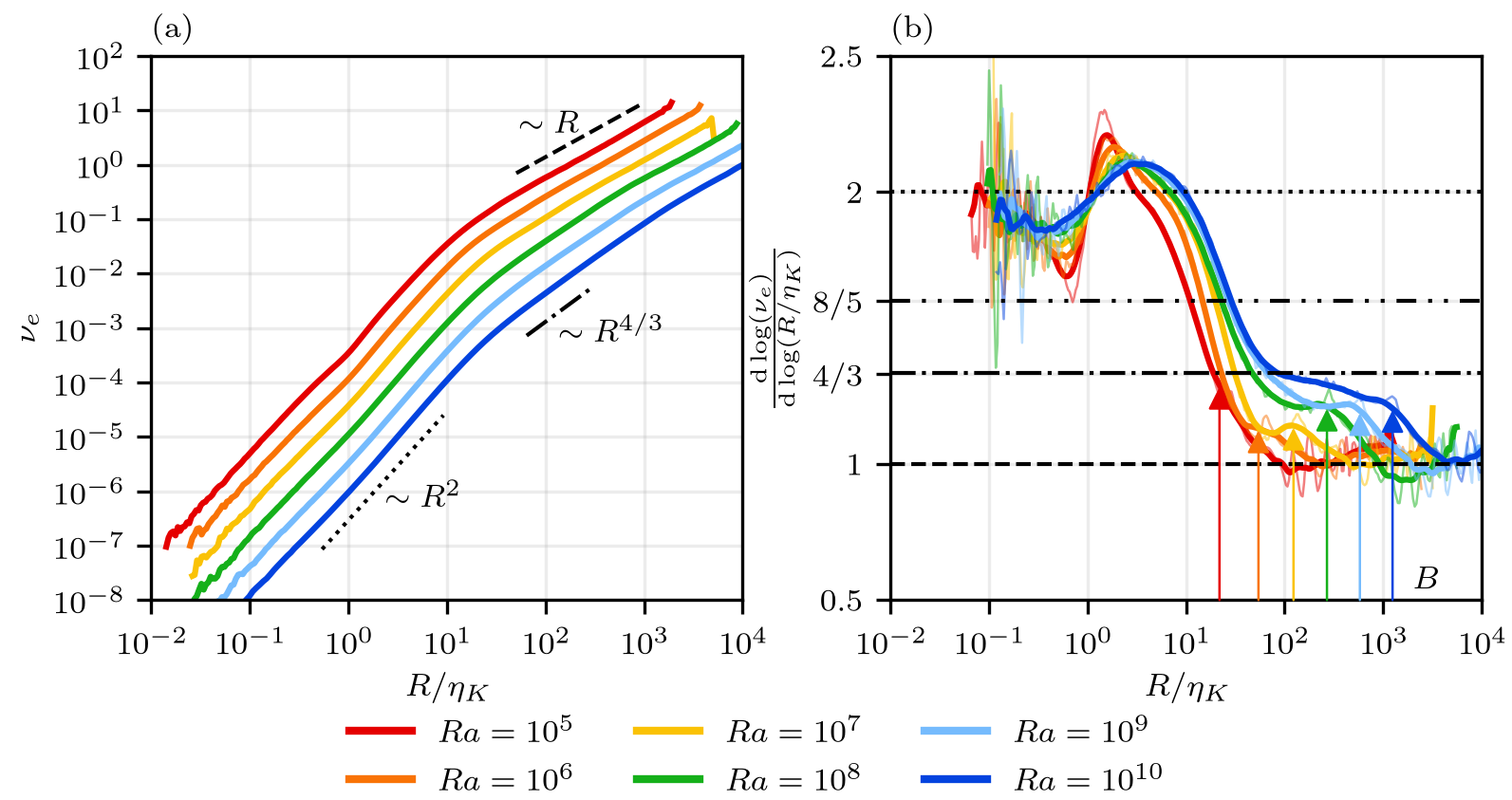}
\caption{\justifying{
Scale-dependent Lagrangian eddy viscosity $\nu_e ( | \bm{R}(t) |) := \left \langle | \bm{R} (t)  \cdot  (\bm{u}^L_1 - \bm{u}^L_2) | \right \rangle_R$ \rev{as} a direct measure \rev{of scale-dependent} turbulent mixing processes. Conditioned on spatial separation $\rev{R=}|\bm{R}|$, it complements the temporal framework of pair dispersion by probing dispersion regimes \rev{directly in spatial scale space}. (a) The expected \rev{sub-Kolmogorov} ballistic ($\sim R^2$) and macroscopic diffusive ($\sim R^1$) regimes are recovered \rev{across $\Ra \in [10^5, 10^{10}]$}. (b) A local slope analysis (thin lines represent raw data, thick lines a Savitzky--Golay filter) reveals \rev{a sharp transition peak surpassing the smooth kinematic limit $\sim R^2$ directly at the viscous-to-inertial crossover ($2 \lesssim R / \eta_K \lesssim 10$), driven by sharp velocity gradients as the viscous bottleneck is passed. No BO59 scaling plateau ($\sim R^{8/5}$) is observed. Rather, a distinct shoulder asymptotically approaching the Richardson limit ($\sim R^{4/3}$) with increasing $\Ra$ is recovered throughout the entire inertial range, before merging into the macroscopic diffusive limit $\sim R$. Coloured arrows $B / \eta_K$ mark this crossover point.}
}}
\label{fig:nu_e}
\end{figure}
%------------------------------------------------------------------------------------------

\rev{In temporal pair dispersion, particle pairs seeded at a fixed initial separation $R_0$ are sampled at various spatial scales at any given instant $t$, inevitably blending viscous catch-up effects with dispersion in the inertial range. To eliminate this dependence on $R_0$ and isolate these regimes directly in scale space}, we define the Lagrangian eddy viscosity as
\begin{equation}
    \label{eq:nu_e}
    \nu_e ( | \bm{R}(t) |) := \left \langle | \bm{R} (t)  \cdot (\bm{u}^L_1 - \bm{u}^L_2) | \right \rangle_R\,.
\end{equation}
Here, $\langle \cdot \rangle_R$ denotes the ensemble average conditioned on the instantaneous spatial separation $R = | \bm{R} |$ and the subscripts $1$ and $2$ denote the individual particles of a given pair. \rev{Conditioning on instantaneous separation reorganises the data directly in scale space and reduces the explicit dependence on elapsed time and on the chosen $R_0$. The positive quantity $\nu_e(R)$ should be interpreted here as a scale-conditioned magnitude of instantaneous radial separation activity, rather than as a signed eddy diffusivity such as the distance-neighbour diffusivity of \cite{Richardson1926} and the Eulerian definition of an eddy viscosity \citep{Bhattacharya2026}. Nevertheless, we will retain the notion of a Lagrangian eddy viscosity.} Following an analogous dimensional framework presented in Section \ref{subsec:Single-particle_and_pair_dispersion}, the shear-driven Richardson scaling \rev{in scale space is} recovered as
\begin{equation}
    \label{eq:Richardson_scaling_R}
    \nu_e (R) \sim \epsilon_{\nu}^{1/3} R^{4/3} \,,
\end{equation}
\rev{whereas a} buoyancy-driven Bolgiano--Obukhov \rev{cascade would imply a steeper spatial scaling according to}
\begin{equation}
    \label{eq:Bolgiano_Obukhov_scaling_R}
    \nu_e (R) \sim \beta^{2/5} \epsilon_{T}^{1/5} R^{8/5}\,.
\end{equation}
We \rev{compute} the scale-dependent Lagrangian eddy viscosity \rev{across} $\Ra \in [10^5, 10^{10}]$, taking statistics only after $t = 1 \tauf$ to eliminate an initial seeding bias. Figure \ref{fig:nu_e}(a) illustrates these results. \rev{At sub-Kolmogorov separations ($R \lesssim \eta_K$), particle pairs reside within the Batchelor regime. Because the locally smooth velocity field is governed by a linear first-order Taylor expansion ($\delta v(R) \sim R$) \citep{Bourgoin2015}, the kinematic scaling $\nu_e(R) \sim R^2$ is recovered.} In contrast, at very large separations ($R \gg 100\eta_K$), the velocities of paired particles become completely uncorrelated, leading to the linear macroscopic diffusive scaling in $R$.

Figure \ref{fig:nu_e}(b) \rev{shows the corresponding} local slope analysis, $\mathrm{d} \log (\nu_e) / \mathrm{d} \log (R / \eta_K)$, with thin lines \rev{denoting} the raw data and thick lines a Savitzky--Golay filter applied to reveal the dominant trend. \rev{Previous studies have attempted to remove initial seeding bias in temporal space via exit-time statistics \citep{Boffetta2002, Biferale2005a}. Scale conditioning, as done here, substantially reduces the explicit initial separation bias, although finite-memory effects can retain some $R_0$ dependence at fixed $R$. Contrasting $\nu_e(R)$ and its local slope for different initial separations $R_0 \in \{1\eta_K, 2\eta_K, 4\eta_K, 8\eta_K, 16\eta_K\}$  at $\Ra=10^9$ in figure \ref{fig:Initial_separations}(b,d) addresses this behaviour. While varying $R_0$ in temporal pair dispersion (see figure \ref{fig:Initial_separations}a,c) significantly influences peak scaling exponents at the onset of the inertial subrange, finite-memory effects in $\nu_e(R)$ are confined to a narrow shell near the initial separation ($R \lesssim 1.3 R_0$), within which randomly oriented particle pairs align with the local Lagrangian strain field. For $R \ge 1.3 R_0$, the curves and local slopes of all initial separations collapse onto a single curve across the inertial and diffusive ranges. At sub-Kolmogorov scales, differentiability gives the asymptotic behaviour $\nu_e\sim R^2$. The observed effective slope above 2 over the finite viscous-to-inertial crossover for $2 \lesssim R / \eta_K \lesssim 10$ \citep{Effinger1987, Falkovich1994} should therefore be interpreted as a rapid departure from this local-linear asymptote, associated with finite-range curvature, intermittency and plume/shear-layer conditioning, rather than as a test for velocity discontinuity.}

Following this \rev{bottleneck crossover}, the \rev{local slopes decrease monotonically across the inertial range. Crucially, the spatial conditioning of $\nu_e(R)$ clearly shows the absence of BO59 scaling in three-dimensional RBC, according to which the local slope should rise upward from the Richardson limit of $n = 4/3$ at inertial scales below the Bolgiano length $R < L_B$ toward a buoyancy-driven exponent of $n=8/5$ at scales $L_B < R<B$. As quantified in table \ref{tab:Scale_separation}, the nominal global Bolgiano crossover scale $L_B / \eta_K$ strictly resides below the macroscopic bulk scale $B/\eta_K$ for all $\Ra$. However, contrary to the BO59 theory, no transient increase or extended plateau toward $n=8/5$ is observed. Instead, as the spatial scale separation $B / \eta_K$ rises, a progressively wider shoulder, asymptotically approaching the Richardson limit of $n = 4/3$ with increasing $Ra$, forms and exists almost throughout the entire macroscopic bulk depth $R\approx B$ before relaxing into the diffusive regime ($\sim R^1$).}

\rev{By removing the temporal $R_0$ catch-up effect via strict spatial scale conditioning, the scale-dependent Lagrangian eddy viscosity $\nu_e(R)$ unambiguously unveils that turbulent mixing across the inertial subrange in three-dimensional RBC is governed by the classical K41--Richardson cascade rather than a BO59 regime. However, $\nu_e(R)$  by definition remains a scalar projection, prohibiting a geometric distinction between one-dimensional plume stretching, planar shearing or isotropic three-dimensional dispersion. To resolve the complete three-dimensional spatio-temporal deformation of fluid elements, we extend our analysis to dense multi-particle clouds in Section \ref{subsec:Spatio-temporal_evolution_of_cloud_ensembles}.}

%------------------------------------------------------------------------------------------
\section{Lagrangian multi-particle dynamics}
\label{sec:Lagrangian_multi-particle_dynamics}

\subsection{PCA-based metrics of Lagrangian particle clouds}
\label{subsec:PCA-based_metrics_of_Lagrangian_particle_clouds}

Pair dispersion reduces a two-particle configuration to a single scalar metric. While sufficient to measure the rate of separation, it is unable to determine whether a material volume is stretched into a filament, flattened into a sheet, severed by a plume, or isotropically dispersed. Principal component analysis (PCA) is a statistical tool that transforms data into a new coordinate system, reducing it to its essential features, the \textit{principal components} \citep{Greenacre2022}. 

\rev{It restores the missing geometric information by representing the evolving particle cloud via a time-dependent second-moment (mass/gyration) ellipsoid whose principal axes align with the PCA directions. In the terminology of \citet{Chertkov2026}, this defines a bulk second-moment ellipsoid, distinct from a minimum-area outer-envelope ellipsoid. Tracking dense clouds of $N_p \gg 4$ particles significantly reduces sensitivity to the particular four-particle sampling of minimal tetrad configurations as done by \citet{Pumir2000, Biferale2005b} and \citet{Watanabe2016}, providing statistical robustness in RBC where thermal plumes can pierce, fold, or partially entrain a cloud (although second moments remain sensitive to particles far from the centroid).} Mathematically, the gyration tensor of particle positions
\begin{equation}
    \label{eq:covariance_matrix}
    \bm{G} = \frac{1}{N_p-1} \sum_{i=1}^{N_p} \bm{x}_i'\bm{x}_i'^T
\end{equation}
is calculated, where $\bm{x}_i'$ is the position of a particle $i$ with respect to the centre of the cloud $\bar{\bm{x}} = N_p^{-1} \sum_{i=1}^{N_p}\bm{x}_i$. Afterwards, the eigenvalue problem
\begin{equation}
    \label{eq:eigenvalues_PCA}
    \bm{G} \bm{v}_k = \lambda_k \bm{v}_k, \quad k \in \{1, 2, 3\}
\end{equation}
is solved to obtain the \rev{ordered} eigenvalues $\lambda_1 \geq \lambda_2 \geq \lambda_3$ and the corresponding orthogonal eigenvectors $\bm{v}_k$. Here, $\bm{v}_1$ aligns with the direction of maximum variance (stretching), while $\bm{v}_3$ aligns with the direction of maximum compression \citep{Pope2000}. In addition to analysing the pure eigenvalues $\lambda_k$, one can compute several metrics to gain insights into the spatio-temporal evolution of a particle cloud and therefore the underlying flow dynamics.

\rev{\emph{Mean squared dispersion} is given by}
\begin{equation}
    \label{eq:MSD}
    \MSD = \mathrm{tr} ( \bm{G} ) = \lambda_1 + \lambda_2 + \lambda_3
\end{equation}
\rev{and represents a direct multi-particle extension of particle pair dispersion $| \bm{R} |^2 (t)  = | \bm{x}_2 - \bm{x}_1 |^2 $ to clouds of $N_p$ particles with the identity}
\begin{equation}
    \label{eq:R_cloud}
    \rev{R_{c}^2 (t) = \frac{1}{N_p^2} \sum_{i=1}^{N_p} \sum_{j=1}^{N_p} | \bm{x}_i(t) - \bm{x}_j(t) |^2 = 2\,\frac{N_p-1}{N_p}\,\MSD(t) \approx 2\,\MSD(t)\,.}
\end{equation}
\rev{Analogous to pair dispersion, one can subtract the initial seeding variance $\MSD_0 = $ \\ $\MSD(t=0)$ and compute an ensemble average over $N_c$ clouds, $\langle \MSD(t) - \MSD_0 \rangle_{N_c}$, where a uniformly seeded sphere of radius $r_0$ obeys the continuum expectation $\MSD_0 = \frac{3}{5} r_0^2$.}

\rev{Because $\MSD$ is an arithmetic sum of ordered eigenvalues, it is fundamentally dominated by growth of the primary eigenvalue $\lambda_1$. As a variance metric, it disproportionally weights particles that are distant from the cloud centroid. Consequently, a particle cloud that is merely stretched into a one-dimensional filament via local shear will experience an increase in $\MSD$, even if its material volume is conserved. Conversely, the \emph{geometric mean radius} and its logarithmic counterpart,}
\begin{align}
    \label{eq:geometric_mean_radius}
    \rev{r= (\lambda_1 \lambda_2 \lambda_3)^{1/6} \qquad \mathrm{and} \qquad v = \frac{1}{6}\log ( \det \bm{G}  ) \,,} 
\end{align}
\rev{separate volumetric second-moment expansion from purely anisotropic stretching. Specifically, if a finite cloud undergoes a uniform affine incompressible deformation $\bm{x}'(t) = \bm{F}(t)\bm{x}'(0)$ with $\det\bm{F} = 1$, the gyration tensor evolves as $\bm{G}(t) = \bm{F}(t)\bm{G}(0)\bm{F}^T(t)$, leaving $\det\bm{G}(t) = \det\bm{G}(0)$ -- and thus $r$ and $v$ -- strictly invariant even as the cloud becomes strongly anisotropic. Growth of $v$ therefore diagnoses the breakdown of a single volume-preserving affine description across the cloud, driven by non-affine turbulent deformation, differential stretching, and folding across velocity-gradient structures. In regimes where the normalized cloud shape becomes approximately self-similar ($\lambda_k \sim t^n$), $\MSD \sim t^n$ implies $r \sim t^{n/2}$. Consequently, the local logarithmic derivative $\mathrm{d} \log r / \mathrm{d} \log t = \mathrm{d} v / \mathrm{d} \log t$ yields scaling exponents of $n/2$.}

\rev{Normalising the eigenvalues by the trace, one can define the \emph{linear shape variables},}
\begin{equation}
    \label{eq:linear_shape_variables}
    L = \frac{\lambda_1 - \lambda_2}{\lambda_1+\lambda_2+\lambda_3}\,,\qquad 
    P = \frac{2 \left( \lambda_2-\lambda_3 \right)}{\lambda_1+\lambda_2+\lambda_3}\,, \qquad 
    S = \frac{3 \lambda_3}{\lambda_1+\lambda_2+\lambda_3}\,,
\end{equation}
where $L+S+P=1$. Large $L$ (linearity) signals cigar- or tube-like structures ($\lambda_1 \gg \left( \lambda_2 \approx \lambda_3 \right)$), whereas large $P$ (planarity) indicates a pancake-like shape as $\left( \lambda_1 \approx \lambda_2 \right) \gg \lambda_3$. \rev{However, because $L$, $P$, and $S$ represent bounded normalised coordinates in $[0, 1]$, they rapidly approach their limiting values under exponential turbulent stretching. As a result, they progressively lose resolution of continued deformation once a cloud is strongly anisotropic. To retain resolution across arbitrary stretching amplitudes, we introduce unbound \emph{logarithmic shape variables} via the dimensionless, traceless logarithmic shape tensor}
\begin{equation}
    \label{eq:B_tensor}
    \rev{\bm{B} = \frac{1}{2}\log\!\left(\frac{\bm{G}}{r^2}\right) = \frac{1}{2}\log(\bm{G}) - v\bm{I} \,,}
\end{equation}
\rev{where $\bm{I}$ is the identity matrix and $\mathrm{tr}(\bm{B}) = 0$. By factoring out the volumetric scale $r^2 = (\det\bm{G})^{1/3}$, its eigenvalues $\beta_k$ quantify purely anisotropic logarithmic stretching along the principal axes $\bm{v}_k$, defining the eigenvalue problem}
\begin{equation}
    \label{eq:log_shape_variables}
    \rev{\bm{B} \bm{v}_k = \beta_k \bm{v}_k \qquad \mathrm{with} \qquad \beta_k = \frac{1}{2}\log\!\left(\frac{\lambda_k}{r^2}\right) = \frac{1}{2}\log ( \lambda_k ) - v \,.}
\end{equation}
\rev{The second invariant of $\bm{B}$,}
\begin{equation}
    \label{eq:I_2}
    \rev{I_2 = \mathrm{tr}(\bm{B}^2) = \beta_1^2 + \beta_2^2 + \beta_3^2 \,,}
\end{equation}
\rev{quantifies the overall magnitude of anisotropy, measuring how strongly the ellipsoid has departed from a spherical shape ($I_2=0$). Its third invariant,}
\begin{equation}
    \label{eq:I_3}
    \rev{I_3 = \mathrm{tr}(\bm{B}^3) = \beta_1^3 + \beta_2^3 + \beta_3^3 \,,}
\end{equation}
\rev{represents the mode of deformation, where $I_3 > 0$ indicates linear (cigar-like) and $I_3 < 0$ planar (pancake-like) deformation. It incorporates all three linear shape variables into a single, unbound logarithmic metric that does not saturate. $I_2$ and $I_3$ permit a distinction between early, one-dimensional plume-driven stretching from three-dimensional isotropic expansion in the bulk and later-state planar compression in the boundary layers.}

\rev{Finally, the \emph{alignment angle}}
\begin{equation}
    \label{eq:alignment_angle}
    \Theta = \arccos{ \left( | \bm{v}_1 \cdot \bm{e}_z | \right)}
\end{equation}
between the first principal axis and the vertical unit vector $\bm{e}_z$ reveals the orientation of the cloud's primary extent. Vertical alignment ($\Theta \rightarrow 0^\circ$) may indicate the presence of rising or falling plumes, whereas horizontal alignment ($\Theta \rightarrow 90^\circ$) signals shear flow or large-scale circulation transport \citep{Pumir2011}.

%------------------------------------------------------------------------------------------

\subsection{Particle seeding and particle cloud evolution}
\label{subsec:Particle_seeding_and_particle_cloud_evolution}

%------------------------------------------------------------------------------------------
\begin{figure}
\centering
\includegraphics[width = \textwidth]{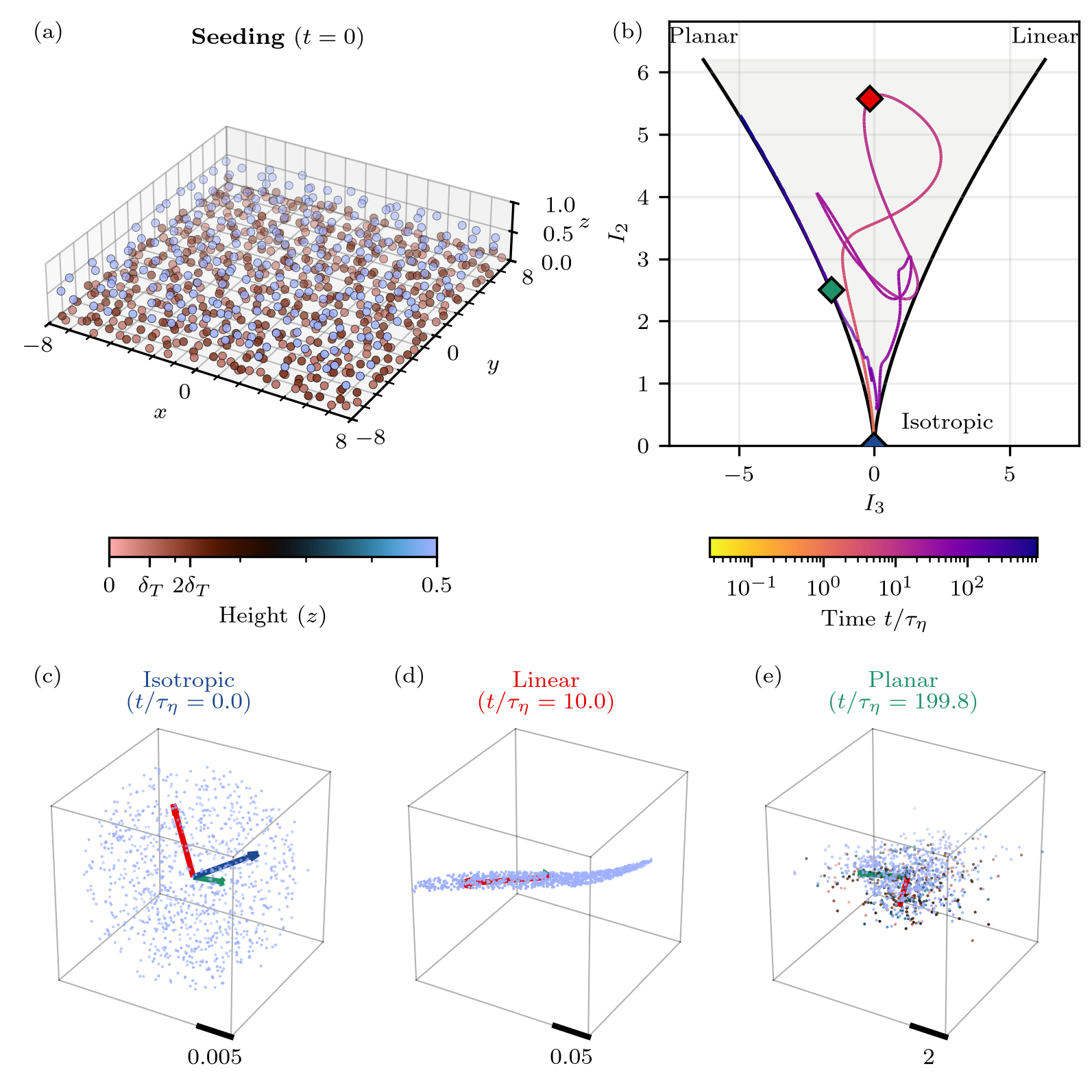}
\caption{\justifying{
Geometric overview of particle clouds at \rev{$\Ra=10^6$ for $r_0 = 4\eta_K$}. (a) Seeding strategy in the laterally periodic domain ($\Gamma=16$). The domain is partitioned into \rev{$16 \times 16$} sectors to prevent spatial aliasing. Clouds are seeded in three distinct layers: \rev{the thermal boundary layer ($z_0 = \delta_T$, salmon), the plume root zone ($z_0 = 2\delta_T$, oxblood) and the core ($z_0 = 0.5$, blue).} Note that initial particle cloud radii are exaggerated for illustration purposes. (b) \rev{Logarithmic invariant map ($I_2$--$I_3$)} characterising the temporal shape evolution of a representative cloud. \rev{The origin and bounding branches} define the limiting geometric states: linear \rev{($I_3=+I_2^{3/2}/\sqrt{6}$)}, planar \rev{($I_3=-I_2^{3/2}/\sqrt{6}$)} and isotropic \rev{($I_2 = I_3 = 0$)}. (c--e) Instantaneous volumetric renders of these three states, corresponding to the coloured markers in (b). The gyration tensor eigenvectors are displayed in the corresponding cloud centroids: $\bm{v}_1$ (red), $\bm{v}_2$ (green) and $\bm{v}_3$ (blue). Time instants are indicated above the particle clouds. Thick lines in the bottom right corners represent spatial scale bars in units of the height $H$ of the layer.
}}
\label{fig:Cloud_Overview}
\end{figure}
%------------------------------------------------------------------------------------------

To quantify the spatio-temporal deformation of particle clouds, we initialised distinct spherical clusters of $N_p = 1,\!000$ particles each. This ensemble size ensures converged statistics for the gyration tensor $\bm{G}$. \rev{As demonstrated for pair dispersion in Section \ref{subsec:Single-particle_and_pair_dispersion}, initial particle separation plays an outsize role in temporal scaling behaviour. To probe this sensitivity for multi-particle clouds, we present two simulation series, C$_1$ and C$_4$, which differ only in the initial cloud radius $r_0\in \{1\eta_K, 4\eta_K\}$ (see table \ref{tab:Series_scales}). This choice permits a direct investigation of viscous `catch-up' effects by contrasting seeding at the dissipative scale ($r_0 = \eta_K$, subject to smooth velocity gradients of the Batchelor regime) against seeding beyond the viscous bottleneck ($r_0 = 4\eta_K$, at the onset of the inertial subrange) while maintaining sufficient macroscopic scale separation \citep{Falkovich1994}.} Seeding was performed in three distinct horizontal planes -- the \rev{thermal boundary layer ($z_0 = \delta_T$),} the plume \rev{root} zone ($z_0 = 2 \delta_T$), and the \rev{core} ($z_0 = 0.5$) -- to \rev{probe} height-dependent transport physics.

The laterally periodic domain ($\Gamma=16$) was partitioned into \rev{$16 \times 16$} sectors in the horizontal direction. \rev{Within} each sector, one independent cloud was seeded at a random \rev{horizontal} position \rev{while maintaining a safety margin of $2r_0$ from sector boundaries} to prevent spatial aliasing effects due to large-scale circulation. In summary, we track the spatio-temporal evolution of \rev{$3 \times 256 \times 1,\!000=768,\!000$} particles over \rev{$1,\!000 \tau_{\eta}$} across Rayleigh numbers spanning four orders of magnitude, $\Ra \in \left[ 10^4, 10^8 \right]$. Figure \ref{fig:Cloud_Overview}(a) illustrates this seeding \rev{configuration} for \rev{$\Ra=10^6$ with $r_0=4\eta_K$}, highlighting the \rev{quasi-}random distribution across the \rev{$256$} sectors per layer and three seeding heights.

\citet{Lumley1977} originally introduced the anisotropy invariant map to characterize Reynolds stresses\rev{, which was later linearised into the `barycentric map' by \citet{Banerjee2007}. While these maps are well-suited for Eulerian statistics such as Reynolds stresses, the exponential nature of Lagrangian turbulent stretching causes linear shape variables to rapidly saturate and lose resolution near the map boundaries. To overcome this limitation, we present an unbounded phase space of the invariants $I_2$ and $I_3$ of the logarithmic traceless shape tensor $\bm{B}$.} Figure \ref{fig:Cloud_Overview}(b) visualises the trajectory of a representative cloud at \rev{$\Ra=10^6$} within this \rev{logarithmic $I_2$--$I_3$ invariant space}. \rev{Unlike closed linear maps, the logarithmic phase space forms a funnel bounded by the axisymmetric limits}: linear\rev{, cigar-like stretching} \rev{($I_3=+I_2^{3/2}/\sqrt{6}$)}, planar\rev{, pancake-like flattening} \rev{($I_3=-I_2^{3/2}/\sqrt{6}$)} and \rev{the spherical, isotropic state at the origin ($I_2 = I_3 = 0$)}. After being initially seeded in a spherical shape, the particle cloud is advected by the flow and undergoes intense stretching and compression, altering its shape and exploring the phase space within these limits. The highlighted markers in figure \ref{fig:Cloud_Overview}(b) correspond to the instantaneous volumetric renders shown in figure \ref{fig:Cloud_Overview}(c--e), illustrating the transformation from the initial sphere \rev{into an elongated cigar-like filament and subsequently into a flattened pancake-like sheet.}

%------------------------------------------------------------------------------------------
\subsection{\rev{Spatio-temporal evolution of cloud ensembles}}
\label{subsec:Spatio-temporal_evolution_of_cloud_ensembles}
%------------------------------------------------------------------------------------------

%------------------------------------------------------------------------------------------
\begin{figure}
\centering
\includegraphics[width = \textwidth]{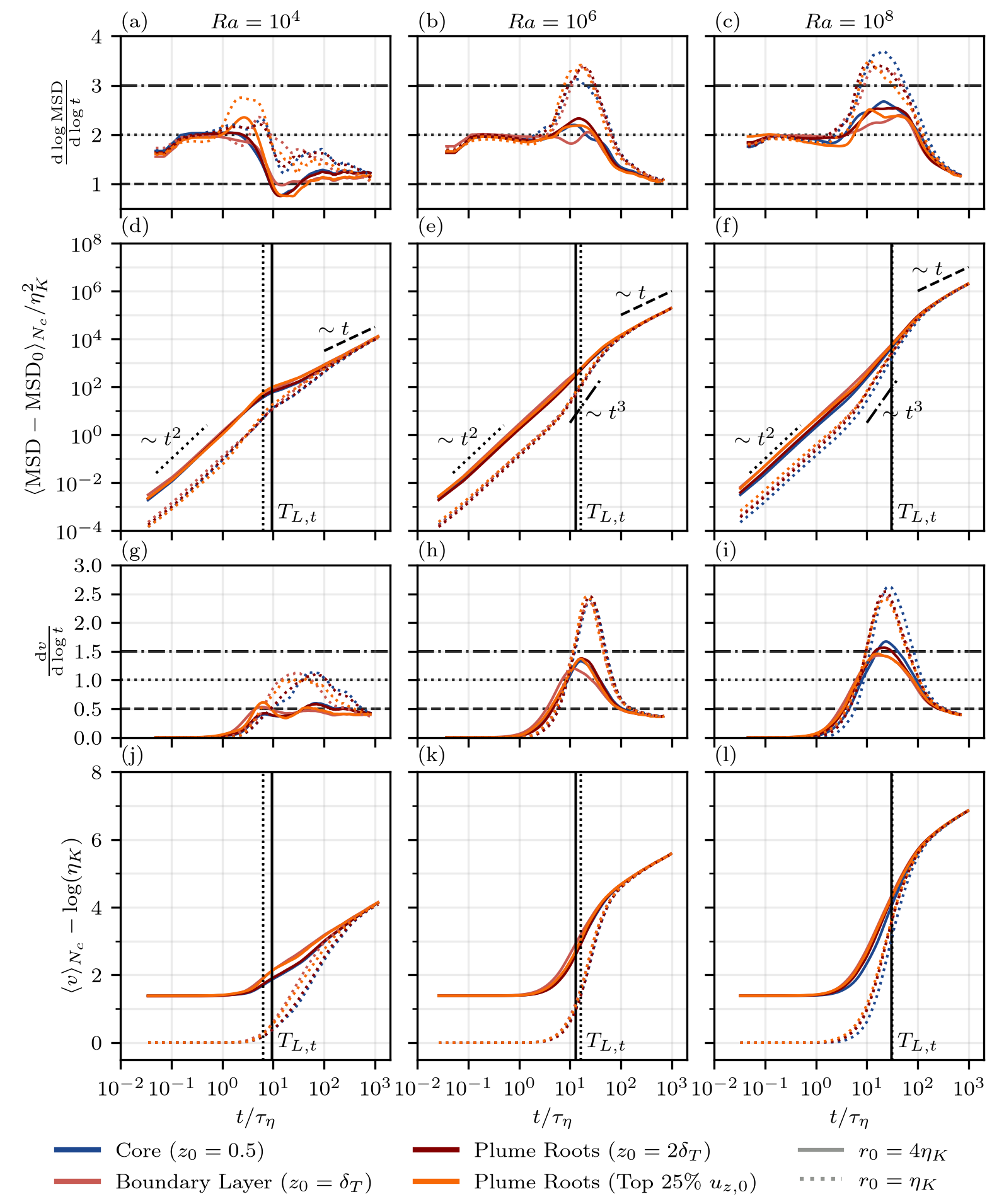}
\caption{\justifying{
\rev{Multi-particle cloud dispersion. Ensemble averages of normalised mean squared dispersion $\langle \MSD(t) - \MSD_0 \rangle_{N_c} / \eta_K^2$ (d--f) and normalised logarithmic geometric mean radius $\langle v \rangle_{N_c} - \log (\eta_K)$ (j--l) over $N_c=256$ clouds of $N_p=1,\!000$ particles are shown alongside local slope analyses (a--c, g--i) across $\Ra \in \left \{ 10^4, 10^6, 10^8 \right \}$. Solid lines represent initial seeding radii of $r_0=4\eta_K$ beyond the dissipative range within the viscous--inertial crossover, while dotted lines denote $r_0=\eta_K$ at the viscous scale. Colours represent different seeding heights $z_0 \in \{\delta_T, 2\delta_T, 0.5\}$ as well as the top quartile of $u_{z,0}$-conditioned clouds in the plume root zone ($z_0 = 2\delta_T$, orange). Vertical lines represent the total Lagrangian integral time scales $T_{L,t}$. As a trace metric, $\MSD$ is dominated by growth of the largest principal variance and particles distant from the cloud centroid, whereas the logarithmic geometric mean radius $v$ remains invariant under volume-preserving affine stretching and isolates non-affine volumetric expansion. Vigorous, transient spikes in the local slope analysis of clouds seeded at $r_0=\eta_K$ are attributed to kinematic `catch-up' effects rather than sustained, buoyancy-driven Bolgiano--Obukhov scaling.}
}}
\label{fig:Cloud_Dispersion}
\end{figure}

%------------------------------------------------------------------------------------------

%------------------------------------------------------------------------------------------
\begin{figure}
\centering
\includegraphics[width = \textwidth]{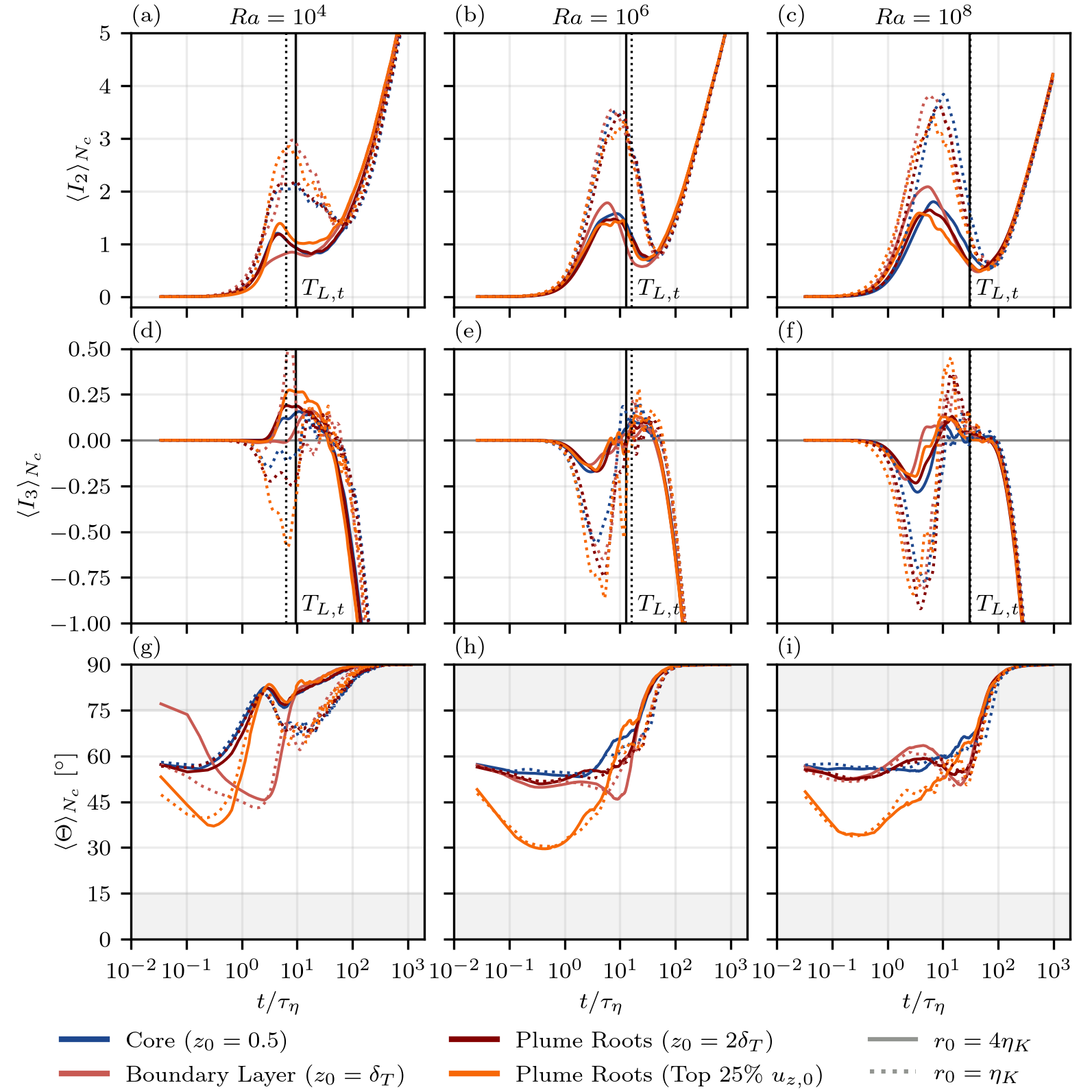}
\caption{\justifying{
\rev{Multi-particle cloud deformation. Ensemble averages of the invariants $\langle I_2 \rangle_{N_c}$ (a--c) and $\langle I_3 \rangle_{N_c}$ (d--f) alongside the alignment angle $\langle \Theta \rangle_{N_c}$ (g--i) are shown for the identical dataset as in figure \ref{fig:Cloud_Dispersion}. Vertical lines represent the total Lagrangian integral time scales $T_{L,t}$. $u_{z,0}$-conditioned clouds exhibit a premature drop in $\Theta$ as a consequence of plume entrainment. The combined trajectory of the invariants $I_2$ and $I_3$ illustrates a characteristic three-stage evolution of convective fluid elements in the present ensembles: initial local strain and plume entrainment drive local maxima in $I_2$ and $I_3>0$ before reverting toward isotropy near the Lagrangian integral time scale $T_{L,t}$. Finally, particle clouds impinge upon the boundary layers and are compressed into pancake-like sheets ($I_3 < 0$), causing $I_2$ to rise continuously over the observed late-time interval.}
}}
\label{fig:Cloud_Shape}
\end{figure}
%------------------------------------------------------------------------------------------

\rev{To probe the macroscopic deformation of particle clouds, figure \ref{fig:Cloud_Dispersion} reports the normalised mean squared dispersion $\langle \MSD(t) - \MSD_0 \rangle_{N_c} / \eta_K^2$ in panels (d--f) and the normalised loga\-rithmic geometric mean radius $\langle v \rangle_{N_c} - \log (\eta_K)$ in panels (j--l) across $\Ra \in \left \{ 10^4, 10^6, 10^8 \right \}$. The corresponding local slope analyses for $\MSD$ and $v$ are shown in figure \ref{fig:Cloud_Dispersion}(a--c) and (g--i), respectively. Here, we only show a Savitzky--Golay filter for visual purposes. Solid lines represent the global ensemble average over $N_c=256$ clouds seeded with an initial radius of $r_0=4\eta_K$, while dotted lines denote $r_0=\eta_K$. To isolate the kinematic impact of thermal plumes, we perform conditional sampling based on the initial vertical velocity field $u_{z,0}=u_z(x,y,z=2\delta_T,t=t_{\rm seed})$. We extract the top quartile ($64$ of $256$) of clouds seeded in the region of highest $u_{z,0}$ values within the plume root zone ($z=2\delta_T$, orange lines). We compared this approach with conditional sampling based on the initial temperature field $T_0=T(x,y,z=2\delta_T,t=t_{\rm seed})$, but found that $u_{z,0}$-sampling produced superior scaling results. This is likely due to vertical velocity having an immediate kinematic effect on particle dispersion, whereas high local temperature anomaly may involve a temporal lag -- indicating the formation of a thermal structure that has not yet accelerated into a fully developed rising plume. A sensitivity analysis across different conditioning thresholds is provided in Appendix \ref{subsec:Appendix_Sensitivity_of_initial_updraft_conditioning}.}

\rev{A direct comparison between the local slope analyses of $\MSD$ and $v$ separates two distinct aspects of cloud growth: because the trace $\MSD = \lambda_1 + \lambda_2 + \lambda_3$ is dominated by an increase of the largest principal variance and particles far from the centroid, its local slopes are inherently volatile and heavily reflect the extreme outer extent of the cloud. Conversely, the geometric mean radius $r = (\det\bm{G})^{1/6}$ and its logarithmic equivalent $v$ remain strictly invariant under uniform affine incompressible stretching ($\det\bm{F} = 1$). Consequently, $v$ suppresses simple one-dimensional elongation in the smooth Batchelor regime and provides a more robust measure of non-affine finite-cloud expansion driven by differential stretching and folding across velocity-gradient structures.}

\rev{Crucially, the initial seeding radius fundamentally dictates scaling behaviour. Clouds seeded at the dissipative scale $r_0=\eta_K$ exhibit heavy, transient overshoots in the local slope analysis that far exceed Richardson scaling ($\sim t^3$). This phenomenon is not attributable to buoyancy-driven Bolgiano--Obukhov scaling ($\sim t^5$), but rather to the kinematic `catch-up' effect: clouds seeded in the Batchelor regime ($r_0 \lesssim \eta_K$) are initially located in a locally smooth velocity field where their local volume is conserved and particle separation is kinematically delayed ($\sim t^2$). As these clouds finally move beyond the Kolmogorov scale into the inertial subrange, they accelerate violently to catch up to their ambient turbulent scaling, resulting in a transient scaling spike in the local slope rather than a sustained scaling plateau. Conversely, clouds with an initial separation beyond this viscous length scale (i.e., $r_0=4\eta_K$) start expanding right as they cross the Kolmogorov time scale $\tau_{\eta}$. Figure \ref{fig:Cloud_Dispersion} demonstrates this: clouds with $r_0=\eta_K$ remain in the ballistic regime for some time beyond $t=\tau_{\eta}$. The compensation for slow initial dispersion as the clouds traverse the viscous bottleneck at the bottom of the inertial subrange ($r\lesssim 10\eta_K$) is reflected in sharp transient peaks of scaling exponents which surpass Richardson scaling ($\sim t^3$) \citep{Falkovich1994}. Because larger initial separations remove this effect, clouds of this size ($r_0=4\eta_K$) authentically reveal turbulent scaling without transient overshoot.}

\rev{At the sub-turbulent $\Ra=10^4$, particle dispersion is merely driven by large-scale structures, and super-diffusive scaling is absent. Importantly, as quantified in table \ref{tab:Scale_separation}, the spatial scale separation $B/r_0$ between the initial cloud size ($r_0$) and the macroscopic bulk depth ($B = 1 - 2\delta_T$) is merely $1.8$ for $r_0=4\eta_K$, while the temporal scale separation $T_{L,t} / \tau_{\eta}=9.3 \lesssim 10$ is also insufficient. These spatio-temporal restrictions directly bypass the inertial subrange and lead to diffusive scaling as $\tau_{\eta}$ is surpassed (figure \ref{fig:Cloud_Dispersion}a,d,g,j). In contrast, at $\Ra=10^8$, the large scale separation provides sufficient runway for the inertial subrange, allowing clouds to manifest sustained Richardson-like super-diffusive scaling (figure \ref{fig:Cloud_Dispersion}i) in absence of the `catch-up' effect.}

\rev{Figure \ref{fig:Cloud_Shape} extends this analysis to the mode of deformation, showing the invariants $\langle I_2 \rangle_{N_c}$ in panels (a--c) and $\langle I_3 \rangle_{N_c}$ in panels (d--f) alongside the alignment angle $\langle \Theta \rangle_{N_c}$ in panels (g--i). For the global ensemble, the influence of different seeding heights vanishes as $\Ra$ is increased. This reveals universal deformation statistics in the fully turbulent regime and corroborates the results of pair dispersion (see Section \ref{subsec:Single-particle_and_pair_dispersion}). However, the $u_{z,0}$-conditioned updraft clouds (orange lines) exhibit a distinct, early drop in $\Theta$ across all $\Ra$. Crucially, this leads to lower peak scaling exponents of $\MSD$ and $v$ in the turbulent regime at $\Ra=10^8$ (figure \ref{fig:Cloud_Dispersion}c,i), and highlights the catch-up effect once more: plume-conditioned clouds experience turbulent dispersion more promptly due to the effect of early thermal plumes than clouds seeded in the isotropic bulk (blue lines), which are compact longer and subsequently undergo a catch-up acceleration.}

\rev{The combined trajectory of the invariants $I_2$ and $I_3$ reveals a characteristic three-stage evolution of convective fluid elements in the present ensembles. Exiting the initial ballistic regime ($t < \tau_\eta$), clouds are stretched into elongated filaments ($I_3 > 0$) by local strain and thermal plumes, driving a local maximum in deformation magnitude $I_2$. As clouds enter the bulk, multi-axial turbulent mixing drives a return toward a more isotropic shape ($I_2 \rightarrow 0$) near the total Lagrangian integral time scale $T_{L,t}$. Finally, as particle clouds impinge upon the boundary layers, they are compressed into pancake-like sheets ($I_3 < 0$), causing $I_2$ to increase continuously over the observed late-time interval.}

\rev{While thermal plumes directly influence dispersion behaviour, transient scaling exponents exceeding the Richardson limit ($\sim t^3$) must be attributed to effects of initial separation rather than sustained Bolgiano--Obukhov scaling ($\sim t^5$).}

%------------------------------------------------------------------------------------------
\section{Discussion and perspective}
\label{sec:Discussion_and_perspective}

In the present work, we have numerically investigated Lagrangian \rev{single-particle and pair} statistics in plane-layer Rayleigh--Bénard convection (RBC) \rev{at an aspect ratio of $\Gamma=4$}, spanning five orders of magnitude from moderate values at $\Ra=10^5$ to the high-Rayleigh-number regime at $\Ra=10^{10}$. \rev{This analysis is complemented by two multi-particle cloud simulation series spanning $\Ra \in [10^4, 10^8 ]$ in a wider domain of $\Gamma=16$, tracking a statistically converged ensemble of dense particle clusters}. Using the GPU-accelerated spectral element solver \textit{nekRS}, we probed single-, pair-, and multi-particle dynamics with high spatial fidelity. By computing Lagrangian particle accelerations at every integration time step and applying \rev{PSD-calibrated} cubic smoothing splines, \rev{we decoupled} physical intermittency from temporal aliasing signatures that typically afflict Lagrangian particle tracking across $C^0$-continuous element boundaries. \rev{This framework, combined with a cumulative moment-integral analysis and batched standard-error estimation, provides} a clean, unamplified view of Lagrangian intermittency at high Rayleigh numbers.

Acceleration PDFs exhibit a distinct symmetric, but highly fat-tailed behaviour. A sharp peak near $a_k/a_{k,\mathrm{rms}}=0$ confirms that particles spend the majority of their lifetime passively advected in a relatively quiescent flow, while strongly stretched exponential tails represent rare, intense acceleration events\rev{, whose prevalence rises monotonically with $\Ra$. We restrict higher-order moment calculations to the statistically converged regime ($|a_k| \le 50 \, a_{k,\mathrm{rms}}$), obtained via evaluation of cumulative fourth-order moment integrals to address the intrinsic convergence difficulties associated with this metric \citep{Yakhot2005}. Beyond $\Ra \gtrsim 10^7$, the horizontal accelerations exhibit systematically wider tails than their vertical counterparts ($K_z / K_h < 1$) as a direct consequence of the laterally unrestricted periodic domain, whereas the variance ratio ($\sigma_z^2 / \sigma_h^2$) approaches unity.} 

\rev{By sampling directional particle accelerations in distinct flow regions,} we observe a monotonic increase of the Heisenberg--Yaglom constant $a_{0,k}$ \rev{with $\Ra$ and $\Rlambda$ across all regions except for $a_{0,z}$ within the wall-confined thermal boundary layers. A collapse of horizontal and vertical components across all $\Ra$ in the core of the domain ($z \in [0.4, 0.6]$) reveals isotropic, HIT-like behaviour and follows the empirical relation of \citet{Sawford2003}, providing quantitative proof of local small-scale isotropy within this region. To explain the directional divergence and the breakdown of purely viscous K41 scaling in the plume-dominated and near-wall regions, we generalised the viscous--thermal mixed model of \citet{Li2021} by introducing a continuous mixing ratio $r \in [0, 1]$ weighting the exact local viscous and thermal dissipation rates, $\accvar = C \Ra^{\gamma} \epsnu^{\frac{3}{2}r} \epsT^{\frac{3}{4}(1-r)}$. Using a profile-likelihood grid search reveals with $95 \%$ confidence that the thermal dissipation contributes significantly to the vertical accelerations in the bulk ($r = 0.82$, driven by thermal plumes) and dominates the horizontal accelerations within the thermal boundary layers ($r = 0.18$, driven by intense near-wall temperature gradients $\epsT \sim |\nabla T|^2$).}

Heat transfer statistics, expressed in terms of the Lagrangian Nusselt number $\NuL$, are profoundly \rev{influenced by this structural intermittency}; a single fluid particle caught in a detaching thermal plume can result in local heat transfer exceeding the Eulerian mean by up to \rev{$250$} times \rev{at $\Ra=10^{10}$. Crucially, we advance prior Lagrangian heat-transfer studies \citep{Gasteuil2007, Schumacher2009, Emran2010, Liot2016a} by evaluating the exact vertical temperature gradient $- \partial T^L / \partial z$ along all $2 \times 2^{20}$ trajectories rather than relying on the estimation of linear conduction $- \partial T^L / \partial z = 1$ and provide the first systematic analysis of higher-order Lagrangian heat-flux moments across $\Ra \in [10^5, 10^{10}]$. Resolving the exact diffusive contribution of Lagrangian heat flux strictly ensures global heat flux conservation, $\langle \NuL \rangle / \NuVt = 1$, across all $\Ra$ and reveals how this microscopic thermal contribution blunts the sharp peaks of the purely convective heat flux $J_c$ at low $\Ra$. Conversely, the progressively thinner thermal boundary layers result in a collapse of the normalised mean and higher-order moments of $\NuL$ and $J_c$ at high $\Ra$. Like acceleration, PDFs become increasingly fat-tailed with rising $\Ra$, quantifying the results of previous studies \citep{Shang2003, Gasteuil2007, Schumacher2009, Liot2016a, Käufer2024} via monotonically increasing normalised variance $\sigma^2 / \NuVt^2$ and kurtosis $K$. Skewness $S$ remains strictly positive and rises with $\Ra$ as the majority of heat is transported in the plume-aligned direction \citep{Käufer2024}. The progressively intermittent nature of Lagrangian heat transfer} can also be connected to enhanced turbulent mixing as plumes rise or fall into the bulk, as shown by velocity fluctuation statistics \rev{and the concentration of heat transport into an ever finer-grained hierarchical network of thermal plumes \citep{Samuel2025, Shevkar2025b}}.

The fundamental origin of these \rev{extreme single-particle acceleration and heat-transfer events} is rooted in the local \rev{velocity-gradient} topology. \rev{Evaluating the invariants of the velocity gradient tensor, $Q$ and $R$, spectrally along Lagrangian trajectories recovers} the classical teardrop shape and Vieillefosse tail across \rev{the global domain, while conditioning to the thermal boundary layers results in} a pronounced skewness toward the upper-left quadrant (vortex-stretching, $Q>0, R<0$). \rev{By introducing a four-criterion spatio-temporal filtering algorithm at $\Ra=10^{10}$ -- reflecting near-wall locality, vertical binormal alignment, 99th-percentile centripetal acceleration and temporal persistence -- we achieved the first Lagrangian isolation of dust-devil-like convective vortices in the $Q$--$R$ plane. Tracking associated fluid parcels throughout their approach, helical swirling and vertical ejection phases reveal deep excursions into the upper-left quadrant ($Q>0, R<0$) while generating intense, localised heat transfer events that exceed the Eulerian mean ($\NuVt$) by more than an order of magnitude. A full-factorial Analysis-of-Variance decomposition with Sobol indices and a bootstrap analysis across $9^4$ parameter permutations confirm that these extracted vortices represent robust, spatially compact physical structures carrying an ensemble-averaged convective heat flux of nearly $8\,\NuVt$ and up to $30\,\NuVt$ in individual vortices. Seeding a dense tracer cloud directly within the eye of a dust devil demonstrates how these coherent near-wall vortices helically stretch and tilt finite fluid volumes during vertical ejection, motivating our multi-particle dispersion analysis.}

Expanding from individual particles to particle pair dispersion \rev{directly links this local velocity-gradient topology to multi-scale dispersion}. We recovered the well-known initial ballistic ($\sim t^2$) and long-term diffusive ($\sim t$) regimes both forward and backward in time\rev{, with the Lagrangian integral time scale $T_L$ accurately bounding the onset of velocity decorrelation}. Faster separation observed in the backward direction is a hallmark of the irreversibility of the turbulent energy cascade, driven by a statistical bias toward compressed particles that is directly linked to the dominance of vortex stretching over compression expressed in the $Q$--$R$ diagram. \rev{Although RBC is a buoyancy-driven flow, the long-standing hypothesis that it sustains a global Bolgiano--Obukhov (BO59) cascade \citep{Procaccia1989, Lvov1991, Lvov1992} has been refuted via direct energy-flux spectra analyses in the Eulerian frame \citep{Kumar2014, Verma2017}. In this study, we address the Lagrangian perspective of this debate. By systematically varying the initial pair separations $R_0$, we demonstrate that transient scaling exponents exceeding the Richardson limit ($n > 3$) for particles seeded at the Kolmogorov scale ($R_0 = \eta_K$) do not reflect a BO59 cascade; rather, they originate from a viscous `catch-up' effect as particles escape the locally smooth Batchelor regime and traverse the viscous--inertial crossover into the inertial subrange \citep{Falkovich1994, Boffetta2002, Sawford2008, Bitane2012, Liot2019}. Seeding particles in pairs directly within the inertial subrange ($R_0 \gtrsim 4 \eta_K$) eliminates this overshoot entirely.} 

\rev{By conditioning particle pairs on their initial seeding region, we find that pairs originating strictly within the thermal boundary layers exhibit enhanced vertical kinematic catch-up as $\Ra$ increases. Consequently, the transient super-diffusive peaks in scaling exponents identified via temporal local slope analyses are fundamentally linked to this kinematic delay and one-dimensional thermal plume ejection rather than a three-dimensional BO59 cascade. To suppress this initial separation memory, we investigated the scale-dependent Lagrangian eddy viscosity $\nu_e (R)$ as a spatial counterpart to temporal pair dispersion. Conditioning on instantaneous separation removes the explicit $R_0$ dependence once particle pairs align with the local strain field ($R \gtrsim 1.3 R_0$) and exposes the viscous-to-inertial crossover ($\eta_K \lesssim R \lesssim 10\eta_K$) \citep{Effinger1987} directly in scale space, where the local slope transiently exceeds the smooth kinematic limit of $n=2$. Crucially, above the nominal Bolgiano length ($R > L_B$), no plateau approaching the BO59 prediction ($\sim R^{8/5}$) is observed. Rather, as spatial scale separation ($B / \eta_K$) widens with $\Ra$, an extended intermediate-scale shoulder emerges that becomes increasingly consistent with the classical K41--Richardson limit ($\sim R^{4/3}$) across the macroscopic bulk scale ($R \approx B$).}

\rev{However, both temporal pair dispersion and the scale-dependent Lagrangian eddy viscosity are scalar metrics that are unable to distinguish between one-dimensional thermal plume stretching, planar shearing, and isotropic three-dimensional dispersion. To overcome this geometric limitation, we tracked the spatio-temporal evolution of dense multi-particle clouds ($N_c = 256$ clouds of $N_p = 1,\!000$ particles per horizontal plane in a $\Gamma = 16$ domain across $10^4 \le \Ra \le 10^8$) using principal component analysis (PCA). We use mean squared dispersion $\MSD$ as the multi-particle analogue to pair dispersion and the unbounded invariants $I_2$ and $I_3$ of the traceless logarithmic shape tensor $\bm{B}$ to distinguish between linear one-dimensional stretching ($I_3 > 0$), two-dimensional planar shearing ($I_3 < 0$), and isotropic dispersion ($I_3 = 0$). Simultaneously, the alignment angle $\Theta$ between the vertical axis $\bm{e}_z$ and the primary eigenvector $\bm{v}_1$ connects intense one-dimensional stretching at the onset of the inertial range directly to thermal plumes. Moreover, we find that the logarithmic geometric mean radius $v = \frac{1}{6}\log(\det\bm{G})$ is fundamentally superior to the arithmetic mean squared dispersion ($\MSD$) in measuring the dispersive behaviour of the particle cloud. While $\MSD$ is strongly weighted by outlier particles and inherently volatile as a variance-based metric, $v$ natively reflects local fluid incompressibility and remains strictly invariant under uniform volume-preserving affine stretching ($\det\bm{F}=1$) in the smooth Batchelor regime, responding selectively to non-affine multi-scale expansion.}

\rev{Contrasting clouds seeded at the dissipative scale ($r_0 = \eta_K$, Series C$_1$) against those seeded beyond the immediate dissipative range within the viscous--inertial crossover ($r_0 = 4\eta_K$, Series C$_4$) via a local slope analysis of $v$ directly corroborates the role of initial-scale memory and viscous catch-up, as well as the impact of macroscopic temporal ($T_L / \tau_{\eta}$) and spatial ($B / r_0$) scale separation. While Kolmogorov-seeded clouds ($r_0 = \eta_K$) exhibit a pronounced catch-up overshoot and low-$\Ra$ clouds at $\Ra=10^4$ directly enter the diffusive regime due to domain confinement ($B / r_0 = 1.8$), clouds seeded at $r_0 = 4\eta_K$ strongly suppress near-viscous catch-up and exhibit a sustained Richardson-like volumetric scaling regime with $2\,\mathrm{d}v/\mathrm{d}\log t \approx 3$ at high $\Ra$ once normalised cloud shapes become approximately self-similar. Clouds seeded closer to the origin of thermal plumes ($z_0=2\delta_T$) -- particularly when conditioned to the top quartile ($25 \%$) of initial updraft velocities $u_{z,0}$ -- experience more immediate plume entrainment (reflected by a prompt drop in $\Theta \to 0^{\circ}$), earlier volumetric expansion, and consequentially lower peak scaling exponents by reducing kinematic lag. This stands directly opposed to BO59 assumptions, according to which buoyant plume driving would induce steeper scaling exponents ($\sim t^5$). Consequently, our results provide no evidence for a sustained three-dimensional BO59 regime; instead, they support a picture where buoyancy in three-dimensional RBC acts in a localised, one-dimensional manner rather than a three-dimensional cascade.}

\rev{The interplay of the invariants $I_2$, $I_3$ and vertical alignment $\Theta$ provides a cohesive perspective on the spatio-temporal evolution of fluid volumes in convection: upon leaving the ballistic regime at $t \approx \tau_\eta$, clouds undergo vigorous uniaxial stretching via thermal plumes or turbulent strain, reflected in a local maximum of $I_2$ and $I_3 > 0$. Subsequent turbulent mixing causes reversion toward isotropy ($I_2, I_3 \to 0$) near the Lagrangian integral time scale $T_L$. Finally, clouds impinge upon the boundary layers and compress these clouds into two-dimensional, pancake-like sheets ($I_3 <0$), causing $I_2$ to increase continuously over the observed late-time interval.}

\rev{While being powerful standalone diagnostics, the combination of single-, pair- and multi-particle diagnostics establishes a cohesive, scale-bridging framework that tracks the entire life cycle of convective transport, where each individual metric complements the others. Single-particle ($a_k$, $\NuL$, $Q$--$R$) statistics isolate the origin of intense buoyancy-driven intermittency: intense thermal dissipation in the thermal boundary layers causes extreme accelerations and permits the formation of dust-devil-like convective vortices that eject fluid parcels carrying up to $250\,\NuVt$ into the bulk. Particle pair statistics ($R^2(t)$, $\nu_e(R)$) illustrate dispersion behaviour across different scales from both a spatial and temporal point of view. Here, the $Q$--$R$ bias toward vortex stretching over compression and the faster backward-in-time dispersion are consistent manifestations of irreversible velocity-gradient dynamics, while plume ejections amplify the viscous catch-up effect and produce a rapid departure from the smooth-flow $R^2$ scaling across the viscous--inertial crossover ($n > 2$ in $\nu_e(R)$). Finally, multi-particle cloud topology ($v$, $I_2$, $I_3$, $\Theta$) recovers the three-dimensional geometric information that is absent in scalar pair metrics and reveals that thermal plumes (driven by the far tails of $\NuL$ and $a_z$) deform fluid volumes via intense one-dimensional stretching ($\Theta \to 0^\circ$, $I_3 > 0$) before shear-driven K41--Richardson dispersion governs volumetric expansion in the bulk. This ensemble of trajectory-based Lagrangian diagnostics crucially complements the classical Eulerian framework. Whereas global quantities, such as $\Nu(\Ra,\Pr)$ and $\Re(\Ra,\Pr)$, fixed-point probes, and instantaneous Eulerian structure functions compress complex dynamics into system-level scaling laws or spatial averages over coexisting flow regimes, the Lagrangian representation directly retains the continuous material history of the same fluid parcel or cloud as it traverses boundary layers, plumes, and bulk regions, preserving the coupled spatio-temporal evolution of heat, momentum, and material geometry.}

\rev{Although} computational constraints currently prohibit Lagrangian DNS at even higher Rayleigh numbers, our statistics provide a clear perspective on the high-Rayleigh-number regime \rev{and establish a distinct dynamical dichotomy in three-dimensional RBC. On the one hand, the convergence of acceleration variance ratio toward unity, the collapse of the Heisenberg constants $a_{0,k}$ onto HIT scaling in the core region, the emergence of an extended K41--Richardson shoulder ($\sim R^{4/3}$) in the scale-dependent eddy viscosity, and the volumetric cloud expansion with $\sim t^3$ all demonstrate that background turbulent mixing in the bulk increasingly resembles homogeneous isotropic turbulence as $\Ra$ increases. On the other hand, the monotonic growth of acceleration and heat-transfer kurtosis, the dominance of thermal dissipation in the thermal boundary layers, the emergence of vigorous dust-devil-like convective vortices, and the immediate uniaxial stretching of plume-entrained particle clouds reveal that buoyancy acts increasingly intermittently, injecting energy at the thermal boundary layers and feeding both localised heat fluxes and one-dimensional structures.}

%------------------------------------------------------------------------------------------
\backsection[Acknowledgements]{
We thank Luca Moriconi, Katepalli R. Sreenivasan, and Michael Wilczek for helpful discussions.
}

\backsection[Funding]
{
M.E. is funded by the Carl Zeiss Foundation within the PollenNet consortium (project number P2022-08-006). M.C. acknowledges support of the Alexander von Humboldt Foundation with a Humboldt Research Award. R.J.S. and J.S. are supported by the European Union (ERC, MesoComp, 101052786). Views and opinions expressed however are those of the authors only and do not necessarily reflect those of the European Union or the European Research Council. The authors gratefully acknowledge the Gauss Centre for Supercomuting e.V. (\href{www.gauss-centre.eu}{\texttt{www.gauss-centre.eu}}) for funding this work by providing computing resources 
through the John von Neumann Institute for Computing (NIC) on the GCS supercomputers JUWELS and JUPITER at Jülich Supercomputing Center (JSC) within projects nonbou and ExaRBC. They further acknowledge the computing centre of the Technische Universität Ilmenau for providing access and research data management resources on its compute cluster MaPaCC24.
}

\backsection[Declaration of interests]{The authors report no conflict of interest.}

\backsection[Author ORCIDs]{\\
M. Ettel, https://orcid.org/0009-0008-6962-3587      \\
R.J. Samuel, https://orcid.org/0000-0002-1280-9881   \\
M. Chertkov, https://orcid.org/0000-0002-6758-515X \\
J. Schumacher, https://orcid.org/0000-0002-1359-4536 }

\backsection[Author contributions]{All authors designed the research. M.E. and R.J.S. conducted the direct numerical simulations and generated the data. All authors discussed and analysed the results and wrote the manuscript.}

\appendix
\section{Evaluation of Lagrangian particle accelerations}
\label{sec:Appendix_Evaluation_of_Lagrangian_particle_accelerations}

\subsection{\rev{Acceleration filtering via cubic smoothing splines}}
\label{sec:Appendix_Acceleration_filtering_via_cubic_smoothing_splines}

%------------------------------------------------------------------------------------------
\begin{figure}
\centering
\includegraphics[width = \textwidth]{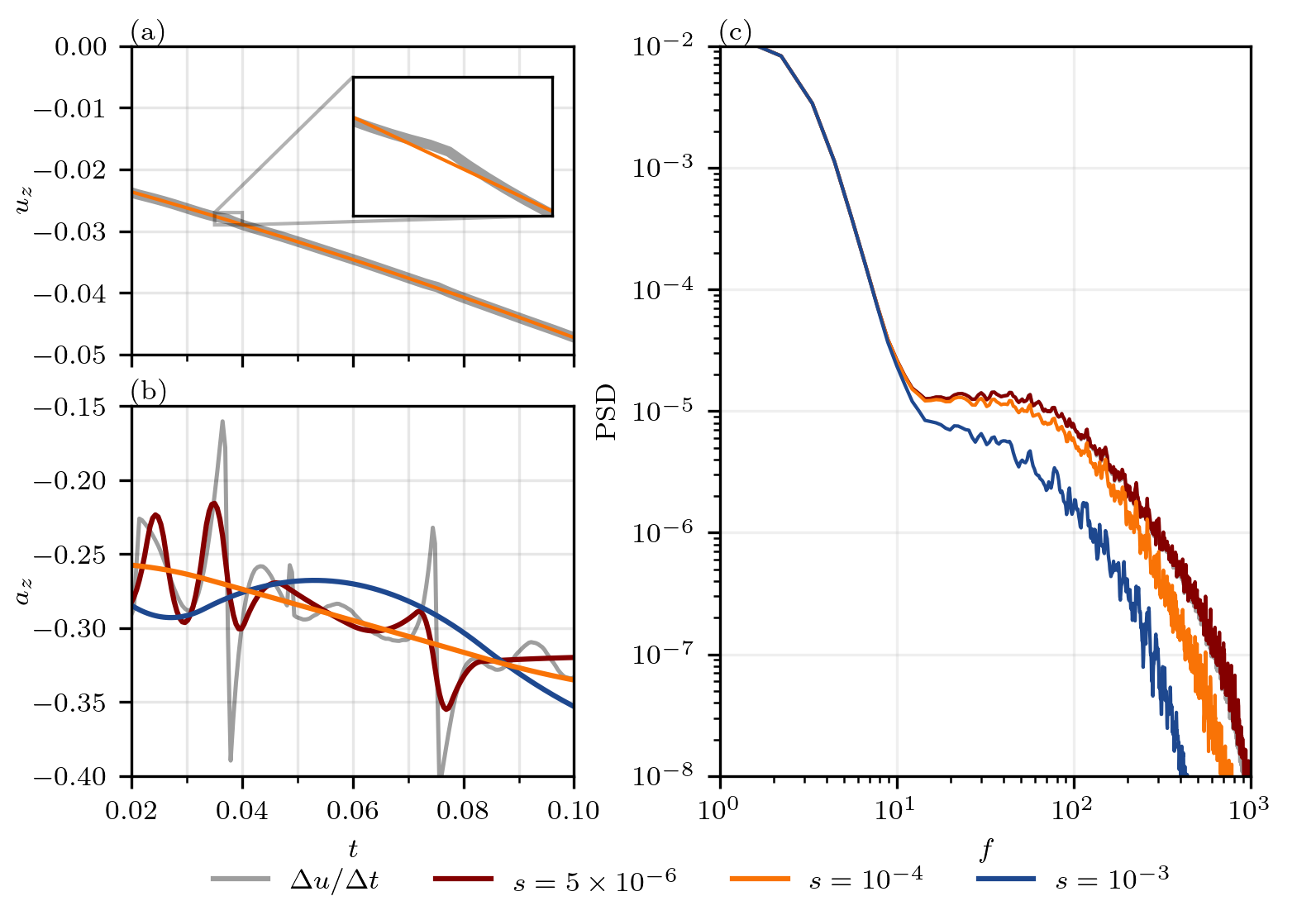}
\caption{\justifying{
Cubic smoothing splines for Lagrangian particle accelerations at $\Ra=10^9$. The Galerkin formulation provides only $C^0$-continuity of the particle velocities (a), which propagate into severe, unphysical spikes in discrete particle acceleration (grey, b). Applying a cubic smoothing spline to the particle velocities (orange, a) effectively eliminates these numerical aliasing effects. (c) Ensemble power spectral density (PSD) is used to calibrate the smoothness parameter $s$. Choosing $s$ too small ($s=5\times10^{-6}$, oxblood) overfits the data and fails to reduce the high-frequency numerical spikes, while choosing too large ($s=10^{-3}$, blue) underfits the data, dampening true physical intermittency. The optimal smoothing parameter ($s=10^{-4}$, orange) truncates the numerical aliasing effects while preserving the true, underlying physics.
}}
\label{fig:u_a_PSD}
\end{figure}
%------------------------------------------------------------------------------------------

Tracking accelerations of Lagrangian particles using the spectral element framework introduces inherent numerical challenges. Because the underlying Galerkin formulation provides only $C^0$-continuity across element boundaries, spatial velocity gradients are fundamentally discontinuous at these interfaces. Consequently, as fluid particles traverse the element boundaries, a simple interpolation and discrete temporal differentiation ($\bm{a}=\Delta \bm{u} / \Delta t$) would result in severe, unphysical acceleration spikes. Higher polynomial orders $p$ exacerbate this issue due to steeper polynomial gradients between tighter clusters of Gauss--Lobatto--Legendre (GLL) nodes towards the spectral element boundaries. 

To decouple signatures of the numerical method from true physical intermittency, a specific temporal tracking and filtering approach is required for the determination of Lagrangian accelerations that is discussed in the following. \rev{Experimental studies of Lagrangian acceleration demand similar approaches, e.g., the differentiating filter-kernel introduced by \citet{Mordant2004}}. Notably, an $L_2$ projection within the Eulerian solver enforces global continuity for the spatial gradients ($\nabla \bm{u}$), yielding mathematically robust fields required for the topological $Q$--$R$ analysis.

To circumvent resolution-based effects across different $\Ra$, we mapped the fields first onto grids of approximately the same total number of GLL points, strictly maintaining a uniform polynomial order of $p=5$. Particle velocities were instantaneously saved at every integration time step to capture the true underlying physical dynamics. To decouple true physical intermittency from the signatures by the spectral element method, we regularise particle velocities by fitting a $C^2$-continuous cubic smoothing spline through each trajectory. We found this approach to be superior to local polynomial regressions (e.g., Savitzky--Golay filters), which overfit these numerical discontinuities, and low-pass filtering, which introduces spectral ringing (Gibbs phenomenon) at the element boundaries. The cubic smoothing spline employed here minimises the cost function
\begin{equation}
    \label{eq:univariate_spline}
    L = \sum_{i=1}^{N} \left( u(t_i) - S(t_i) \right)^2 + s \int_{t=t_1}^{t_N} \left( \frac{d^2S}{dt^2}\right)^2dt.
\end{equation}
The first term represents the residual sum of squares, penalising deviations from the raw velocity data $u(t_i)$. The second term penalises large, unphysical bending of the trajectory. The smoothness parameter $s$ weights the two terms, dictating the trade-off between data fidelity and trajectory smoothness. 

% Splines
Figure \ref{fig:u_a_PSD} illustrates the application of this filtering approach for $\Ra=10^{9}$ using cubic smoothing splines of varying $s$. As an example particle crosses an element boundary at $t \approx 0.035$, its velocity trajectory exhibits a distinct kink (grey, figure \ref{fig:u_a_PSD}a), which propagates into a strong, non-physical numerical spike in its discrete acceleration (grey, figure \ref{fig:u_a_PSD}b). While the applied cubic smoothing spline of $s=5 \times 10^{-6}$ (oxblood) closely follows the original data, the induced numerical spike is dampened, but still present. In contrast, $s=10^{-3}$ (blue) smoothens the peak out entirely and thus fails to capture the physical reality of the data.

%------------------------------------------------------------------------------------------
\begin{table}
\begin{center}
\def~{\hphantom{0}}
\begin{tabular}{lcc}
$\Ra$ & $\langle \Delta t \rangle \times 10^{-3}$ & $s$ \\[3pt]
\hline
$10^5$    & $3.75$ & $10^{-7}$ \\
$10^6$    & $2.11$ & $10^{-7}$ \\
$10^7$    & $1.11$ & $10^{-7}$ \\
$10^8$    & $0.61$ & $10^{-5}$ \\
$10^9$    & $0.44$ & $10^{-4}$ \\
$10^{10}$ & $0.19$ & $10^{-4}$ \\
\end{tabular}
\caption{Spline parameters for Lagrangian acceleration. The average integration time step size $\langle \Delta t \rangle$ and the calibrated cubic smoothing spline parameter $s$ are reported.}
\label{tab:splines}
\end{center}
\end{table}
%------------------------------------------------------------------------------------------

% PSD
Setting the smoothness parameter $s$ properly is thus fundamental to obtaining physically correct data. To achieve this proper trade-off, we transformed the data into the frequency domain and computed the ensemble-averaged power spectral density (PSD) of the raw data and splines of different $s$ for $500$ randomly chosen particles. Here, it is crucial that lower frequencies (representing true, large-scale acceleration events) and moderate frequencies (representing physical intermittent acceleration events) closely follow the spectrally obtained numerical data, while very high frequencies (representing the numerical spikes) are cut off. Figure \ref{fig:u_a_PSD}(c) illustrates this for $\Ra=10^{9}$: using a stiff spline of $s=10^{-3}$ (blue) fails to capture the true intermittency of turbulence by underfitting the data. In contrast, $s=5\times10^{-6}$ (oxblood) follows the numerical spikes too closely, overfitting the data. Here, $s=10^{-4}$ (orange) optimally strikes the balance, tracking the true data closely at smaller frequencies while truncating the numerical artifacts. Table \ref{tab:splines} summarises the calibrated smoothness parameters $s$ used for all $\Ra$.

Subsequently, particle acceleration is computed analytically by taking the first derivative of the spline, $\bm{a}=d \bm{S}/dt$. This eliminates noise amplification inherent to discrete finite-difference calculations over variable time steps. We also excluded the initial $10$ integration time steps from the data to eliminate seeding biases at $t=0$, where particles initialised with $\bm{u}=\bm{0}$ cause unphysical accelerations. The result is a set of robust, physically rigorous, and continuous particle acceleration trajectories, which are required for well-resolved acceleration statistics.

\subsection{\rev{Higher-order moments of Lagrangian acceleration}}
\label{sec:Appendix_Higher-order_moments_of_Lagrangian_acceleration}

%------------------------------------------------------------------------------------------
\begin{figure}
\centering
\includegraphics[width = \textwidth]{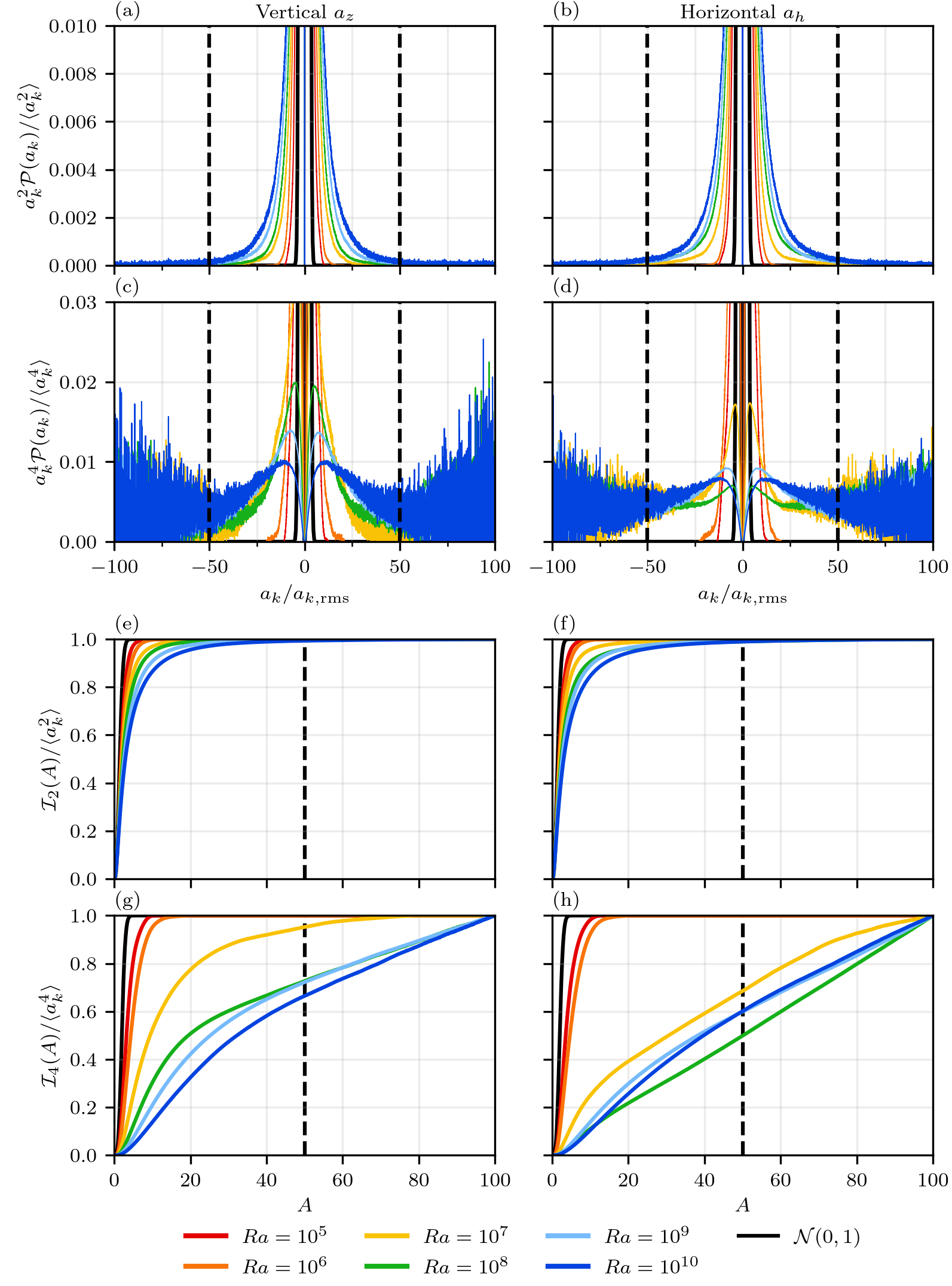}
\caption{\justifying{
\rev{Moment contributions and cumulative moment integrals of Lagrangian accelerations. The relative contributions of variance $a_k^2 \mathcal{P}(a_k) /\langle a_k^2 \rangle$ and kurtosis $a_k^4 \mathcal{P}(a_k) /\langle a_k^4 \rangle$ are shown in vertical ($z$) and horizontal ($h$) directions in panels (a,c) and (b,d), respectively. While the variance is dominated by the centre of the PDFs ($a_k \lesssim 30 \, a_{\mathrm{rms}}$), kurtosis fails to converge and is dominated by the far tails $a_k \gtrsim 50 \, a_{\mathrm{rms}}$ for $\Ra \gtrsim 10^8$. Normalised cumulative moment integrals $\mathcal{I}_n (A) = \int_{0}^A{ a_k^n \mathcal{P}(a_k) da_k}$ in panels (e--h) confirm this lack of convergence. Thus, moment calculation is restricted to $|a_k| \leq 50a_{k,\mathrm{rms}}$ (dashed black line).
}}}
\label{fig:accPDFs_moment_contributions}
\end{figure}
%------------------------------------------------------------------------------------------

% Introduction
\rev{Because the applied filtering algorithm ensures $C^2$-continuous cubic velocity smoothing splines $\bm{S}$, thus eliminating non-physical numerical spikes in Lagrangian particle acceleration ($\bm{a}=d \bm{S}/dt$), the far tails of the PDFs in figure \ref{fig:accPDFs} represent real, highly intermittent acceleration events. Although their rate of occurrence is small, their sheer magnitude is enough to dominate higher-order moments. For example, an isolated spike of $100 \, a_{\mathrm{rms}}$ contributes to the fourth-order sum by a factor of $10^8$. Investigating the relative contributions of these magnitudes to the acceleration PDFs is therefore necessary to gauge statistical convergence.}

% Contributions
\rev{Figure \ref{fig:accPDFs_moment_contributions} shows the relative contributions of variance $a_k^2 \mathcal{P}(a_k) /\langle a_k^2 \rangle$ and kurtosis $a_k^4 \mathcal{P}(a_k) /\langle a_k^4 \rangle$ in vertical ($z$, figure \ref{fig:accPDFs_moment_contributions}a,c), and horizontal ($h$, figure \ref{fig:accPDFs_moment_contributions}b,d) directions. While variance is robustly captured by the centre of the PDFs ($a_k \lesssim 30 \, a_{\mathrm{rms}}$) for all $\Ra$, kurtosis exhibits strong divergence: while the contribution decays initially, it develops a local minimum at $a_k \approx 50 \, a_{\mathrm{rms}}$ before rising sharply again for $\Ra \gtrsim 10^8$. A similar inflection point has also been observed in the kurtosis contribution in the work of \citet{Schumacher2009} at $\Ra=1.2 \times 10^8$. Computing the normalised cumulative moment integrals $\mathcal{I}_n (A) = \int_{0}^A{ a_k^n \mathcal{P}(a_k) da_k}$ demonstrates this further in figure \ref{fig:accPDFs_moment_contributions}(e--h): while the lower $\Ra \lesssim  10^6$ plateau quickly, $\Ra \gtrsim 10^8$ fail to converge entirely -- indicating that the moment is dominated by the far tails of the PDFs. Because Lagrangian accelerations are extremely intermittent, convergence of kurtosis is notoriously difficult to obtain. \citet{Voth2002} similarly failed to achieve moment convergence for $5\times10^6$ events, accompanying their resulting values for kurtosis with large error bars. In the present work, even an ensemble of more than $2 \times 10^{10}$ events is insufficient to achieve full convergence. \citet{Yakhot2005} claim that convergence of Lagrangian acceleration kurtosis may be fundamentally unobtainable.}

% Kurtosis values
\rev{Because finite ensembles cannot statistically resolve the far-tail asymptote, calculating the global kurtosis over the entire domain introduces a severe statistical, sample-size-dependent bias. In the present study, we thus decided to truncate the calculation of higher-order moments at the point where the tails of the acceleration PDFs become increasingly dominated by these rare, extreme events, at $|a_k| \le 50 \, a_{k,\mathrm{rms}}$ (dashed black line in figure \ref{fig:accPDFs_moment_contributions}). The necessity of this approach is exemplified by the case of $\Ra=10^{10}$: computing kurtosis for the global domain -- including the extreme far-tail events -- leads to $K_h(\Ra=10^{10}) = 1466 \pm 716$, compared to $K_h(\Ra=10^{10}) = 78.5 \pm 0.3$ for the restricted domain, highlighting not just the magnitude with which these rare events contribute, but also the standard error they produce.}

% Vertical vs. horizontal
\rev{Finally, the kurtosis contributions in vertical and horizontal direction (figure \ref{fig:accPDFs_moment_contributions}c,d) highlight stronger far-tail weighting in the horizontal direction. This underlines the finding of \citet{Schumacher2008}, whereby horizontal accelerations are more intermittent than vertical accelerations in a laterally unrestricted domain.}

\section{\rev{Sensitivity analyses}}
\label{sec:Appendix_sensitivity_analyses}

\subsection{\rev{Sensitivity of dust-devil-like convective vortices}}
\label{subsec:Appendix_Sensitivity_of_dust-devil-like_convective_vortices}

\begin{figure}
\centering
\includegraphics[width = \textwidth]{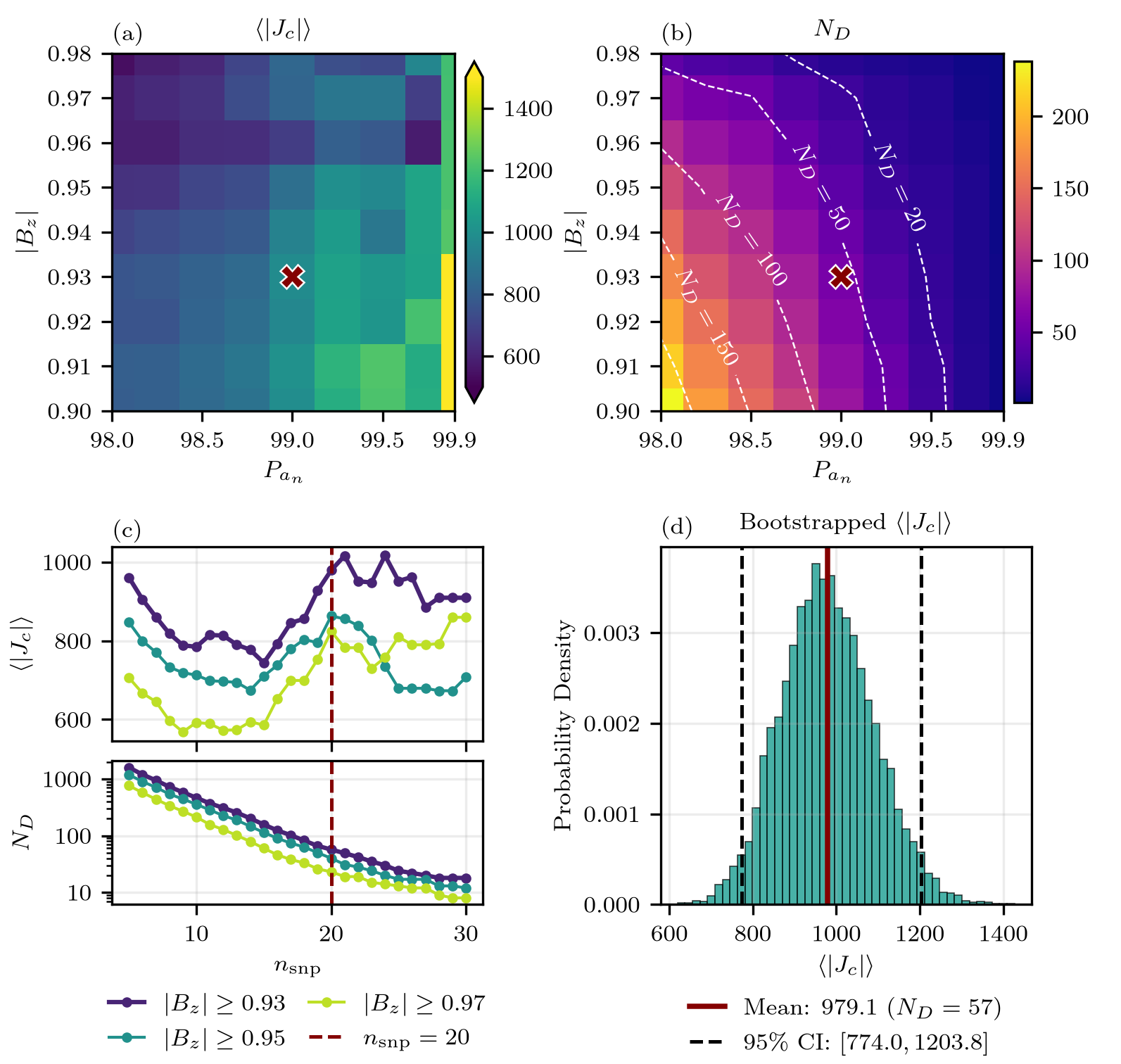}
\caption{\justifying{
\rev{Sensitivity analysis of dust-devil-like convective vortices at $\Ra=10^{10}$. The chosen parameter combination ($\Pan = 99.0\%$, $\Bz \geq 0.93$, $\nsnp \geq 20$ and $\rfac=4$) is marked in oxblood. (a,b) Heatmaps of $\Jc$ and $N_D$: very strict alignment thresholds ($\Bz \gtrsim 0.96$) reduce $\Jc$ by excluding tilted dust devils, whereas extreme acceleration thresholds ($\Pan \gtrsim 99.5 \%$) drastically reduce the ensemble ($N_D<20$). (c) The required number of consecutive snapshots $\nsnp$ (spaced by $0.01 \tauf$) acts as a high-pass temporal threshold: $\Jc$ exhibits a coherent dust-devil plateau at $\nsnp \gtrsim 20$, excluding both violent short-term fluctuations ($\nsnp \lesssim 10$) and transitional dynamics ($10 \lesssim \nsnp \lesssim 20$). (d) Bootstrapped probability density across $10,\!000$ resamples of $N_D=57$ events at the chosen parameter combination exhibits a Gaussian distribution (Mean $\Jc = 979.1$, $95\%$ CI: $[774.0, 1203.8]$) and confirms that convective heat fluxes may reach nearly $8 \NuVt$, representing a robust physical signature of dust devils rather than statistical outliers.
}}}
\label{fig:SA_dust_devils}
\end{figure}

\rev{The extraction of dust-devil-like convective vortices (`dust devils') requires balancing the \emph{quantity} (logarithmic number of extracted structures, $\log_{10}(N_D)$) and \emph{quality} (ensemble-averaged absolute convective heat flux, $\Jc$) of the identified events. To quantify the relative importance and non-linear coupling of the four extraction criteria introduced in Section \ref{subsec:Velocity_gradient_invariant_analysis_and_dust-devil-like_vortices}, we perform a functional Analysis-of-Variance (ANOVA) decomposition of the total variance $\mathrm{Var}(Y)$. For each input parameter $X_i$, we compute the first-order Sobol index $S_{1,i}$ (measuring the isolated main effect $X_i$ on $Y$) and the total-order Sobol index $S_{T,i}$ (measuring the main effect of $X_i$ alongside all interactions with the remaining parameters $\bm{X}_{\sim i}$ on $Y$) \citep{Sobol2001}:}  
\begin{equation}
\label{eq:sobol_indices}
\rev{S_{1,i} = \frac{\mathrm{Var}\left(\mathbb{E}[Y \mid X_i]\right)}{\mathrm{Var}(Y)}\,, \qquad S_{T,i} = 1 - \frac{\mathrm{Var}\left(\mathbb{E}[Y \mid \bm{X}_{\sim i}]\right)}{\mathrm{Var}(Y)}\,,}
\end{equation}
\rev{where the difference $S_{T,i} - S_{1,i}$ measures cross-parameter interaction effects and $\mathbb{E}$ represents the expectation value. We evaluate the four extraction criteria across a balanced full-factorial Cartesian grid of $9^4 = 6,\!561$ parameter permutations, guaranteeing exactness of the ANOVA decomposition:}
\begin{itemize}
    \item \rev{$\Pan \in \{98.0, 98.25, 98.5, 98.75, 99.0, 99.25, 99.5, 99.75, 99.9\}$}
    \item \rev{$\Bz \in \{0.90, 0.91, 0.92, 0.93, 0.94, 0.95, 0.96, 0.97, 0.98\}$}
    \item \rev{$\nsnp \in \{5, 8, 11, 14, 17, 20, 23, 26, 29\}$}
    \item \rev{$\rfac \in \{2.0, 2.5, 3.0, 3.5, 4.0, 4.5, 5.0, 5.5, 6.0\}$}
\end{itemize}
\rev{We note that the criterion \emph{Locality} ($z \le 2 \delta_T$) remains a fixed parameter as dust devils originate close to the thermal boundary layer by nature. Table \ref{tab:sobol_indices} reports the Sobol indices, complemented by a sensitivity analysis in figure \ref{fig:SA_dust_devils}, where markers in oxblood reflect our chosen parameter combination ($\Pan = 99.0\%$, $\Bz \geq 0.93$, $\nsnp \geq 20$ and $\rfac=4$). Interaction effects for the logarithmic yield $\log_{10}(N_D)$ are negligible across all parameters given very small $S_T - S_1 \leq 0.012$. This confirms that these criteria exhibit a multiplicative effect, resulting in diagonal iso-contours in figure \ref{fig:SA_dust_devils}(b) and an exponential decay in figure \ref{fig:SA_dust_devils}(c, bottom), where temporal persistence $\nsnp$ acts as the primary filter ($S_1 = 0.506$). Crucially, the spatial clustering radius $\rfac$ exhibits nearly no influence on either $N_D$ ($S_T = 0.006$) or $\Jc$ ($S_T = 0.023$), proving that the extracted vortices are spatially compact and insensitive to our choice of $\rfac$.}

%------------------------------------------------------------------------------------------
\begin{table}
  \color{black}
  \begin{center}
  \def~{\hphantom{0}}
  \begin{tabular}{lcccccc}
  \hline
  Parameter & \multicolumn{3}{c}{$\langle |J_c| \rangle$} & \multicolumn{3}{c}{$\log_{10}(N_D)$} \\
  \cmidrule(lr){2-4} \cmidrule(lr){5-7}
   & $S_1$ & $S_T$ & $S_T - S_1$ & $S_1$ & $S_T$ & $S_T - S_1$ \\
  \midrule
  $P_{a_n}$ & 0.499 & 0.770 & 0.271 & 0.415 & 0.427 & 0.012 \\
  $|B_z|$ & 0.167 & 0.333 & 0.166 & 0.061 & 0.066 & 0.005 \\
  $n_{\mathrm{snp}}$ & 0.013 & 0.271 & 0.259 & 0.506 & 0.519 & 0.013 \\
  $r_{\mathrm{fac}}$ & 0.013 & 0.023 & 0.010 & 0.004 & 0.006 & 0.002 \\
  \hline
  \end{tabular}
  \caption{\rev{Analysis of variance (ANOVA) of dust-devil extraction criteria across a $9^4$ full-factorial parameter grid. Exact first-order ($S_1$) and total-order ($S_T$) Sobol indices are reported alongside the interaction index ($S_T - S_1$) to quantify the main effects and non-linear coupling of each parameter on the mean absolute convective heat flux ($\Jc$) and logarithmic event yield ($\log_{10}(N_D)$).}}
  \label{tab:sobol_indices}
  \end{center}
\end{table}
%------------------------------------------------------------------------------------------

%------------------------------------------------------------------------------------------
%------------------------------------------------------------------------------------------

\rev{Centripetal acceleration ($\Pan$, $S_T = 0.770$) and kinematic alignment ($\Bz$, $S_T = 0.333$) mainly drive convective heat flux intensity $\Jc$. Notably, $\nsnp$ exhibits a large interaction effect on $\Jc$ ($S_T - S_1 = 0.259$) while the main effect is small ($S_1 = 0.013$). The top panel of figure \ref{fig:SA_dust_devils}(c) reveals non-monotonic behaviour of $\Jc$ with $\nsnp$, which is shared across multiple $\Bz$. At small values of $\nsnp \lesssim 10$, violent, short-lived events cause large convective heat fluxes, whereas the transitional regime ($10 \lesssim \nsnp \lesssim 18$) reveals smaller $\Jc$. Crucially, a coherent plateau establishes at $\nsnp \gtrsim 20$, reflecting the sustained dust devils.}

\rev{Figure \ref{fig:SA_dust_devils}(a,b) highlights the strong interaction between $\Bz$ and $\Pan$: very strict vertical alignment ($\Bz \gtrsim 0.96$) reduces $\Jc$ by discarding tilted vortices, which, however, represents a key feature of dust devils as illustrated in figure \ref{fig:Tornado_cloud}(c). Conversely, large acceleration thresholds ($\Pan \gtrsim 99.5\%$) fracture trajectories and drastically reduce the ensemble ($N_D<20$).}

\rev{Our chosen parameter combination strikes the balance between these physical constraints: $\Bz \geq 0.93$ (permitting vertical tilts up to $21.6^\circ$) is large enough to capture tilted dust devils while excluding horizontal vortex tubes. $\nsnp \geq 20$ ($0.20\tauf$) is placed directly at the onset of the coherent plateau in figure \ref{fig:SA_dust_devils}(c, top), eliminating short turbulent fluctuations by acting as a high-pass filter without discarding sustained vortices. $\Pan = 99.0 \%$ isolates vigorous rotational events while retaining stability to small parameter variations (figure \ref{fig:SA_dust_devils}a,b). Bootstrapping $\Jc$ across $10,\!000$ resamples of the $N_D = 57$ events, see figure \ref{fig:SA_dust_devils}(d), yields a symmetric Gaussian distribution with a mean of $\Jc = 979.1$ ($95\%$ CI: $[774.0, 1203.8]$). This confirms that the extracted ensemble is not driven by individual outlier-events, but rather captures the physical signature of dust devils -- significantly driving local convective heat fluxes to nearly $8$ times the global Eulerian mean $\NuVt$ (see table \ref{tab:Series_scales}).}

%------------------------------------------------------------------------------------------
\subsection{\rev{Sensitivity of initial updraft conditioning}}
\label{subsec:Appendix_Sensitivity_of_initial_updraft_conditioning}

%------------------------------------------------------------------------------------------
\begin{figure}
\centering
\includegraphics[width = \textwidth]{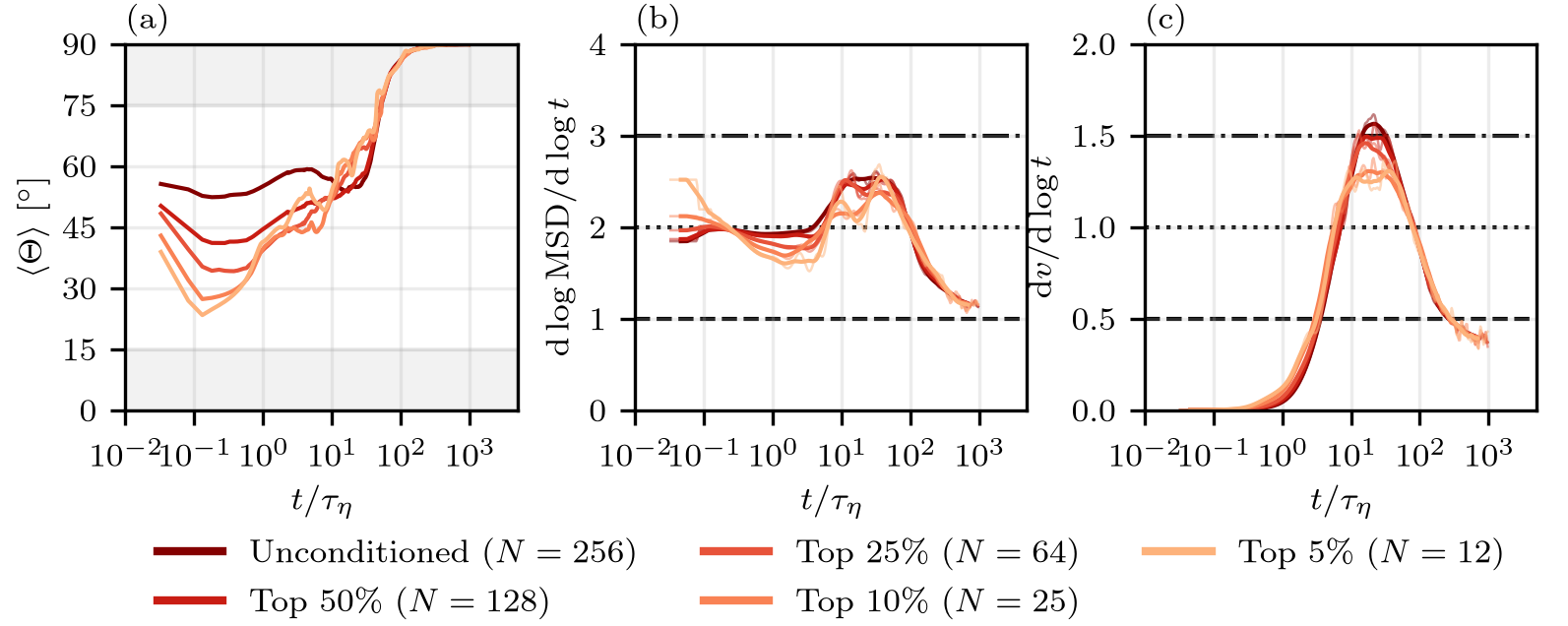}
\caption{\justifying{
\rev{Sensitivity analysis of initial updraft conditioning. Particle clouds seeded at $\Ra=10^8$ with $r_0 = 4\eta_K$ in the plume root zone ($z_0 = 2\delta_T$) are contrasted across different thresholds of $u_{z,0}$-conditioning. More stringent updraft conditioning yields earlier, larger drops in vertical alignment $\Theta$ in panel (a), which leads to lower peak scaling exponents in the local slope analysis of the logarithmic geometric mean radius $v$ in panel (c) --- a consequence of kinematic `catch-up'. A direct comparison between the local slope analyses of mean squared dispersion $\MSD$ in panel (b) and logarithmic geometric mean radius $v$ in panel (c) of the figure underscores the superiority of the latter metric as a gauge of the volumetric evolution of a particle cloud.
}}}
\label{fig:Cloud_uz_sensitivity}
\end{figure}
%------------------------------------------------------------------------------------------

\rev{While more stringent initial updraft conditioning ($u_{z,0}$) samples stronger thermal updrafts, it reduces the underlying ensemble size in lockstep. To investigate the implication of different $u_{z,0}$-conditioning thresholds, we perform a sensitivity analysis for clouds seeded at $\Ra=10^8$ with $r_0 = 4 \eta_K$ in the plume root zone ($z_0 = 2\delta_T$) across the unconditioned ensemble and the top percentages $50\%$, $25\%$, $10\%$ and $5\%$ (corresponding to $N_c \in \{256, 128, 64, 25, 12\}$ clouds). Figure \ref{fig:Cloud_uz_sensitivity} contrasts vertical alignment $\Theta$ in panel (a), the local slope analyses of mean squared dispersion $\MSD$ in panel (b), and logarithmic geometric mean radius $v$ in panel (c).}

\rev{More stringent conditioning thresholds lead to earlier, stronger drops in $\Theta$, directly reflecting stronger and immediate thermal plume entrainment. A comparison between the local slope analyses of $\MSD$ and $v$ underscores the superiority of the logarithmic geometric mean radius in capturing the volumetric evolution of a particle cloud: since $\MSD$ is inherently volatile, a clear distinction between conditioning thresholds is elusive. Conversely, the local slope analysis of $v$ reveals earlier onsets of expansion and smaller peak scaling exponents for stricter thresholds. As detailed in Section \ref{subsec:Spatio-temporal_evolution_of_cloud_ensembles}, this behaviour is a direct consequence of kinematic `catch-up' effects.}

\rev{To balance statistical convergence of the ensemble with physically meaningful updraft conditioning thresholds, we select the top quartile ($25\%$ of $u_{z,0}$, $N_c = 64$ clouds) as our reference threshold in Section \ref{subsec:Spatio-temporal_evolution_of_cloud_ensembles}.}

\bibliographystyle{jfm}

\end{document}